\documentclass[11pt]{article}%

\usepackage{amsmath,amssymb,amsfonts}
\usepackage{calc}
\usepackage{dsfont}
\usepackage{hyperref}
\usepackage{slashed}
\usepackage{cite}
\usepackage{epsfig}

\newcommand{\beq}{\begin{eqnarray}}
\newcommand{\eeq}{\end{eqnarray}}

\newcommand{\s}{\newline \vspace*{-3.5mm}}

\makeatletter
\def\section{\@startsection {section}{1}{\z@}{-3.5ex plus -1ex minus -.2ex}{2.3ex plus .2ex}{\large\bf}}
\def\subsection{\@startsection{subsection}{2}{\z@}{-3.25ex plus -1ex
minus -.2ex}{1.5ex plus .2ex}{\normalsize\bf}}
\makeatother
\newcommand{\captionfonts}{\small}
\makeatletter  
\long\def\@makecaption#1#2{%
  \vskip\abovecaptionskip
  \sbox\@tempboxa{{\captionfonts #1: #2}}%
  \ifdim \wd\@tempboxa >\hsize
    {\captionfonts #1: #2\par}
  \else
    \hbox to\hsize{\hfil\box\@tempboxa\hfil}%
  \fi
  \vskip\belowcaptionskip}
\makeatother   

\newcommand{\f}[2]{\frac{#1}{#2}}
\newcommand{\sssty}[1]{\scriptscriptstyle#1}

\graphicspath{{fig/}}

\begin{document}

\thispagestyle{empty}

\begin{center}

\hfill CERN-PH-TH/2012-176 \\
\hfill KA-TP-26-2012 \\
\hfill SFB/CPP-12-43 \\
\hfill ZU-TH 09/12 \\
\hfill LPN12-062

\begin{center}

\vspace{1.7cm}

{\LARGE\bf Higgs Low-Energy Theorem (and its corrections) \\
[3mm] in Composite Models} 
\end{center}

\vspace{.8cm}

{\bf M.~Gillioz$^{\,a}$}, {\bf R.~Gr\"ober$^{\,b}$}, {\bf C.~Grojean$^{\,c}$}, {\bf
  M.M\"uhlleitner$^{\,b}$} and {\bf E.~Salvioni$^{\,c,\,d}$} \\

\vspace{0.8cm}

${}^a\!\!$
{\em {Institut f\"ur Theoretische Physik, Universit\"at Z\"urich, CH-8051 Z\"urich, Switzerland}
}\\
${}^b\!\!$
{\em {Institut f\"ur Theoretische Physik, Karlsruhe Institute of Technology, D-76128 Karlsruhe, Germany}
}\\
${}^c\!\!$
{\em {Theory Division, Physics Department, CERN, CH-1211 Geneva 23, Switzerland}}\\
${}^{d}\!\!$
{\em {Dipartimento di Fisica e Astronomia, Universit\`a di Padova and INFN, Via Marzolo 8, \\I-35131 Padova, Italy}}

\end{center}

\vspace{0.4cm}

\centerline{\bf Abstract}
\vspace{2 mm}
\begin{quote}
The Higgs low-energy theorem gives a simple and elegant way to estimate the couplings of the Higgs boson to massless gluons and photons induced by loops of heavy particles. We extend this theorem to take into account possible nonlinear Higgs interactions as well as new states resulting from a strong dynamics at the origin of the breaking of the electroweak symmetry. We show that, while it approximates with an accuracy of order a few percents single Higgs production, it receives corrections of order $50\%$ for double Higgs production. A full one-loop computation of the $gg\to hh$ cross section is explicitly performed in MCHM5, the minimal composite Higgs model based on the $SO(5)/SO(4)$ coset with the Standard Model fermions embedded into the fundamental representation of $SO(5)$. In particular we take into account the contributions of all fermionic resonances, which give sizeable (negative) corrections to the result obtained considering only the Higgs nonlinearities. Constraints from electroweak precision and flavor data on the top partners are analyzed in detail, as well as direct searches at the LHC for these new fermions called to play a crucial role in the electroweak symmetry breaking dynamics.
\end{quote}

\newpage
\renewcommand{\thefootnote}{\arabic{footnote}}
\setcounter{footnote}{0}

\section{Introduction}

There is growing evidence that a Higgs boson is the agent of electroweak symmetry breaking (EWSB)~\cite{HiggsBirth}. And one of the most pressing questions is to uncover the true nature of this Higgs boson: is it the elementary scalar field of the Standard Model? Is it part of a supermultiplet? Is it a composite scalar emerging as a bound state from a  strongly coupled sector? Theoretical arguments based on naturalness considerations tend to favor one of the two latter scenarios. But the LHC experiments have now opened a new data-driven era and the first experimental indications might come from possible deviations in the measurements of the Higgs couplings~\cite{Higgsfits} compared to the ones predicted by the Standard Model (SM) that are unambiguously fixed by the value of the Higgs mass itself. However, both supersymmetric and composite Higgs bosons  
in the decoupling limits can be arbitrarily close to the SM Higgs at the energy scale currently probed by the LHC and deciphering the different scenarios from one another might require more luminosity than the one currently accumulated.  A more unambiguous answer would come from a direct observation of additional particles: supersymmetric partners of the SM particles or additional resonances of the strong sector.
In both scenarios, new particles in the top sector have a special status: in supersymmetric models, the stops cannot be too heavy without destabilizing the weak scale~\cite{Barbieri:1987fn}; the top partners in composite models are responsible for generating the potential of the would-be Goldstone Higgs boson and, as recently noticed in Refs.~\cite{lighttoppartners}, they have to be lighter than about 700~GeV to naturally accommodate a Higgs boson as light as 125~GeV. Even if the actual numbers are model-dependent, this conclusion is rather generic and certainly calls for improving the ongoing dedicated direct searches for the top partners~\cite{futuresearches}.

In principle, indirect information on these top partners could also be obtained in Higgs physics.
It is indeed quite ironic that, while the Higgs boson is supposed to be at the origin of the masses of all elementary particles,  the currently most sensitive channels are the ones that involve massless particles, {\it i.e.} particles with no direct coupling to the Higgs boson: the gluons for its production and the photons for its decay. Clearly these processes appear only at the loop level in the SM and are therefore potentially sensitive to new states circulating in the loops. The structure of the Higgs couplings to photons and gluons are beautifully captured by the background field method known as Low-Energy Theorem (LET)~\cite{Ellis:1975ap, Kniehl:1995tn} and the effects of the top partners, or any new particles, are encoded by their sole contributions to the QED and QCD beta functions. In composite Higgs models, the corrections due to the  Higgs compositeness to the SM gluon fusion production cross section, $\sigma( gg\to h)$, and to the SM decay width into two photons, $\Gamma (h\to \gamma \gamma)$, were estimated in Ref.~\cite{silh} to be generically of the order of $\xi=(v/f)^2$, where $v\sim$~246~GeV is the weak scale and $f$ is the characteristic scale of the strong sector (the equivalent of $f_\pi$ in QCD). From dimensional arguments, the additional corrections due to the top partners should be of the order of
 $(m_t/m_T)^2$, $m_{t,T}$ being respectively the mass of the  top and the typical mass scale of his partners, {\it i.e.} a correction potentially as big as the one originating from the strong dynamics itself when the top partners are around 700~GeV. However, it was quickly realized~\cite{singlehiggsindep, Low:2010mr, Azatov:2011qy} that these leading-order corrections from the top partners actually cancel and that they only give a contribution that we shall estimate (see Section~\ref{sec:singlehprod}) to scale like $\xi (m_h/m_t)^2/10$, {\it i.e.} one order of magnitude smaller than the strong dynamics contribution, sweeping any hope to learn anything about the top partners from the measurement of the Higgs production by gluon fusion and leaving only the Higgs production in association with a top-antitop pair~\cite{ttbarH} as a place to indirectly look for new physics in the first LHC run. A proper effective description of the top partners/fermionic resonances would be needed to study this promising channel.
 
The second LHC run will certainly increase the sensitivity of the direct searches for top partners.
But it will also open new possibilities to indirectly probe the top partner sector in Higgs physics by exploring multi-Higgs production. A generalization of the LET exists for double Higgs production in the SM, and it is known to give a reasonably good estimate of the total rate within 20\% accuracy for a light Higgs~\cite{Glover:1987nx, Plehn:1996wb} but it badly fails to reproduce the differential distributions~\cite{Baur:2002rb}. The extension of the LET to strong EWSB models is not totally trivial since, in the SM, there is a strong destructive interference between two contributions and therefore strong dynamics  that gives rise to a third new contribution can have order-one effects on the double Higgs production by gluon fusion, $gg\to hh$, as already noticed in Refs.~\cite{Dib:2005re, doubleh2, ContinoETAL}. In an explicit composite Higgs model, we are going to compare the LET results to an explicit one-loop computation taking into account all the contributions from the fermionic resonances in the top sector. We will find that the LET is less accurate than in the SM, and underestimates the full result by up to $50\%$. On the other hand the pure strong dynamics effects arising due to the sole Higgs nonlinearities are overestimating the production rate which receives $\mathcal{O}(30\%)$ corrections from the top partners. 

The outline of our paper is as follows. In Section~\ref{sec:paramet}, we first present two equivalent effective Lagrangians, the linearly realized strongly-interacting light Higgs Lagrangian~\cite{silh} as well as a chiral Lagrangian where the $SU(2)_L\times U(1)_Y$ SM gauge symmetry is nonlinearly realized~\cite{doubleh1}, to describe Higgs physics at the LHC and we relate these effective Lagrangians to explicit composite Higgs models.
In Section~\ref{sec:LET}, we recall the LET in the SM and extend it to composite Higgs models, reproducing the often heard results that  the gluon fusion production cross sections can be significantly reduced by up to 50\%$\div$70\% compared  to the SM production.
We also re-discuss within the LET approximation the cancellation  of the top partner contributions to single Higgs production by gluon fusion and we estimate the size of the corrections to the LET results.
We finally derive the LET for double Higgs production by gluon fusion and show that a factor 2$\div$3 enhancement over the SM can be obtained in specific examples. This LET enhancement factor is nonetheless smaller  than the one obtained considering only the pure Higgs nonlinearities effects.
Section~\ref{sec:partners} is devoted to the top partners in an explicit minimal composite Higgs model, MCHM5, based on the coset space $SO(5)/SO(4)$ with SM fermions embedded into fundamental representations of $SO(5)$.
We first discuss in detail the constraints on the masses and couplings of these new fermions 
coming from electroweak (EW) precision measurements and from flavor physics. We then present exhaustive constraints from direct searches for these top partners using the current Tevatron and LHC data.
In Section~\ref{singleprodMCHM5}, a detailed computation of the contribution of the top partners to $\sigma (gg \to h)$ is presented, confirming that the LET gives a good approximation and confirming the scaling behavior of the corrections obtained in Section~\ref{sec:LET}.
Finally, Section~\ref{double Higgs prod MCHM5} is devoted to double Higgs production.
We demonstrate explicitly that the full loop computation with the top partners can easily give negative corrections of order 30\% to the pure strong dynamics computation where only the Higgs nonlinearities were taken into account \cite{doubleh2}. Contrary to the single Higgs production case, the LET does not capture well these contributions of the top partners, and this discrepancy is further amplified when looking at particular regions of the phase space selected by kinematical cuts. Various technical details are collected in a series of appendices. 

\section{Low-energy effective Lagrangian for a composite Higgs boson \label{sec:paramet}}
%
%
An interesting solution to the hierarchy problem is given by the Higgs
boson being a composite bound state emerging from a new strongly-interacting sector, broadly characterized by a mass scale $m_{\rho}$ and a coupling $g_{\rho}$. If in addition the Higgs emerges as a pseudo-Goldstone boson of a spontaneous symmetry breaking 
$\mathcal{G}/\mathcal{H}\,$ at the scale $f=m_{\rho}/g_{\rho}$, then
it can be naturally lighter than the other resonances of the strong
sector and $v^{2}/f^{2}\ll 1$ can be accommodated. A low-energy, model-independent description of this idea is given by the strongly-interacting light Higgs (SILH) Lagrangian \cite{silh}, which applies to the general scenario where the Higgs is a light pseudo-Goldstone boson, including Little Higgs and Holographic composite Higgs models. At scales much smaller than $m_{\rho}$, deviations from the SM are parameterized in terms of a set of dimension-six operators. The subset that will be relevant in our discussion is  
\begin{align} 
{\cal L}_{\mathrm{SILH}} \,=\,\,& \frac{c_H}{2 f^2} \partial^\mu (H^\dagger
H) \partial_\mu (H^\dagger H) + \frac{c_{r}}{2f^{2}}H^{\dagger}H(D_{\mu}H)^{\dagger}(D^{\mu}H) - \frac{c_6
  \lambda}{f^2} (H^\dagger H)^3 \nonumber \\ \label{SILH}
+\,& \left( \frac{c_y y_f}{f^2} H^\dagger H \bar{f}_L H f_R +
  \mathrm{h.c.} \right)  +  \frac{c_g g_s^2}{16\pi^2 f^{2}} \frac{y_t^2}{g_\rho^2}
H^\dagger H G_{\mu\nu}^a G^{a\, \mu\nu} + \frac{c_{\gamma}g^{\prime\,2}}{16\pi^{2}f^{2}}\frac{g^{2}}{g_{\rho}^{2}}H^{\dagger}H B_{\mu\nu}B^{\mu\nu}
\end{align}
where $g_s, g, g'$ are the SM $SU(3)_c\times SU(2)_{L}\times U(1)_Y$ gauge couplings, whereas $\lambda$ and $y_{f}$ are the Higgs quartic and Yukawa coupling appearing in the SM Lagrangian, respectively. The first four operators are genuinely sensitive to the
strong interaction, whereas the last two parameterize the effective
couplings of the Higgs to gluons and to photons, respectively, mediated
by loops of heavy particles. As the operators proportional to $c_{g}$
and $c_{\gamma}$ do not respect the symmetry under which the Goldstone
Higgs shifts they have to be suppressed by powers of the couplings
which break this symmetry, thus explaining the extra factor
$g_{SM}^{2}/g_{\rho}^{2}$ $(g_{SM}=y_{t}, g)$ appearing in front of
them. These operators are important in the presence of relatively light resonances, \emph{i.e.} when $g_{\rho}\sim g_{SM}$. Recent analyses \cite{lighttoppartners} show that in a large class of models, a Higgs as light as $125\,\mathrm{GeV}$ implies the presence of one or more anomalously light (sub-TeV) fermionic resonances, which thus can contribute sizeably to $c_{g}$ and $c_{\gamma}$. Also notice that by choosing flavor-diagonal
couplings $y_f$ Minimal Flavor Violation (MFV) is automatically
implemented in the Lagrangian so that it complies with flavor bounds.

In Eq.~\eqref{SILH} we have kept explicitly the operator proportional to $c_{r}$, which can be eliminated at $\mathcal{O}(1/f^{2})$ by a field redefinition
\beq \label{reparam}
H\to H + a(H^{\dagger}H)H/f^{2}\,,
\eeq
under which
\beq \label{coeff transform}
c_{H}\to c_{H}+2a\,,\qquad c_{r}\to c_{r}+4a\,,\qquad c_{6}\to c_{6}+4a\,,\qquad c_{y}\to c_{y}-a\,,
\eeq
while $c_g$ and $c_\gamma$ do not change under this transformation.
The choice $c_{r}=0$ corresponds to the `SILH basis', which can be
reached starting from a generic basis where $c_{r}\neq 0$ by applying
the transformation in Eq.~\eqref{reparam} with $a=-c_{r}/4$. We choose
to keep explicitly the operator proportional to $c_{r}$ as the `natural' basis
for nonlinear $\sigma$-models actually corresponds to a non-vanishing
$c_{r}$ \cite{Low:2010mr}. Furthermore, since physical amplitudes have
to be invariant under field redefinitions, Eqs.~(\ref{reparam}) and 
(\ref{coeff transform}) will be used as a consistency check of our
results. In Eq.~\eqref{SILH} we have also omitted the custodial breaking operator
\beq
\frac{c_T}{2 f^2} \left( H^{\dagger}\!\!\stackrel{\longleftrightarrow}{D^{\mu}}\!\!H \right)^2\,  ,
\eeq
where $H^\dagger \!\!
\stackrel{\longleftrightarrow}{D^\mu} \!\! H \equiv
H^{\dagger}(D^{\mu}H)-(D^{\mu}H)^{\dagger}H$, which gives a
contribution to the $T$ parameter \mbox{$\hat{T}=c_{T}v^{2}/f^{2}\,$}
and thus is strongly constrained by electroweak data. This operator does not contribute to the processes that we will be interested in. If the strong sector is invariant under custodial symmetry, as it happens for example in models based on the coset $SO(5)/SO(4)$, then $c_T$ vanishes at \mbox{tree-level}. For a discussion of Higgs physics where the assumption of custodial invariance is relaxed, see Ref.\cite{Farina:2012ea}. 

The SILH Lagrangian represents an expansion in $\xi \equiv (v/f)^2$ and
can be used in the vicinity of the SM, which corresponds to $\xi = 0$. On the other hand, the technicolor limit
($\xi \to 1$) requires the resummation of the full series in $\xi$. Such a
resummation is possible in the Holographic Higgs models of Refs.~\cite{Agashe:2004rs,Contino:2006qr,Contino:2003ve}. These models are based on a five-dimensional gauge theory in Anti-de-Sitter (AdS) space-time.
In the minimal realization, the bulk symmetry $SO(5) \times U(1)$
is broken to the SM group $SU(2)_L \times U(1)_Y$ on the ultraviolet (UV) brane and
to $SO(4)\times U(1)$ on the infrared (IR) brane. The coset $SO(5)/SO(4)$
provides four Goldstone bosons, one of which is the
physical Higgs boson and the three remaining ones are eaten by the
massive SM vector bosons. The Higgs couplings to gauge bosons and
its self-interactions are modified compared to the
SM, and the modification factors can be expressed in terms of
the parameter $\xi$.
The Higgs Yukawa couplings and the form of the
Higgs potential of the low-energy effective theory depend on the way
the SM fermions are embedded into representations of the bulk
symmetry. In the second part of this work we refer to the model MCHM5
\cite{Contino:2006qr} where the fermions transform in the fundamental
representation of $SO(5)$. An alternative realization of the $SO(5)/SO(4)$
composite Higgs, denoted by MCHM4, contains fermions
embedded into the spinorial representation \cite{Agashe:2004rs} (for more details see App.~\ref{AppB:mchm4}). In this case, however, large corrections to the
$Zb_L b_L$ coupling are present and rule out an important part
of the parameter space \cite{Agashe:2005dk}. In contrast, if fermions are embedded into the
fundamental or adjoint representation of $SO(5)$, the custodial symmetry of the strong sector includes a left-right
parity, which protects the $Zb_L b_L$ coupling from receiving tree-level corrections \cite{Agashe:2006at}. 

Another useful description of the low-energy theory is given by an effective chiral
Lagrangian where the $SU(2)\times U(1)_Y$ symmetry is nonlinearly realized. The
Goldstone bosons $\pi^a$ \mbox{($a=1,2,3$)} providing the longitudinal degrees of freedom of the
$W^\pm$ and $Z$ bosons are introduced by means of the field
\beq
\Sigma (x) = e^{i\sigma^a \pi^a(x)/v}  \,,
\eeq 
where $v \simeq 246\,\mathrm{GeV}$ and $\sigma^a$ are the Pauli
matrices. The field $\Sigma$ transforms linearly under
$SU(2)_L \times SU(2)_R$. Introducing a scalar field $h$, assumed to transform as a singlet under the custodial
symmetry, leads to the following effective Lagrangian \cite{doubleh1}
\begin{align} \nonumber
\mathcal{L}\, &=\, \frac{1}{2} (\partial_\mu h)^2 - V(h) + \frac{v^2}{4} {\rm Tr}\left[(D_\mu \Sigma)^\dagger \, D^\mu \Sigma\right] \left( 1 + 2 \, a \, \frac{h}{v} + b \, \frac{h^2}{v^2}  +  b_3 \, \frac{h^3}{v^3} + \cdots \right)  \\ 
\,& - \frac{v}{\sqrt{2}} \, (\bar{u}_L^i \bar{d}_L^i) \, \Sigma \, \left[1 + c \, \frac{h}{v} +  c_2 \, \frac{h^2}{v^2}  + \cdots \right]  
\left(
\begin{array}{c} 
y_{ij}^u \, u_R^j \nonumber \\ 
y_{ij}^d \, d_R^j 
\end{array} \right)  +  \mathrm{h.c.} \,\,+\,\mathcal{L}^{(4)}, \quad \mbox{with} \quad\\ \nonumber
V(h)\, &=\, \frac{1}{2} \, m_h^2 \, h^2 + d_3 \, \left(\frac{m_h^2}{2v} \right) \, h^3 + d_4 \left(\frac{m_h^2}{8v^2} \right) h^4 + \cdots \; ,\\ \label{eff Lagr}
\mathcal{L}^{(4)}\, &=\frac{g_{s}^{2}}{48\pi^{2}}G^{\mu\nu\,a}G_{\mu\nu}^{a}\left(k_{g}\frac{h}{v}+\frac{1}{2}k_{2g}\frac{h^{2}}{v^{2}}+\ldots \right)+\frac{e^{2}}{32\pi^{2}}F_{\mu\nu}F^{\mu\nu}\left(k_{\gamma}\frac{h}{v}+\ldots\right)\,,
\end{align}
with the mass of the scalar given by $m_h$. In Eq.~\eqref{eff Lagr} we have introduced the higher-dimensional couplings $k_{g},k_{2g},k_{\gamma}$, which are mediated at loop level by strong sector resonances. The Higgs couplings to fermions, $c,c_2,...$, are assumed to be
flavor-diagonal, so that MFV is realized. In Table~\ref{coupvalues}
the values of the couplings in the effective Lagrangian Eq.~(\ref{eff
  Lagr}) are listed in the SILH approach and in the holographic Higgs
model MCHM5 (for the latter, only Higgs nonlinearities are considered). 
\begin{table}
\begin{center}
\renewcommand{\arraystretch}{1.5}
\begin{tabular}{ccc}
\hline
Parameters & SILH & MCHM5, pure Higgs nonlinearities \\ \hline
$a$ & $1 -(c_H-c_{r}/2) \,\xi/2$ & $\sqrt{1-\xi}$ \\
$b$ & $1+(c_{r}-2c_H)\, \xi$ & $1-2\xi$ \\
$b_3$ & $(c_{r}-2c_{H}) 2\,\xi/3$ & $-\frac{4}{3} \xi \sqrt{1-\xi}$ \\
$c$ & $1- (c_H/2 + c_y)\,\xi$ & $\frac{1-2\xi}{\sqrt{1-\xi}}$ \\
$c_2$ & $-(c_H+3c_y+c_{r}/4)\,\xi/2$ & $-2 \xi$ \\
$d_3$ & $ 1+(c_6-c_{r}/4-3c_H/2)\,\xi$ & $\frac{1-2\xi}{\sqrt{1-\xi}}$ \\
$d_4$ & $ 1+(6c_6 -25c_H/3-11c_{r}/6)\, \xi$ & $\frac{1-28\xi(1-\xi)/3}{1-\xi}$ \\
$k_{g}=k_{2g}$ & $ 3c_{g}(y_{t}^{2}/g_{\rho}^{2})\xi$ & $0$ \\
$k_{\gamma}$ & $ 2c_{\gamma}(g^{2}/g_{\rho}^{2})\xi$ & $0$ \\
\hline
\end{tabular}
\end{center}
\caption{Values of the couplings of the effective Lagrangian Eq.~(\ref{eff Lagr}) in the SILH framework (with $c_T=0$) and for MCHM5 considering only Higgs nonlinearities (\emph{i.e.} neglecting the effects of resonances). The latter are taken from Ref.~\cite{Espinosa:2012qj}. The values of the SILH parameters in MCHM5 are, in the `natural' basis for the nonlinear $\sigma$-model where $c_{r}=-4\,c_{H}$, $\,c_{H}=1/3,\,c_{r}=-4/3,\,c_{y}=4/3,\,c_{6}=-4/3\,$.} 
\label{coupvalues}
\end{table}
The SM with an elementary Higgs boson corresponds to
$a=b=c=d_3=d_4=1$, $c_2=b_3=k_{g}=k_{2g}=k_{\gamma}=0$ and vanishing higher order terms in
$h$. 

\section{Applying the Higgs Low-Energy Theorem}\label{sec:LET}
In this section we discuss applications of the Higgs low-energy
theorem \cite{Ellis:1975ap,Kniehl:1995tn} in composite
models. The LET allows 
one to obtain the leading interactions of the Higgs boson with gluons and
photons arising from loops of heavy particles. By heavy particles we
mean here both SM states ($W$ and top) and new states belonging to the
composite sector. These couplings are needed in the computation of the
cross sections of single and double Higgs production via gluon fusion at
the LHC as well as of the partial width of the decay $h\to
\gamma\gamma\,$. We will adopt a model-independent approach and
compute these quantities in terms of the parameters of the effective
Lagrangians defined in Section~\ref{sec:paramet}, Eqs.~(\ref{SILH})
and (\ref{eff Lagr}), putting special emphasis on the former, namely the SILH description. Our analysis extends the results of Refs.~\cite{Low:2010mr,Low:2009di} to Higgs pair production in gluon fusion, and also includes a discussion of corrections to the LET approximation arising from higher order terms in the $1/M$ expansion, where $M$ is the mass of the generic heavy particle running in the loops. Notice that the LET can be extended to 2-loop order to include the leading QCD corrections, see for example Ref.~\cite{Kniehl:1995tn} for applications in the SM. However, our discussion will be mainly limited to couplings at the leading 1-loop order.

\subsection{Higgs interactions with gluons \label{sec:lowenergy}}
According to the LET the interactions of the physical Higgs boson with gluons, mediated by loops of heavy
coloured particles, can be obtained by treating the Higgs $H$ as a background field and taking the field-dependent mass of each heavy particle as a threshold for the running of the QCD gauge coupling.\footnote{Throughout the paper, we will denote by $H$ both the Higgs doublet and the scalar field with $\left\langle H\right\rangle \neq 0\,$, as it will always be clear from the context which one we are referring to. On the other hand, $h$ denotes the physical Higgs scalar.} Assuming the heavy particles to transform in the fundamental representation of $SU(3)_{c}$ one obtains the following effective
Lagrangian
\beq
\mathcal{L}_{eff}=\frac{g_{s}^{2}}{64\pi^{2}}G_{\mu\nu}^{a}G^{a\,\mu\nu}\sum_{p_{i}}\delta b_{p_{i}} \log m_{p_{i}}^{2}(H)\,,
\eeq
where $\delta b=2/3$ if particle $p_{i}$ is a Dirac fermion, and
$\delta b = 1/6$ if it is a complex scalar. In this paper we will
focus only on the effects of the heavy fermion sector, which in
composite Higgs models typically includes new states beyond the
top quark. By expanding the field-dependent masses of the heavy particles
around the vacuum expectation value (VEV) 
$\langle H \rangle$ we obtain the couplings of the
Higgs boson to gluons mediated by loops of heavy fermions

\beq \label{hngg}
{\cal L}_{h^n gg} = \frac{g_s^2}{96\pi^2} G_{\mu\nu}^a G^{a\, \mu\nu}\left( A_{1} h +\frac{1}{2}A_{2}h^{2}+\ldots\right)\,,  
\eeq
where we have defined
\beq \label{eq:a1a2def}
A_n \equiv \left( \frac{\partial^{n}}{\partial H^{n}} \log \det {\cal
    M}^{2} (H) \right)_{\langle H \rangle} 
\eeq
with $\mathcal{M}^{2}\equiv \mathcal{M}^{\dagger}\mathcal{M}$, and
$\mathcal{M}$ is the heavy fermion mass matrix. In the SM only the top
quark contributes\footnote{The bottom contribution is non-negligible, but cannot be computed using the low-energy theorem, due to the smallness of the bottom quark mass.} with \mbox{$m_t(H)=y_t H/\sqrt{2}\,$}, so that Eq.~\eqref{hngg} can be rewritten at all orders in $h$ as (see for example Refs.~\cite{Kniehl:1995tn,Hagiwara:1989xx})
\beq
\mathcal{L}_{h^{n}gg}=\frac{g_{s}^{2}}{48\pi^{2}}G_{\mu\nu}^{a}G^{a\,\mu\nu}\,\log\left(1+\frac{h}{v}\right)\,.
\eeq
The corresponding gauge invariant operator is $\,\log(H^{\dagger}H)G_{\mu\nu}^{a}G^{a\,\mu\nu}$, which is associated with a chiral fermion. The lowest-order operator arising from vector-like fermions is instead $\,H^{\dagger}H G_{\mu\nu}^{a}G^{a\,\mu\nu}$. The effects of these two operators on double Higgs production were discussed in Ref.~\cite{Pierce:2006dh}.

Using Eq.~\eqref{hngg} it is straightforward to derive the expression of the $hgg$ and $hhgg$ couplings in the SILH formalism. We refer the reader to App.~\ref{appendix:silh gluon higgses} for a derivation, and simply report here the results. We remark that from now on we will work in the unitary gauge, where the Higgs doublet reads $(0\,,\,H/\sqrt{2})^{T}$.
The effective coupling of the Higgs boson to two gluons reads (see App.~\ref{appendix:silh gluon higgses})
\beq \label{hgg}
\mathcal{L}_{hgg}= \frac{g_{s}^{2}}{48\pi^{2}}G^{a}_{\mu\nu}G^{a\,\mu\nu}\frac{h}{v}\left[\frac{1}{2}\left(\frac{\partial}{\partial\log H}\log \det \mathcal{M}^{2}(H)\right)_{H=v}-\frac{c_{H}}{2}\xi \right]\,.  
\eeq
This coupling governs the rate of single Higgs production via gluon fusion,
and its expression was already obtained in Refs.~\cite{Low:2010mr,Low:2009di}. The production rate normalized to the SM one is given by the square of the expression in square brackets in Eq.~\eqref{hgg}. On the other hand the effective coupling of two Higgs bosons to two gluons, which contributes to Higgs pair production via gluon fusion, has the following expression
\beq \label{hhgg}
\mathcal{L}_{hhgg}=\,\frac{g_{s}^{2}}{96\pi^{2}}G^{a}_{\mu\nu}G^{a\,\mu\nu}\frac{h^{2}}{v^{2}}\left[\frac{1}{2}\left(\left(\frac{\partial^{2}}{\partial (\log H)^{2}}-\frac{\partial}{\partial \log H}\right)\log\det \mathcal{M}^{2}(H)\right)_{H=v}-\frac{c_{r}}{4}\xi \right]\,.
\eeq
In terms of the effective Lagrangian in Eq.~\eqref{eff Lagr}, the couplings read 
\beq \label{eff couplings eff lagr}
\mathcal{L}_{hgg}= \frac{g_{s}^{2}}{48\pi^{2}}G^{a}_{\mu\nu}G^{a\,\mu\nu}\frac{h}{v}\,(c+k_{g})\,,\quad\qquad  \mathcal{L}_{hhgg}=\,\frac{g_{s}^{2}}{96\pi^{2}}G^{a}_{\mu\nu}G^{a\,\mu\nu}\frac{h^{2}}{v^{2}}\,(2c_{2}-c^{2}+k_{2g})\,. 
\eeq
In the expression of the $hhgg$ coupling in Eq.~\eqref{eff couplings eff lagr}, the first term comes from the triangle top loop involving the $t\bar{t}hh$ vertex, whereas the second is the contribution of top box diagrams, see Fig.~\ref{fig:diaghhprod}. On the other hand, $k_g$ and $k_{2g}$ are parameterizing the contributions from integrated-out heavy particles.

%
%
\subsection{Higgs interaction with photons}
Although the main focus of this paper is on gluon fusion, we give here
the expression for the coupling of the Higgs boson to photons, as it
is another loop process of crucial relevance for Higgs phenomenology
at the LHC. This coupling receives contributions both from loops of
heavy fermions and from the $W$ boson loop. Application of the LET leads to the following effective Lagrangian \cite{Kniehl:1995tn}
\beq
\mathcal{L}_{eff}=\frac{e^{2}}{16\pi^{2}}F_{\mu\nu}F^{\mu\nu}\Big(\sum_{f} Q_{f}^{2}\log m_{f}^{2}(H)-\frac{7}{4}\log m_{W}^{2}(H)\Big)\,,
\eeq
which is valid for $m_{h}\lesssim 2m_{W},\,2m_{f}\,$, and where we have assumed
that the heavy fermions transform in the fundamental representation of $SU(3)_{c}\,$. Expanding around the VEV we obtain the $h\gamma\gamma$ interaction
\beq \label{hgaga starting point}
\mathcal{L}_{h\gamma\gamma} = \frac{e^{2}}{16\pi^{2}}F_{\mu\nu}F^{\mu\nu} h \left( Q_{t}^{2}A_{1}-\frac{7}{4}\left(\frac{\partial}{\partial H}\log m_{W}^{2}(H)\right)_{\left\langle H\right\rangle}\right)\,,
\eeq
where we have assumed that all fermions have electric charge equal to that of the top quark\footnote{In all models considered in this paper, only top-like resonances contribute to the $h\gamma\gamma$ coupling. The extension to heavy states with different electric charge is straightforward.}, \mbox{$Q_{f}=Q_{t}=2/3\,$}, and $A_{1}$ was defined in Eq.~\eqref{eq:a1a2def}. By performing simple manipulations we obtain (see App.~\ref{appendix:silh gluon higgses}) 
\begin{align} \nonumber
\mathcal{L}_{h\gamma\gamma}=\, \frac{e^{2}}{32\pi^{2}}F_{\mu\nu}F^{\mu\nu}\frac{h}{v} &\Bigg[4 Q_{t}^{2}\Bigg(\frac{1}{2}\left(\frac{\partial}{\partial \log H} \log \det \mathcal{M}^{2}(H)\right)_{H=v} -\frac{c_{H}}{2}\xi \Bigg) \\ \label{hgammagamma} &- J_{\gamma}(4m_{W}^{2}/m_{h}^{2}) \left(1+\xi\left(\frac{c_{r}}{4}-\frac{c_{H}}{2}\right)\right)\Bigg]\,,
\end{align}
where we have replaced the LET approximation for the $W$ loop with the full result encoded by the function
\beq
J_{\gamma}(x)=F_{1}(x)\,,\qquad F_{1}(x)=2+3x[1+(2-x)f(x)]\,,\qquad f(x)=\arcsin^{2}(x^{-1/2})\,,
\eeq 
which tends for large $x$ to $J_{\gamma}(\infty)=7 = 22/3-1/3\,$, where the first term comes from the transverse polarizations of the $W$ and is precisely equal to the gauge contribution to the $\beta$ function of the $SU(2)_{L}$ coupling, while the second term arises from the eaten Goldstone bosons. The use of the full expression for the $W$ loop implies that the formal validity of Eq.~\eqref{hgammagamma} is extended to $m_{h}\lesssim 2 m_{f}\,$. The rescaling of the decay width \mbox{$\Gamma(h\to \gamma\gamma)/\Gamma(h\to \gamma\gamma)_{SM}$} is obtained by comparing the square of the expression multiplying \mbox{$hF^{\mu\nu}F_{\mu\nu}$} in Eq.~\eqref{hgammagamma} in the two cases. In terms of the parameters of the effective Lagrangian in Eq.~\eqref{eff Lagr} the coupling reads
\beq
\mathcal{L}_{h\gamma\gamma}=\frac{e^{2}}{32\pi^{2}}F^{\mu\nu}F_{\mu\nu}\frac{h}{v}\Big(4Q_{t}^{2}c+k_{\gamma} - aJ_{\gamma}(4m_{W}^{2}/m_{h}^{2})\Big)\,.
\eeq
\subsection{Single Higgs production via gluon fusion}\label{sec:singlehprod}
\begin{figure}[h]
\begin{center}
   \includegraphics[width=.2\textwidth,angle=-90]{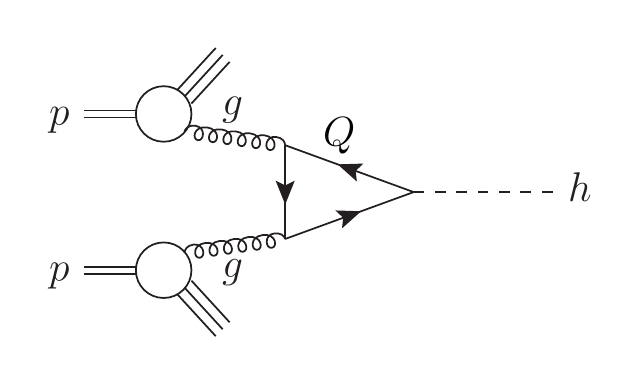}
\end{center}
\vspace*{-0.4cm}
\caption{\label{fig:hfeyndiag} Generic diagram contributing to Higgs production 
in gluon fusion. In the SM case we have $Q\equiv t,b$. In composite
Higgs models with additional heavy fermionic resonances these add to
the particles $Q$ running in the loop.}
\end{figure}

For a SM Higgs boson, the gluon fusion process \cite{georgi} gives the
dominant production cross section at the LHC, see Refs.~\cite{SMhiggs}
for reviews. At leading order (LO), the process proceeds via a top
loop, with a subleading contribution from the bottom loop, see Fig.~\ref{fig:hfeyndiag}. 
In composite Higgs models extra heavy, colored fermions with sizeable
couplings to the Higgs boson are typically present, whose
contributions to the gluon fusion process should in general be taken
into account. It has been shown
\cite{singlehiggsindep,Low:2010mr,Azatov:2011qy,comphiggsnnlo} that in
explicit constructions the $gg\to h$ cross section (computed in the
LET approximation) is insensitive to the details of the heavy fermion
spectrum, \emph{i.e.} it does not depend on the couplings and masses
of composites, but only on the ratio $v/f$, where $f$ is the overall
scale of the strong sector. This was found to be true both in models
with partial compositeness and in Little Higgs theories. In fact,
although the top Yukawa coupling receives a correction due to the mixing with resonances which depends on composite couplings, this contribution is exactly canceled by the loops of extra fermions, leading to a dependence of the $gg\to h$ rate only on $v/f\,$. This also implies that the cross section can be obtained by simply multiplying the SM one by $c^{2}$, where $c$ is the rescaling of the top Yukawa coming only from the nonlinearity of the $\sigma$-model, and neglecting corrections due to fermionic resonances.

Let us review how this cancellation arises. It is due to the fact that the determinant of the heavy fermion mass matrix takes the form
\beq \label{indep condition}
\det \mathcal{M}^{2}(H)\,=\,F(H/f)\times P(\lambda_{i},M_{i},f)\,,
\label{eq:matrixrel}
\eeq
where $F$ is a function satisfying $F(0)=0$ since the top becomes
massless in the limit of unbroken electroweak symmetry, and $P$ is a
function of the composite couplings $\lambda_{i}$ and masses
$M_{i}$, but independent of $H$. It is then immediate to see that the
$hgg$ coupling in Eq.~\eqref{hgg} does not depend on the masses and
couplings of the resonances.\footnote{The coefficient $c_{H}$ does not
  receive contributions from the heavy fermion sector. It is, however,
  generated by integrating out heavy scalars or vectors
  \cite{Low:2009di}, see App.~\ref{AppB:LH} for an explicit example.} The origin of the factorization in
Eq.~\eqref{indep condition} was explained in the context of partial
compositeness in Ref.~\cite{Azatov:2011qy}, by means of a spurion
analysis. There it was also pointed out that such a factorization can break down if the top mixes with more than one composite operator, leading to a dependence of the $hgg$ vertex on composite couplings. Nevertheless in many explicit constructions, including Little Higgs models, the factorization in Eq.~\eqref{indep condition} takes place. Still, the independence of the $hgg$ vertex on the composite couplings (collectively denoted by $\lambda_{i}$) holds exactly only in the LET approximation, and corrections due to finite fermion mass effects are expected. We can estimate the residual dependence on the $\lambda_{i}$ due to finite fermion mass effects in a simple way. Assuming for simplicity the presence of only one top partner $T$, the mass eigenvalues can be written at $\mathcal{O}(1/f^{2})$ as
\beq \label{field-dep masses}
m_{t}(H)=\frac{y_{t}H}{\sqrt{2}}\left(1-\frac{c_{y}^{(t)}}{2}\frac{H^{2}}{f^{2}}\right)\,,\qquad m_{T}(H)=\lambda_{T}f\left(1+a_{T}\frac{H^{2}}{f^{2}}\right)\,,
\eeq
where $a_{T}$ is a parameter dependent on the couplings $\lambda_{i}=\{y_{t},\lambda_{T}\}$ as $a_{T}=\mathcal{O}(y_{t}^{2}/\lambda_{T}^{2})$.\footnote{If the quadratic divergence in the Higgs mass due to the top is cancelled by $T$, then the absence of an $\mathcal{O}(H^{2})$ term in $\mathrm{Tr}\,\mathcal{M}^{2}(H)=m_{t}^{2}(H)+m_{T}^{2}(H)$ implies $a_{T}=-y_{t}^{2}/(4\lambda_{T}^{2})$. See for example the explicit values of $c_{y}^{(t)}$ and $a_{T}$ in the Littlest Higgs model, reported in App.~\ref{AppB:LH}, Eq.~\eqref{cyt and aT in LH}.} On the other hand, we can write $c_{y}^{(t)}=c_{y}^{(\sigma)}+\mathcal{O}(y_{t}^{2}/\lambda_{T}^{2})$, where $c_{y}^{(\sigma)}$ is a constant arising from the pure nonlinearity of the $\sigma$-model. The LET result for the $hgg$ coupling reads, taking the effect of $c_{H}$ into account, 
\beq \label{hggLETgeneral}
\mathcal{L}_{hgg}=\frac{g_{s}^{2}}{48\pi^{2}}G_{\mu\nu}^{a}G^{a\,\mu\nu}\frac{h}{v}\left(1-\left(c_{y}^{(t)}-2a_{T}+\frac{c_{H}}{2}\right)\xi \right)\,.
\eeq
Notice that in the limit where $T$ is heavy, corresponding to large $\lambda_{T}$, the effects of the heavy resonance on the $hgg$ coupling vanish. In fact, $a_{T}$ goes to zero, whereas $c_{y}^{(t)}\to c_{y}^{(\sigma)}$, implying that only the nonlinearity in the top Yukawa arising from the nonlinear $\sigma$-model is relevant.
 
By using the expression of the top Yukawa coupling $(m_{t}/v)(1-(c_{y}^{(t)}+c_{H}/2)\xi)$ we can compute explicitly the top loop diagram, retaining the first subleading term in the $1/m_t^2$ expansion. This is the leading correction to the LET coupling, given that $m_{T}\gg m_{t}$. Thus Eq.~\eqref{hggLETgeneral} is improved to
\beq \label{corrections}
\mathcal{L}_{hgg}=\frac{g_{s}^{2}}{48\pi^{2}}G_{\mu\nu}^{a}G^{a\,\mu\nu}\frac{h}{v}\left[1-\left(c_{y}^{(t)}-2a_{T}+\frac{c_{H}}{2}\right)\xi+\frac{7}{120}\frac{m_{h}^{2}}{m_{t}^{2}}\left(1-\xi (c_{y}^{(t)}+ c_{H}/2)\right)+\ldots \right]\,,
\eeq
where we have used $\hat{s}= m_{h}^{2}$, and the ellipses stand for subleading corrections (including terms of order $1/m_{T}^{2}$). The independence of the LET $hgg$ vertex of the composite couplings $\lambda_{i}$ is equivalent to the statement that $c_{y}^{(t)}-2a_{T}$ is a constant, $c_{y}^{(t)}-2a_{T}=c_{y}^{(\sigma)}$.\footnote{Notice that by using Eq.~\eqref{field-dep masses} one finds $\det\mathcal{M}^{2}=y_{t}^{2}\lambda_{T}^{2}f^{2}H^{2}(1-(c_{y}^{(t)}-2a_{T})H^{2}/f^{2})/2$. So if the factorization in Eq.~\eqref{indep condition} holds then $c_{y}^{(t)}-2a_{T}$ is a constant.} If this is the case then the dependence on the $\lambda_{i}$ of Eq.~\eqref{corrections} is due to the last term, and we can estimate the sensitivity of the cross section to the $\lambda_{i}$ to be, for a light top partner $\lambda_{T}\sim y_{t}$,
\begin{equation}   
\frac{\delta \sigma(gg\to h)}{\sigma(gg\to h)_{SM}}\sim \frac{7}{60}\frac{m_{h}^{2}}{m_{t}^{2}}\,\xi \simeq 0.06 \,\xi\, \; ,
\end{equation}
where in the last equality we assumed $m_{h}=125\,\mathrm{GeV}$. Thus corrections are expected to be very small even for large $\xi$. This estimate will be confirmed in Section~\ref{singleprodMCHM5}, where the $gg\to h$ cross section will be computed in MCHM5 retaining the full mass dependence. We note that in this model the cross section can be strongly suppressed compared to the SM, reaching $\sigma/\sigma_{SM}=1/3$ for a low compositeness scale $\xi = 0.25$. However, the Higgs branching ratios into $\gamma\gamma,WW$ and $ZZ$ are enhanced compared to the SM, so MCHM5 can still be compatible with the excesses observed by ATLAS and CMS at $m_{h}\sim 125\,\mathrm{GeV}\,$ even for values of $\xi$ as large as those considered in this paper \cite{Espinosa:2012qj}.
%
%
%
%
%
\subsection{Double Higgs production via gluon fusion}\label{sec:doublehprod}
\begin{figure}[h]
\begin{center}
\includegraphics[width=0.25\linewidth,angle=-90]{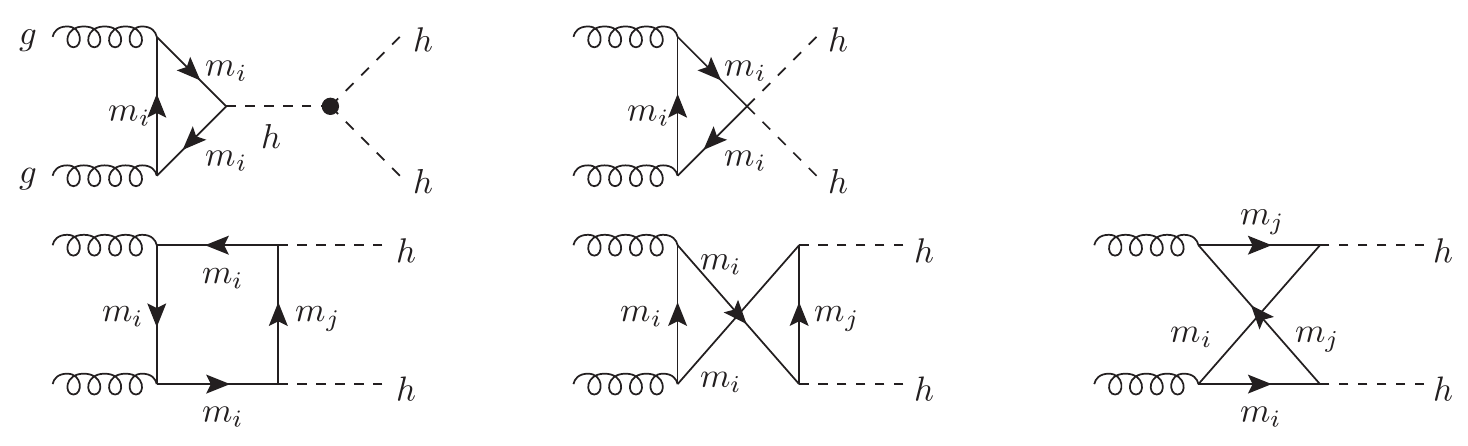}
\caption{Generic diagrams contributing to double Higgs
  production via gluon fusion in composite Higgs models with $n_f$ novel fermionic resonances
  of mass $m_{i}$ ($i=1,...,n_f$). The index $j$ is introduced to
indicate that different fermions can contribute to each box diagram.}
\label{fig:diaghhprod}
\end{center}
\end{figure}

Within the SM, double Higgs production via gluon fusion received interest mainly because it is sensitive to the trilinear Higgs self-coupling \cite{triplehiggs}, see the first diagram in Fig.~\ref{fig:diaghhprod}. 
In composite Higgs models, the process $gg\to hh$ is affected
essentially in two main ways. First, the nonlinearity of the strong sector
gives rise to a $f\bar{f}hh$ coupling (which vanishes in the SM) and
thus to a genuinely new contribution to the amplitude, see the second
diagram in Fig.~\ref{fig:diaghhprod}. Second, one should take into
account the effects of top partners, which include also new box
diagrams involving off-diagonal Yukawa couplings\footnote{Note that
  these Yukawa couplings only involve the top quark and its charge $2/3$ heavy composite partners.} (shown in the second line of Fig.~\ref{fig:diaghhprod}). A first study of $gg\to hh$ in composite Higgs models, neglecting top partners, was performed in Ref.~\cite{doubleh2}, where it was found that a large enhancement of the cross section is possible due to the new $t\bar{t}hh$ coupling (see also Ref.~\cite{Dib:2005re} for an earlier study in the context of Little Higgs models). For example, in MCHM5 with $\xi = 0.25$, which corresponds to $f\simeq 500\,\mathrm{GeV}$, the cross section was found to be about 3.6 times larger than in the SM. Recently, Ref.~\cite{ContinoETAL} performed a model-independent study of the process, making reference to the effective Lagrangian in Eq.~\eqref{eff Lagr} and again neglecting the effects of top partners, and found a large sensitivity of the cross section to the $c_{2}$ coefficient parameterizing the $t\bar{t}hh$ coupling.  

In this paper we include for the first time the effects of top partners in double Higgs production via gluon fusion. This is especially interesting in the light of the results of Refs.~\cite{lighttoppartners}, where a naturally light composite Higgs was shown to be tightly correlated with the presence of light top partners, as such light resonances can in principle affect the $gg\to hh$ cross section in a sizeable way. Our analysis will confirm that this is indeed the case.

We start by discussing the cross section in the LET approximation,
which greatly simplifies the computation. In this limit, the amplitude
is simply the sum of two diagrams, one with the effective
$hgg$ coupling followed by a trilinear Higgs coupling and the other
involving the effective $hhgg$ coupling. Adopting the SILH formalism, and recalling the expressions of the relevant Feynman rules
\beq
\begin{array}{lll}
hgg& :\, i\,\frac{\alpha_{s}}{3\pi
  v}\delta^{ab}(p_{1}^{\nu}p_{2}^{\mu}-p_{1}\cdot p_{2}\, g^{\mu\nu})\left[\frac{1}{2}\left(\frac{\partial}{\partial\log H}\log \det \mathcal{M}^{2}(H)\right)_{H=v}-\frac{c_{H}}{2}\xi\right]\,, \\[0.1cm]
hhgg& :\, i\, \frac{\alpha_{s}}{3\pi
  v^{2}}\delta^{ab}(p_{1}^{\nu}p_{2}^{\mu}-p_{1}\cdot p_{2}\, g^{\mu\nu})\left[\frac{1}{2}\left(\left(\frac{\partial^{2}}{\partial (\log H)^{2}}-\frac{\partial}{\partial \log H}\right)\log\det \mathcal{M}^{2}(H)\right)_{H=v}-\frac{c_{r}}{4}\xi\right]\,, \\[0.1cm]
hhh& :\, -i\,3\frac{m_{h}^{2}}{v}\left[1+\xi\left(c_{6}-\frac{3}{2}c_{H}-\frac{1}{4}c_{r}\right)\right]\,
\end{array}
\eeq
(where $p_{1,2}$ denote the momenta of the incoming gluons), we can write the amplitude as
\beq 
\mathcal{A}_{\textrm{LET}}\,(gg\to hh) = \frac{\alpha_{s}}{3\pi
  v^{2}}\delta^{ab} (p_{1}^{\nu}p_{2}^{\mu}-p_{1}\cdot p_{2}\, g^{\mu\nu})\, C_{\textrm{LET}}(\hat{s})\,,
\eeq
where
\begin{align} \nonumber
C_{\textrm{LET}} (\hat{s}) 
=\,\,& \frac{3m_h^2}{\hat{s}-m_h^2}
 \left[ \frac{1}{2}\left(\frac{\partial}{\partial\log H}\log \det \mathcal{M}^{2}(H)\right)_{H=v} + \xi \,(c_6 - 2 c_H - \frac{c_r}{4} ) \right] \nonumber \\ \label{gghh ampl} &\qquad \quad \,+\, \frac{1}{2}\left(\left(\frac{\partial^{2}}{\partial (\log H)^{2}}-\frac{\partial}{\partial \log H}\right)\log\det \mathcal{M}^{2}(H)\right)_{H=v} - \frac{c_r}{4} \xi 
 \\ \nonumber
=\,\,& \frac{3m_h^2}{\hat{s}-m_h^2}
\left( 1 -\xi (c_y^{(t)} - c_6 + 2 c_H + \frac{c_r}{4} - 3 c_g
  \frac{y_t^2}{g_\rho^2} ) \right) -\left(1+ \xi (c_y^{(t)} + \frac{c_r}{4}
  - 3c_g \frac{y_t^2}{g_\rho^2}) \right) \,,
\end{align}
with $\hat{s} \equiv (p_1+p_2)^2$ denoting the partonic center-of-mass
(c.m.)~energy. To obtain the second equality in Eq.~\eqref{gghh ampl} we used Eqs.~\eqref{A1explicit} and \eqref{A2explicit} contained in App.~\ref{appendix:silh gluon higgses}. It is immediate to check that the combinations
$c_{y}^{(t)}-c_{6}+2c_{H}+c_{r}/4$ and $c_{y}^{(t)}+c_{r}/4$ are invariant under
the reparameterization in Eqs.~(\ref{reparam}) and (\ref{coeff
  transform}). For completeness, we also give the result in terms of the coefficients of the effective Lagrangian in Eq.~(\ref{eff Lagr}):
\beq \label{C efflagr}
C_{\textrm{LET}}^{{\cal L}_{\textrm{eff}}} (\hat{s}) =  \frac{3m_h^2}{\hat{s}-m_h^2} \, (c+k_{g}) \, d_3 +
2 c_2 - c^2+k_{2g} \;.
\eeq
The partonic cross section reads
\beq
\hat{\sigma}_{gg\to hh} = \frac{G_F^2 \alpha_s^2 (\mu) \hat{s}}{128 (2\pi)^3}
\, \frac{1}{9}\,\sqrt{1-\frac{4m_{h}^{2}}{\hat{s}}} \, C_{\textrm{LET}}^2 (\hat{s}) \;.
\label{eq:partoniccxn}
\eeq 
The hadronic cross section is obtained by convolution with the parton distribution function $f_{g/P}$ of the gluon in the proton, 
\beq \label{hadr cross sect}
\sigma = \int_{4 m_h^2/s}^1 d \tau \int_\tau^1 \frac{dx}{x}
f_{g/P}( x, Q) \, f_{g/P} (\tau/x, Q) \, \hat{\sigma}_{gg\to hh} (\tau s) \,,
\eeq
with the collider c.m. energy $s$ related to $\hat{s}$ by $\hat{s} = \tau s$. The renormalization scale $\mu$ and the factorization scale $Q$ are chosen equal to the invariant mass of the Higgs boson
pair, $\mu=Q=\sqrt{\hat{s}}$. Throughout the paper, the parton
distribution functions of MSTW2008 \cite{mstw08} are employed. For $\xi \to 0$, Eq.~(\ref{gghh ampl}) correctly reproduces the SM result in the limit of large top mass \cite{Glover:1987nx,Plehn:1996wb}
\beq
C_{\textrm{LET}}^{SM}(\hat{s})= \frac{3m_h^2}{\hat{s}-m_h^2} -1 \,.
\eeq
In the SM the $m_{t}\to \infty$ limit gives a cross section in agreement with the full result only within $20\%$ for $m_{h}\lesssim 200\,\mathrm{GeV}$ (for $m_{h}=125\,\mathrm{GeV}$ we find $\sigma^{SM}_{\textrm{LET}}=14.6\,\mathrm{fb}$ and $\sigma_{full}^{SM}=17.9\,\mathrm{fb}$) and moreover it produces incorrect kinematic distributions, as noticed in Ref.~\cite{Baur:2002rb}. Thus we expect the LET to be in general less accurate in $gg\to hh$ than in single Higgs production.

\begin{figure}[ht]
\centering
   \includegraphics[width=.47\textwidth]{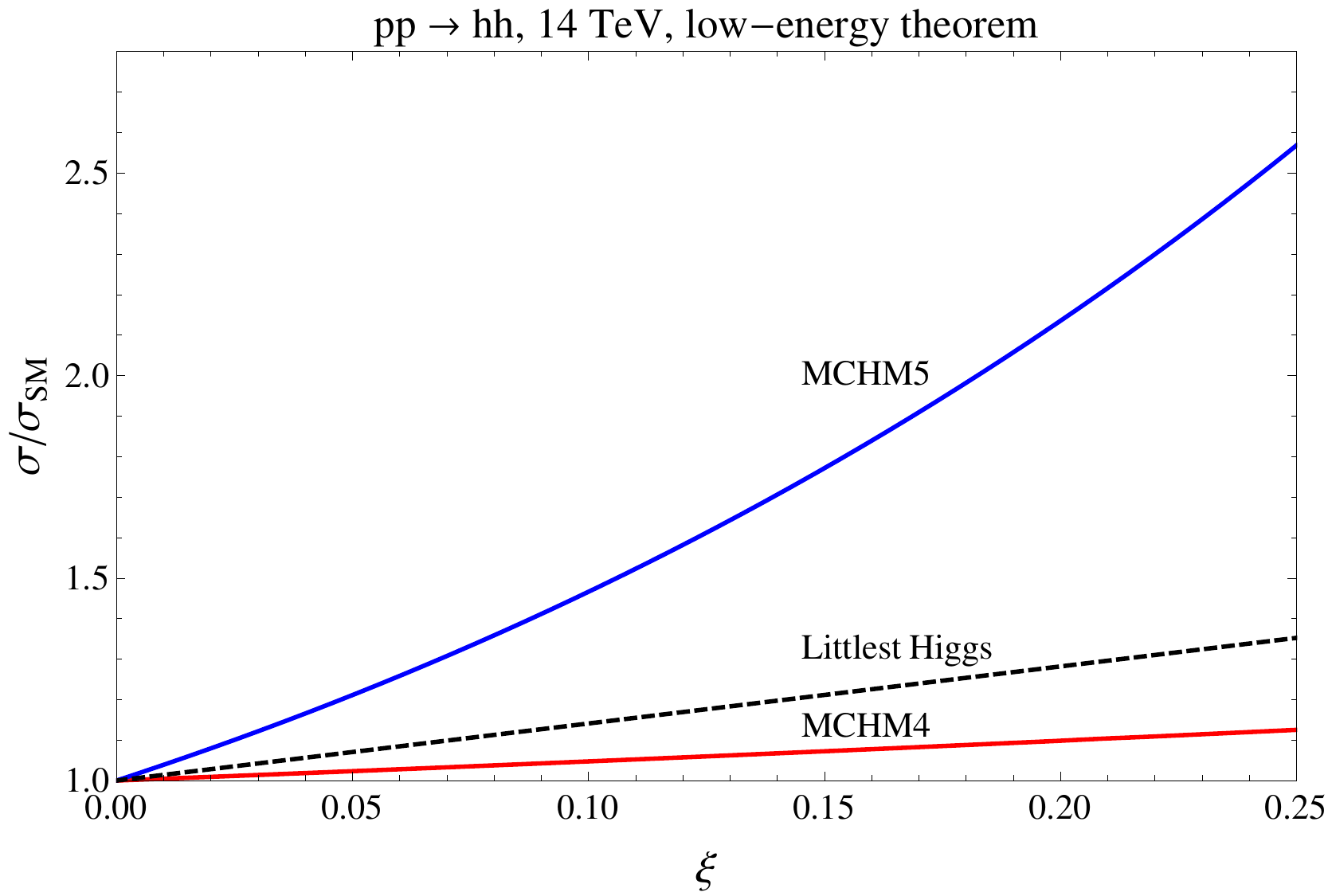} 
\hspace{1mm}
   \includegraphics[width=.47\textwidth]{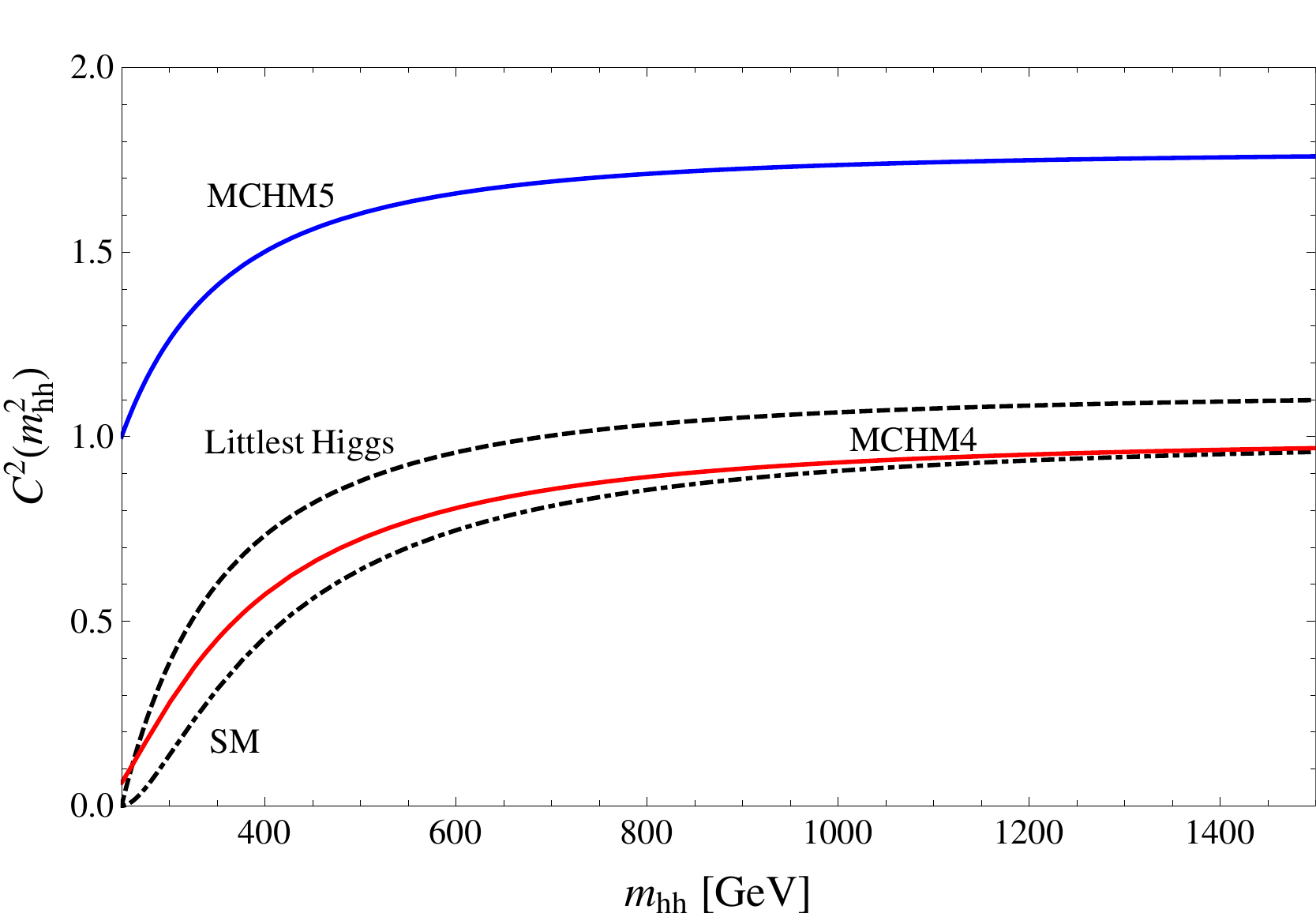}
\caption{(\emph{Left panel}) The $pp\to hh$ cross section for $m_h=125$~GeV at  LHC14,
  computed using the LET, normalized to the SM cross section also
  computed in the $m_{t}\to \infty$ limit. MHCM5 is discussed in
  detail in the text, whereas the $gg\to hh$ amplitudes for MCHM4 and
  for the Littlest Higgs model are given in App.~\ref{LH and
    MCHM4}. (\emph{Right panel}) Square of the function $C(m_{hh}^2)$,
  which was defined in Eq.~\eqref{gghh ampl} and is proportional to
  the LET $gg\to hh$ amplitude, in the three models under consideration (for $\xi = 0.25$) and in the SM, as a function of $m_{hh}=\sqrt{\hat{s}}$.}
\label{fig:let_models}
\end{figure}

From Eq.~\eqref{gghh ampl} we read off that in models where the
factorization Eq.~\eqref{indep condition} of $\det\mathcal{M}^{2}$
holds, the $gg\to hh$ LET cross section is insensitive to composite
couplings, due to a cancellation completely analogous to the one that we discussed
for single Higgs production. In the left panel of Fig.~\ref{fig:let_models} we show for
$m_h=125$~GeV and a c.m. energy of \mbox{14 TeV} the $pp\to hh$ cross section normalized to the SM cross section (both were computed applying the LET) as a function of $\xi$ for some well-known models, in all of which the cancellation holds.  
%
%
We note that in MCHM5 the enhancement of the cross section is
striking. This can be traced back to the behavior of the function
$C_{\textrm{LET}}(\hat{s})$, which is proportional to the LET amplitude and is shown in the right panel of Fig.~\ref{fig:let_models} for the three models under consideration and for the SM. The enhancement for MCHM5 is evident. As pointed out for the first time in Ref.~\cite{doubleh2}, where the $gg\to hh$ process
was studied in MCHM5 considering only Higgs nonlinearities (or equivalently in the limit of heavy top partners) but keeping
the full dependence on $m_{t}$, the dramatic increase of the $gg\to hh$ cross section compared to the SM is mostly due to the presence of a new $t\bar{t}hh$ coupling. The large enhancement of $gg\to hh$ in MCHM5 is in contrast with the strong suppression in the same model of the single Higgs production cross section, which for
$\xi=0.25$ equals $1/3$ of the SM value (see Section~\ref{singleprodMCHM5}).
  
By comparison with Ref.~\cite{doubleh2} we find that when fermionic resonances are above the cutoff, the LET
underestimates the ratio $\sigma^{MCHM5}/\sigma^{SM}$ by about $30\%\,$: for example for
$\xi = 0.25$, application of the LET gives a cross section of 2.6
times the SM, whereas Ref.~\cite{doubleh2} found an enhancement factor
of 3.6. This difference is due to the fact that in the former case
$m_{h}\ll m_{t}$ is assumed, whereas in the latter the full $m_{t}$
dependence was retained. Notice that the best estimate of the cross
section that can be obtained using the LET is $\sigma^{MCHM5} =
(\sigma^{MCHM5}_{\textrm{LET}}/\sigma^{SM}_{\textrm{LET}})\times \sigma^{SM}_{full}$, because part of
the corrections due to the finite top mass should cancel in the ratio of LET cross sections. In fact, in terms of cross sections the disagreement between the LET and the result obtained taking into account only Higgs nonlinearities is larger. For $\xi = 0.25$ we obtain $\sigma^{MCHM5}_{\textrm{LET}}= 37\,\mathrm{fb}$, whereas
Ref.~\cite{doubleh2} found $\sigma^{MCHM5}=64\,\mathrm{fb}$,
\emph{i.e.} the difference is of order $50\%$. 

In order to understand this behavior we investigated in more detail
the validity of the LET both for single and double Higgs
production. In single production the expansion parameter is
$m_h^2/(4m_t^2)$ and the series converges very quickly. In double Higgs
production on the other hand, one needs to expand in $\hat{s}/(4 m_t^2)$ with $\hat{s} \ge 4m_{h}^{2}$,
which is not small, so in general the expansion does not work as well as for
the single Higgs case. In MCHM5, the validity of the expansion gets
even worse. The reason why the LET is less accurate for MCHM5 than for
the SM (where it underestimates the cross section by about $20\%$) is
mainly the presence of the new triangle diagram containing the
$t\bar{t}hh$ coupling, which contrarily to the triangle diagram
involving the virtual Higgs exchange does not vanish at large
$\hat{s}$.  This is confirmed by taking into account the corrections
of $\mathcal{O}(1/m_{t}^{2})$ to the LET result, which are reported in
App.~\ref{app:expansion}. Compared to the SM we have an additional
contribution $\sim \xi$ from the two-Higgs two-fermion coupling which goes
like $\sim \hat{s}/m_t^2$, and which in contrast to the triangle diagram with virtual Higgs exchange is not
suppressed by the Higgs propagator, see Eqs.~(\ref{eq:exp1}) and (\ref{eq:exp2}). Therefore in MCHM5 for
large $\xi$, where the coupling $c_{2}$ is sizeable (see
Table~\ref{coupvalues}), the corrections do not improve at all the LET
result. For a model-independent study of the $gg\to hh$ process
including only modifications to the top couplings, see
Ref.~\cite{ContinoETAL}.   

While in the LET approximation the contributions of loops of top partners to the
$gg\to hh$ amplitude exactly cancel out with that coming from the
modification of the top Yukawa due to mixing with resonances, the sensitivity to
composite couplings of the full double Higgs production cross section
(computed retaining all dependence on masses) is expected to be much
larger than for $gg\to h$, where it was shown to be negligible. By
direct computation in MCHM5, we will see in Section~\ref{double Higgs
  prod MCHM5} that this is the case, \emph{i.e.} the full $gg\to hh$
cross section has a sizeable sensitivity to the details of the
spectrum of the top partners. This effect is not captured by the simple LET result, which is completely determined by $\xi$. Therefore, while the low-energy theorem provides a useful tool to obtain a rough estimate of the cross section, a complete loop computation is needed to describe correctly the effects of top partners in the $gg\to hh$ process.

\section{Composite Higgs model with extra fermionic resonances}\label{sec:partners}
We consider a composite Higgs scenario with the symmetry group
of the strongly-interacting sector given by $SO(5)$,
which is spontaneously broken down to $SO(4)$ at the scale $f$. In
order to correctly reproduce the fermion charges an additional local
symmetry $U(1)_X$ is introduced, leading to the symmetry breaking
pattern $SO(5)\times U(1)_{X}/SO(4)\times U(1)_{X}$. This is the minimal realisation
including custodial symmetry and which implies four Goldstone bosons
(GBs) transforming as a $\mathbf{4}$ of $SO(4)\sim SU(2)_{L}\times
SU(2)_{R}$. The SM electroweak group $SU(2)_L \times U(1)_Y$ is
embedded into $SO(4)\times U(1)_X$ and the hypercharge $Y$ is then given by
$Y=T_{R}^{3}+X$ \cite{Agashe:2004rs ,Contino:2006qr}. The GBs are
parameterized in terms of the field
\begin{equation}
\Sigma = \Sigma_{0}e^{\Pi/f}\,,\qquad \Pi = -i\sqrt{2}\,T^{\hat{a}}h^{\hat{a}}\,,\qquad \Sigma_{0}=\,(\,0\,,\,0\,,\,0\,,\,0\,,\,1\,)\,,
\end{equation} 
where $T^{\hat{a}}\,\,(\hat{a}=1,\ldots,4)$ are the generators of
$SO(5)/SO(4)\,$, and $h^{\hat{a}}$ are the 4 real GBs. Using the
explicit expressions of the $SO(4)$ and $SO(5)/SO(4)$ generators ($i,j=1,...,5$) 
\begin{align}
T^{a_{L,R}}_{ij}\,=\,&
-\frac{i}{2}\left[\frac{1}{2}\epsilon^{abc}(\delta_{i}^{\,b}\delta_{j}^{\,c}-\delta_{j}^{\,b}\delta_{i}^{\,c})\pm
  (\delta_{i}^{\,a}\delta_{j}^{\,4}-\delta_{j}^{\,a}\delta_{i}^{\,4})\right]\,,\quad
a_{L,R} = 1,2,3 \;, \; a,b,c=1,2,3 \\
T_{ij}^{\hat{a}}\,=\,& -\frac{i}{\sqrt{2}}(\delta_{i}^{\,\hat{a}}\delta_{j}^{\,5}-\delta_{j}^{\,\hat{a}}\delta_{i}^{\,5})\,, \quad \hat{a} = 1,\ldots, 4
\end{align} 
one obtains
\begin{equation}
\Sigma 
=\,\frac{\sin(h/f)}{h}\begin{pmatrix} h_{1}, & h_{2}, & h_{3}, & h_{4}, & h\,\mathrm{cotan}(h/f)  \end{pmatrix}\,,\qquad h = \sqrt{\sum_{\hat{a}} h_{\hat{a}}^{2}}\,. 
\end{equation}
The two-derivative Lagrangian reads
\begin{equation}
\mathcal{L_{\mathrm{kin}}}=\frac{f^{2}}{2}(D_{\mu}\Sigma)(D^{\mu}\Sigma)^{T}\,,\qquad D_{\mu}\Sigma = \partial_{\mu}\Sigma + ig\,W_{\mu}^{a}\Sigma T_{L}^{a}+ig'\,B_{\mu}\Sigma T_{R}^{3}\,.
\end{equation}
By performing an $SO(4)$ rotation, it is always possible to align the Higgs VEV to the $h_{3}$ direction, thus identifying $H=h_{3}$, where $H$ is the Higgs field (with $\left\langle H\right\rangle \neq 0$). Then in the unitary gauge
\begin{align} \label{xi definition}
\Sigma \,=\,& \Sigma_{0} \begin{pmatrix}
1 & & & & \\
 & 1 & & & \\
 & & \cos(H(x)/f) & & -\sin(H(x)/f) \\
  & & & 1 & \\
   & & \sin(H(x)/f) & & \cos(H(x)/f) \end{pmatrix} \equiv \Sigma_{0}\, \zeta(x) \\[0.1cm]
=\,&\,(\,0\,,\,0\,,\,\sin(H/f)\,,\,0\,,\,\cos(H/f)\,)\,,
%
\end{align}
and therefore 
\begin{equation}
\mathcal{L_{\mathrm{kin}}}= \,\frac{1}{2}\partial_{\mu}H\partial^{\mu}H+\frac{g^{2}f^{2}}{4}\sin^{2}\left(\frac{H}{f}\right)\left[W_{\mu}^{+}W^{-\,\mu}+\frac{1}{2\cos^{2}\theta_{W}}Z_{\mu}Z^{\mu}\right]
\end{equation}
which fixes $\,\,f^{2}\sin^{2}(\left\langle H \right\rangle/f)=v^{2}
\simeq (246\,\mathrm{GeV})^{2}\,$. \s

Fermionic resonances are described using the language of partial
compositeness. We introduce vector-like fermions which have quantum
numbers such that they can mix linearly with the SM fermions $q_{L} =
(t_{L},b_{L})^{T}$ and $t_{R}\,$, and which at the same time have
`proto-Yukawa' interactions with the composite Higgs. We introduce
composite fermions transforming as a complete $\mathbf{5}_{2/3}$ under
$SO(5)\times U(1)_{X}\,$. This representation has the
phenomenologically desirable feature that no tree-level corrections to
the $Z$-$b$-$\overline{b}$ coupling arise, provided a discrete
symmetry $P_{LR}$ exchanging the $SU(2)_{L}$ and $SU(2)_{R}$ factors
is enforced \cite{Contino:2006qr,Carena:2006bn}. The composite multiplets can be written as
\begin{equation}
\psi=\frac{1}{\sqrt{2}}\begin{pmatrix} B-X^{5/3} \\
-i (B+X^{5/3}) \\
T+X^{2/3} \\
i(T-X^{2/3}) \\
\sqrt{2}\tilde{T} \end{pmatrix}\,.%
\end{equation}
Under $SU(2)_{L}\times SU(2)_{R}$, a $\mathbf{5}$ of $SO(5)$
decomposes as $\mathbf{5}\sim (\mathbf{2},\mathbf{2})\oplus
(\mathbf{1},\mathbf{1})\,$. The $SU(2)_{L}$ doublets $Q=(T,B)^{T}$ and
$X=(X^{5/3},X^{2/3})^{T}$ form a bidoublet $(\mathbf{2},\mathbf{2})$
under $SU(2)_L\times SU(2)_R$, while $\tilde{T}$ is a singlet
$(\mathbf{1},\mathbf{1})\,$. The SM quantum numbers of the composite
fields are summarized in Table \ref{tab:fermion charges}.
\begin{table}[ht]
		\centering
		\setlength{\tabcolsep}{5pt}
		\begin{tabular}{c|c|c|c|c} 
	$\mathrm{field}$ & $T_{L}^{3}$   &  $T_{R}^{3}$ & $Y$ &$Q_{el}=T_L^3+Y$  \\   
	\hline
	$T$ & $+1/2$ & $-1/2$ & $1/6$ & $+2/3$  \\
	$B$ & $-1/2$ & $-1/2$ & $1/6$ & $-1/3$ \\
	$X^{5/3}$ & $+1/2$ & $+1/2$ & $7/6$ & $+5/3$ \\
	$X^{2/3}$ & $-1/2$ & $+1/2$ & $7/6$ & $+2/3$ \\
	$\tilde{T}$ & $0$ & $0$ & $2/3$ & $+2/3$ \\
	
		\end{tabular}
		\caption{SM quantum numbers of the composite fermions in
                  $\psi$. The last column denotes the electric charge.}
	\label{tab:fermion charges}
\end{table}
Note that $Q$ has the same quantum numbers as the elementary doublet 
$q_{L}=(t_{L},b_{L})^{T}$, whereas $\tilde{T}$ has the same quantum
numbers as $t_{R}$. The doublet $X$ is
peculiar of the $\mathbf{5}$ representation (it is absent in the most
minimal case of the spinorial representation $\mathbf{4}$). Taking
into account only one set of fermionic composites, the Lagrangian for
the fermion sector then reads 
\beq
\mathcal{L}_{f}\,&=&\,
i\overline{q}_{L}\slashed{D}q_{L}+i\overline{t}_{R}\slashed{D}t_{R}+i\overline{b}_{R}
\slashed{D}b_{R} + i\overline{\psi}_{L}\slashed{D}\psi_{L}+
i\overline{\psi}_{R}\slashed{D}\psi_{R} \nonumber \\ 
&- & yf (\overline{\psi}_{L}\Sigma^{T})(\Sigma\psi_{R}) -
M_{0}\bar{\psi}_{L}\psi_{R} +\mathrm{h.c.} \label{PartComplagrangian}  \\
&- & \Delta_{L}\overline{q}_{L}Q_{R}-
\Delta_{R}\overline{\tilde{T}}_{L}t_{R}+ \mathrm{h.c.} \; ,
\nonumber 
\eeq
where the covariant derivative acting on $\psi$ is given by
\begin{equation}
D_{\mu}\psi=\left[\partial_{\mu}-ig\,W_{\mu}^{a}T_{L}^{a}-ig'\,B_{\mu}(T_{R}^{3}+X)\right]\psi\,
, \qquad X = (2/3)\mathds{1}_{5}\,.  
\end{equation}
For later convenience, we also give the embedding of $q_{L}$ and $t_{R}$ in $SO(5)$ vectors:
\beq
\mathcal{Q}_{L}=\frac{1}{\sqrt{2}}\begin{pmatrix} b_{L} & -ib_{L} & t_{L} & it_{L} & 0 \end{pmatrix}^{T}\,, \qquad 
\mathcal{T}_{R}=\frac{1}{\sqrt{2}}\begin{pmatrix} 0 & 0 & 0 & 0 & \sqrt{2}t_{R} \end{pmatrix}^{T}\,.
\eeq
Using these expressions the linear mixings can be rewritten as
\beq
\mathcal{L}_{mix}= - \Delta_{L}\overline{\mathcal{Q}}_{L}\psi_{R} - \Delta_{R}\overline{\psi}_{L}\mathcal{T}_{R} + \mathrm{h.c.}
\eeq
From the Lagrangian in Eq.~(\ref{PartComplagrangian}) we obtain the mass
terms and the Yukawa couplings. The mass matrix reads
\beq \label{fermionmassmatrix}
-{\cal L}_m = \overline{
\left( \begin{array}{c} t_L \\ T_L \\ X^{2/3}_L \\
  \tilde{T}_L \end{array} \right) }
\left( \begin{array}{cccc} 
0 & \Delta_{L} & 0 & 0 \\
0 & M_{0}+\frac{fys^{2}}{2} & \frac{yfs^{2}}{2} & \frac{yfsc}{\sqrt{2}} \\
0 &  \frac{yfs^{2}}{2} & M_{0}+\frac{fys^{2}}{2} & \frac{yfsc}{\sqrt{2}} \\
\Delta_{R} & \frac{yfsc}{\sqrt{2}} & \frac{yfsc}{\sqrt{2}} & M_{0}+yfc^{2} \end{array} \right) 
\left( \begin{array}{c} t_R \\ T_R \\ X^{2/3}_R \\ \tilde{T}_R \end{array} \right)+\mathrm{h.c.}
\;,
\eeq
where we have introduced the abbreviation $s \equiv \sin (\left\langle H\right\rangle /f) =v/f$ and analogously \mbox{$c\equiv  \cos (\left\langle H\right\rangle /f)$}.
%
%
%
The diagonalization of the matrix, which mixes fundamental fields and
composite states, is immediate before electroweak symmetry breaking,
{\it i.e.} for $v=0$. Then the mass terms become
diagonal after the rotations 
\begin{align} \nonumber
&\,\begin{pmatrix} q_{L} \\ Q_{L} \end{pmatrix} \to \begin{pmatrix} \cos\phi_{L} & \sin\phi_{L} \\
-\sin\phi_{L} & \cos\phi_{L} \end{pmatrix} \begin{pmatrix}q_{L} \\ Q_{L} \end{pmatrix}\,,\qquad \tan\phi_{L}=\frac{\Delta_{L}}{M_{0}} \\ \label{PCrotations}
&\,\begin{pmatrix} t_{R} \\ \tilde{T}_{R} \end{pmatrix} \to \begin{pmatrix} \cos\phi_{R} & \sin\phi_{R} \\
-\sin\phi_{R} & \cos\phi_{R} \end{pmatrix} \begin{pmatrix}t_{R} \\ \tilde{T}_{R} \end{pmatrix}\,,\qquad \tan\phi_{R}=\frac{\Delta_{R}}{M_{0}+yf}\,. 
\end{align}
In this limit the top is massless, whereas the masses of the
composite states are 
\begin{equation} \label{LO masses}
M_{Q}=\frac{M_{0}}{c_{L}}\,,\qquad M_{X}= M_{0}\,,\qquad M_{\tilde{T}}=\frac{yf+M_{0}}{c_{R}}\,.
\end{equation}
Electroweak symmetry breaking effects generate additional
mixings, which also involve $t_{L}$ and $t_{R}$. Thus the top becomes
massive due to its mixing with composite states. At the leading order
in $\xi\equiv v^{2}/f^{2}$ we have
\begin{equation}
m_{t}=y\sin\phi_{L}\sin\phi_{R}\frac{v}{\sqrt{2}}\,.
\end{equation}   
Furthermore, the masses of the composite fermions in Eq.~(\ref{LO masses})
get corrections of order $\xi$. The Lagrangian
Eq.~\eqref{PartComplagrangian}, however,  does not give rise to a mass
for the bottom quark, because there is no composite in $\psi$ that has the
right quantum numbers to mix with $b_{R}$. Rather than introducing
another fermionic multiplet (e.g., a $\mathbf{5}_{-1/3}$) to solve
this issue, we introduce a small breaking of the partial compositeness
pattern, namely a Yukawa coupling of the Higgs to elementary states
\begin{equation}
\mathcal{L}_{b}=\,-\lambda_{b}\overline{q}_{L} H b_{R} + \mathrm{h.c.}\,,
\end{equation} 
where $H$ is the Higgs doublet. We will, however, neglect the small
effects proportional to $\lambda_{b}\,$. 
%
%
%

Expanding the proto-Yukawa term up to second order in the physical Higgs $h$ we obtain the Higgs Yukawa 
couplings and the two-Higgs two-fermion couplings. The Yukawa coupling part of the Lagrangian reads
\begin{equation}
-\mathcal{L}^{hff}=y\, h\,\overline{\begin{pmatrix} t_L \\ T_L \\ X^{2/3}_L \\
  \tilde{T}_L
 \end{pmatrix}} 
 \underbrace{ \begin{pmatrix}
0&0&0&0\\
0&s c & s c & \frac{1 - 2s^{2}}{\sqrt{2}}\\
0&s c & s c & \frac{1- 2s^{2}}{\sqrt{2}}\\
0& \frac{1- 2s^{2}}{\sqrt{2}} & \frac{1- 2s^{2}}{\sqrt{2}} & -2 s c
\end{pmatrix}}_{G_{\sssty{hf\bar{f}}}}
 \begin{pmatrix}
   t_R \\ T_R \\ X^{2/3}_R \\ \tilde{T}_R\end{pmatrix}+\mathrm{h.c.}
\label{eq:yukcoup}
\end{equation}
For the two-Higgs two-fermion interactions we find
\begin{equation}
 -\mathcal{L}^{hhff}=\f{y}{2f}\, h^{2} \,\overline{\begin{pmatrix} t_L \\ T_L \\ X^{2/3}_L \\
  \tilde{T}_L
 \end{pmatrix}} 
 \underbrace{ \begin{pmatrix}
0&0&0&0\\
0&1-2\,s^2 & 1-2\, s^2 & -2\sqrt{2} sc \\
0&1-2\,s^2 & 1-2\,s^2 & -2\sqrt{2} sc\\
 0&-2\sqrt{2} sc & -2\sqrt{2} sc &-2\left(1-2\,s^2\right)	
\end{pmatrix}}_{G_{\sssty{hhf\bar{f}}}}
 \begin{pmatrix}
    t_R \\ T_R \\ X^{2/3}_R \\ \tilde{T}_R\end{pmatrix}+\mathrm{h.c.}
\label{eq:2h2qcoup}
\end{equation}
After the rotation into the mass eigenstate basis, the two matrices $G_{\sssty{hf\bar{f}}}$ and
$G_{\sssty{hhf\bar{f}}}$ yield the single and double Higgs couplings to fermions, respectively,
which will be needed for the calculation of the single and double
composite Higgs production cross sections through gluon fusion. 

\subsection{Constraints from electroweak precision data and flavor physics}
\label{sec:EWPT}

\begin{figure}[ht]
	\centering
	\includegraphics[width=0.49\linewidth]{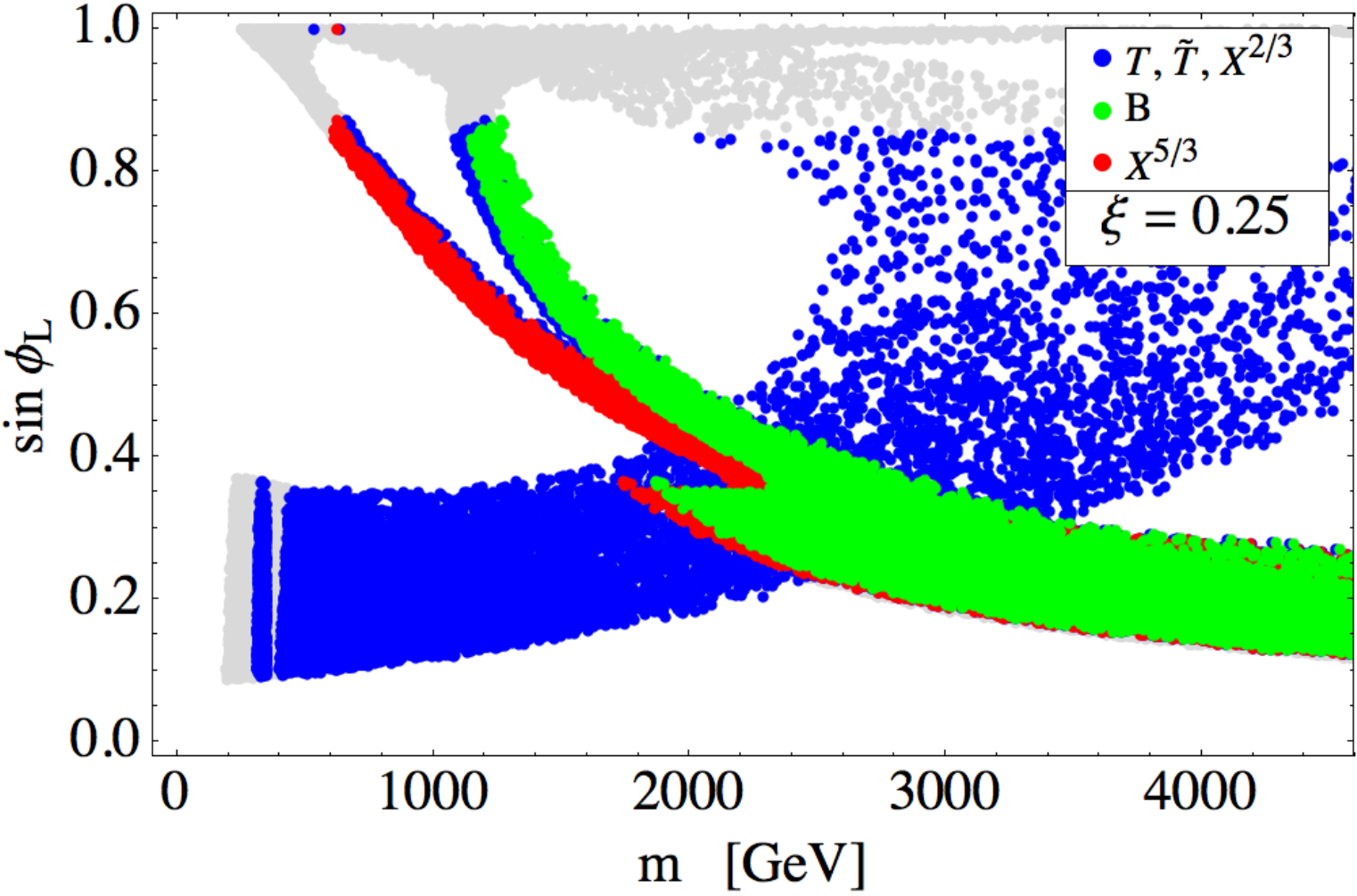}
	\includegraphics[width=0.49\linewidth]{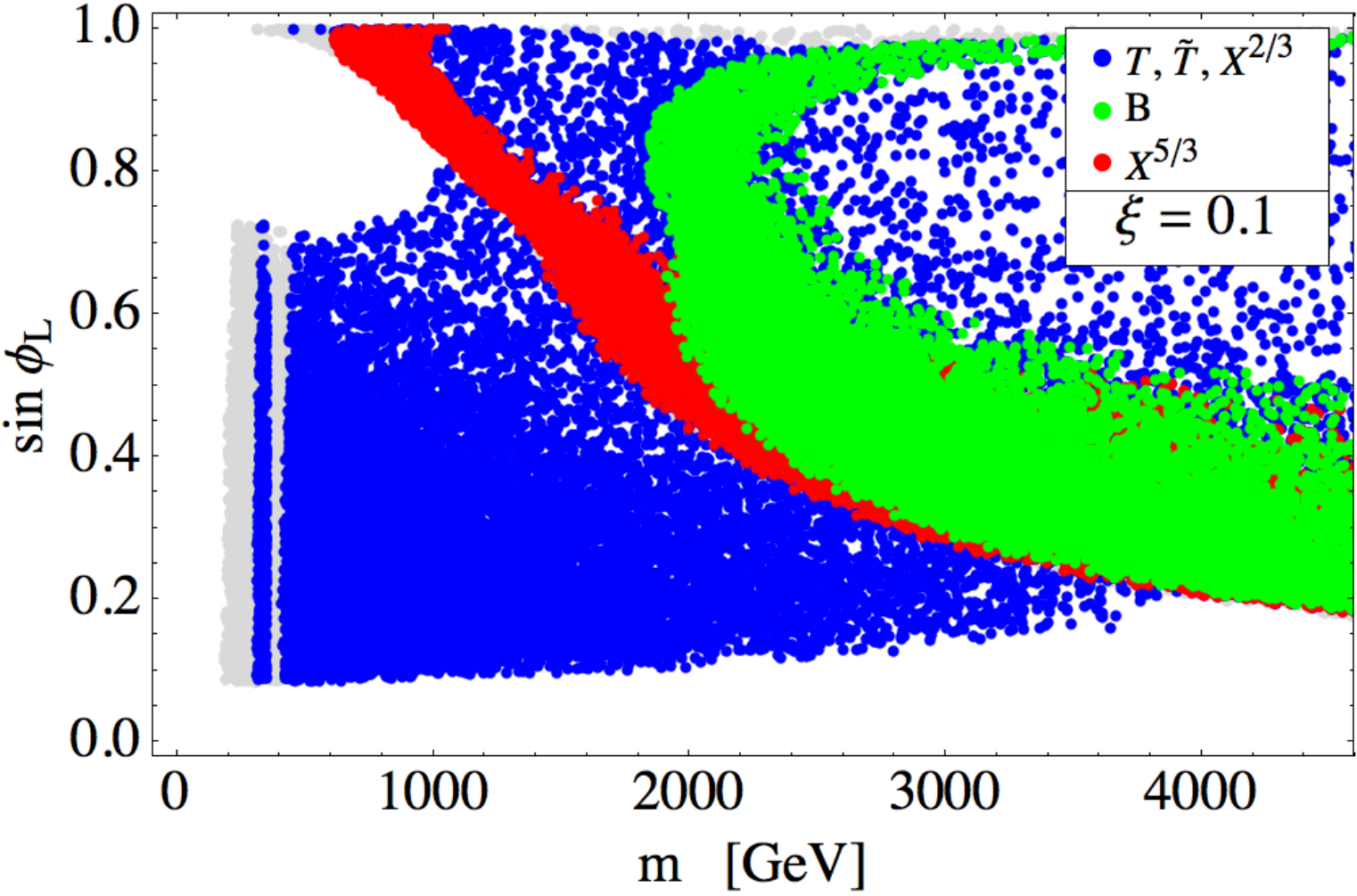}
	\caption{Physical mass spectrum of the composite fermions for
          a sample of points passing the electroweak precision tests,
          as a function of the left-handed top compositeness for $\xi=0.25$ (left) and $\xi=0.1$ (right). The
          blue/dark gray
          points are top-like fermions (charge $+2/3$),
          the green/fair gray
          points bottom-like (charge $-1/3$), and the red/medium gray
          ones correspond to the exotic $X$ (charge
          $+5/3$). Light gray 
          points are excluded by present collider constraints, see Section~\ref{sec:searches}.}
	\label{fig:spectrum}
\end{figure}

The strongest experimental constraints on composite Higgs models still
come from the electroweak precision measurements at the $Z$ pole mass
at LEP. A convenient description of LEP precision data is given in terms of
the parameters $\epsilon_1$, $\epsilon_2$, $\epsilon_3$ and
$\epsilon_b$~\cite{Altarelli:1990zd, Altarelli:1991fk,
  Altarelli:1993sz}. These parameters are on the one hand measured with
high precision~\cite{ALEPH:2005ab}, and on the other hand can easily
be computed theoretically. In addition to the SM contribution present in the decoupling limit $f \to \infty$, the MCHM5
contributes to the $\epsilon$ parameters through three different
effects. The first beyond the SM (BSM) effect arises from the modified
coupling of the Higgs to $W$ and $Z$ gauge bosons, which induces a
logarithmically divergent contribution to the oblique parameters $T$
and $S$, or equivalently to $\epsilon_1$ and $\epsilon_3$. The
contribution is cut-off by the mass $m_\rho$ of the first composite
vector resonance~\cite{Barbieri:2007bh},
\begin{equation}
	\Delta\epsilon_1^{\textrm{IR}} = -\frac{3 \, \alpha(M_Z)}{16 \pi \cos^2\theta_W} \xi \log\left( \frac{m_\rho^2}{m_h^2} \right),
	\hspace{1cm}
	\Delta\epsilon_3^{\textrm{IR}} = \frac{\alpha(M_Z)}{48 \pi \sin^2\theta_W} \xi \log\left( \frac{m_\rho^2}{m_h^2} \right).
\end{equation}
The second effect is the direct contribution of the vector $\rho$ and axial-vector $a$ resonances to the $S$ parameter, which in the MCHM is found to be (see Ref.~\cite{Contino:2010rs})
\begin{equation}
	\Delta\epsilon_3^{\textrm{UV}} = \frac{m_W^2}{m_\rho^2} \left( 1 + \frac{m_\rho^2}{m_a^2} \right) \cong 1.36 \, \frac{m_W^2}{m_\rho^2}\,.
\end{equation}%
\begin{figure}[b]
	\centering
	\includegraphics[width=0.49\linewidth]{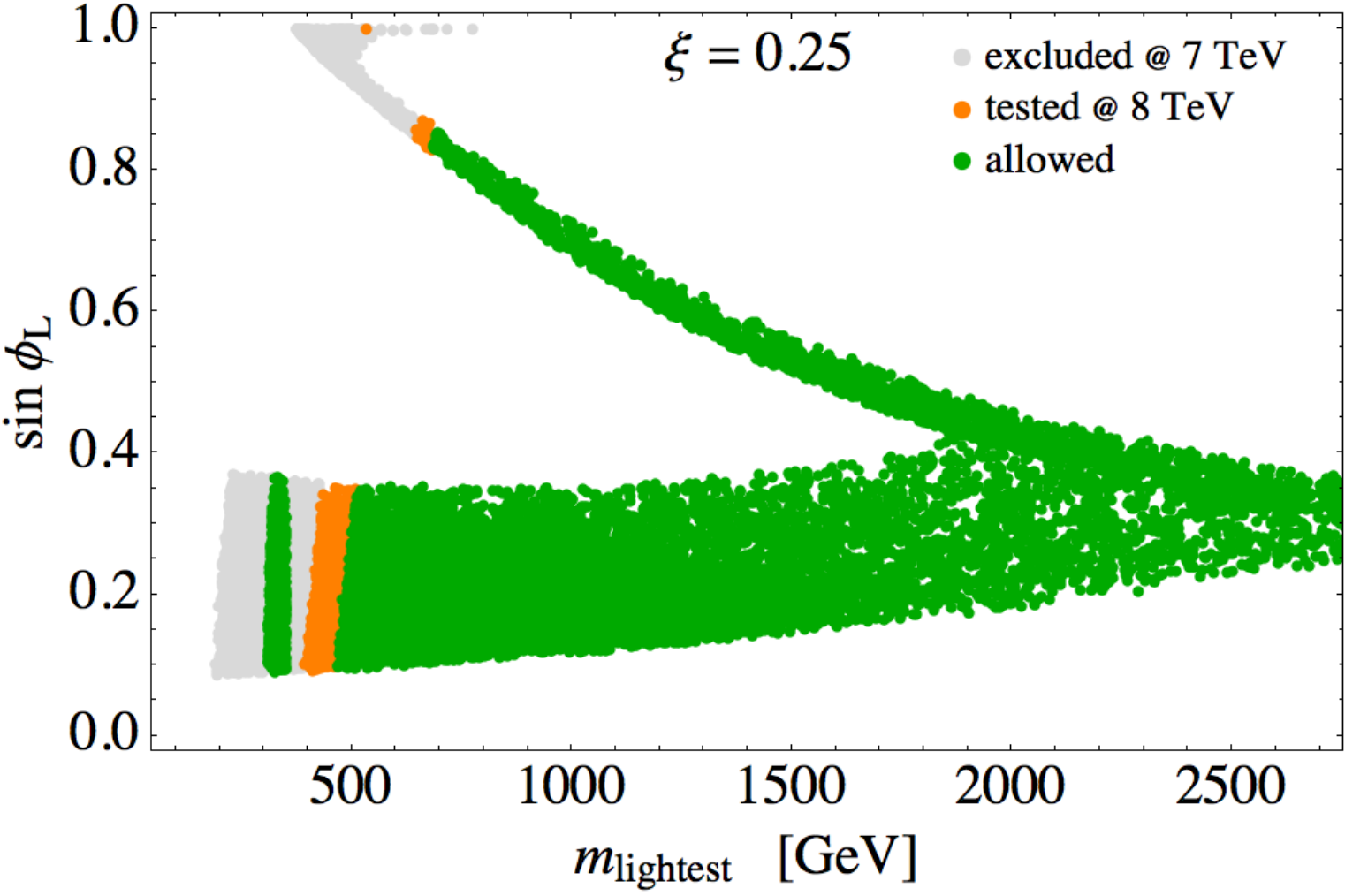}
	\includegraphics[width=0.49\linewidth]{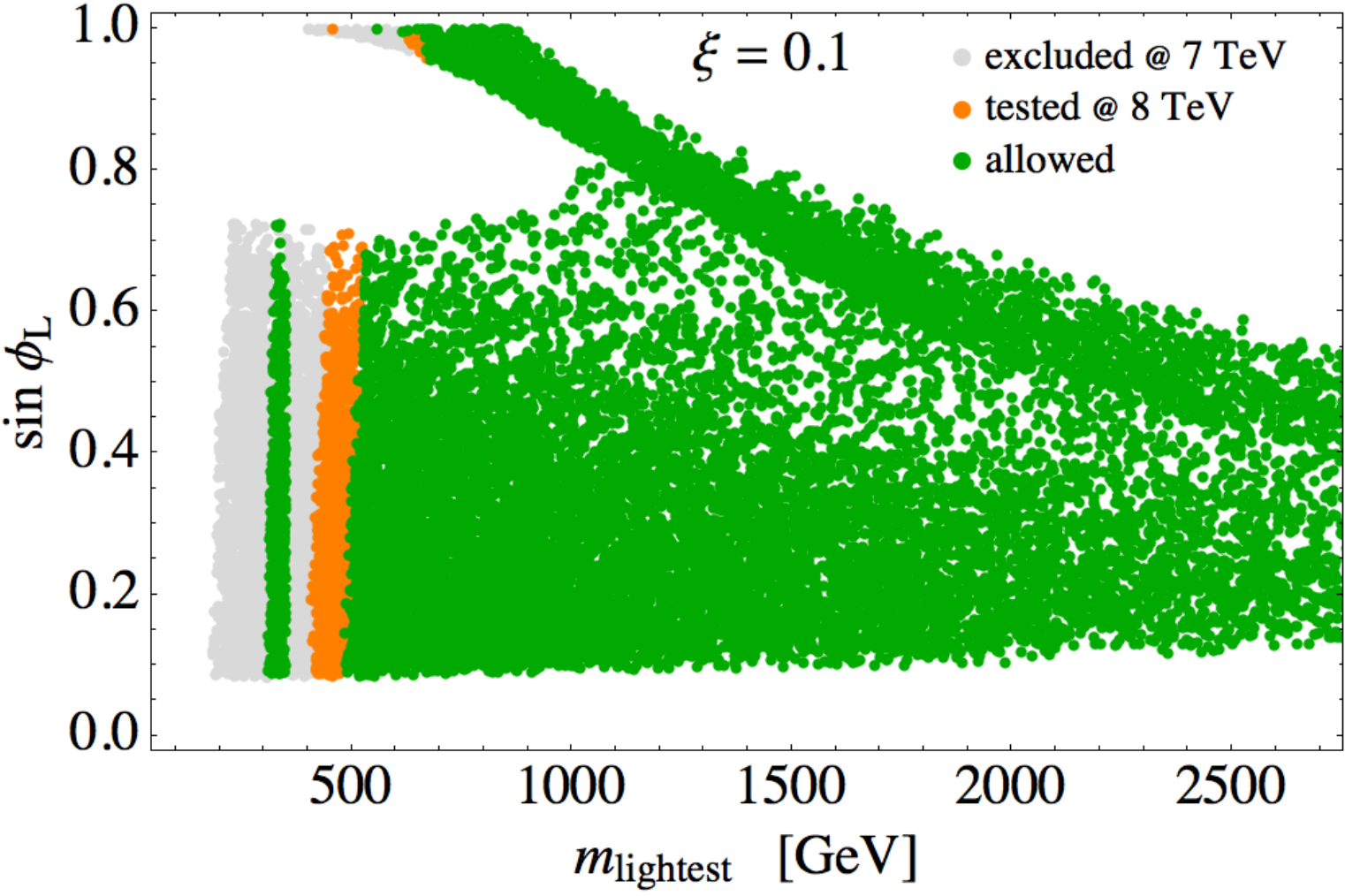}
	\caption{A sample of parameters passing the $\chi^2$-test of
          electroweak precision observables, displaying  the
          compositeness of the left-handed top versus the mass of the
          lightest top partner, for $\xi = 0.25$ (left) and $\xi =
          0.1$ (right). The points in light gray do not pass the
          direct collider constraints. Points in orange/medium gray
          pass the present constraints but will be tested by the LHC
          running at 8~TeV with an integrated luminosity of \mbox{$15
            \,\mathrm{fb}^{-1}$}, see Section~\ref{sec:searches}.}
	\label{fig:EWPT}
\end{figure}%
In the second equality, we have used the relation $m_a / m_\rho \cong
5/3$, obtained in the five-dimensional realization of the 
model~\cite{Agashe:2004rs}. The third and last contribution to
electroweak precision parameters comes from the top partners at one
loop, giving contributions both to the $T$ parameter and the 
$Z$-$\bar{b}$-$b$ vertex, {\it i.e.} respectively to $\epsilon_1$ and 
$\epsilon_b$~\cite{Agashe:2005dk, Lodone:2008yy, Gillioz:2008hs,
Anastasiou:2009rv}. Computing the precise value of these
contributions requires the numerical diagonalisation of the mass matrix of
the top quark and its partners, which depends on the parameters 
$\Delta_L$, $\Delta_R$, $M_0$, $y$ and $f$. The requirement that the
top mass matches the measured value $m_t = 173.3$~GeV allows, however,
to express the corrections to $\epsilon_1$ and $\epsilon_b$ in terms
of four dimensionless parameters,
\begin{equation}
	\Delta\epsilon_1^{fermions} = f_1\left( \xi, \phi_L, \phi_R, R \right),
	\hspace{1cm}
	\Delta\epsilon_b^{fermions} = f_b\left( \xi, \phi_L, \phi_R, R \right),
\end{equation}
where $\xi$, $\phi_L$, $\phi_R$ are defined above and $R = \left( M_0
  + y f \right) / M_0$. The function $f_1$ is computed exactly at one
loop, while for $f_b$ only the longitudinal polarisations of the gauge
bosons are taken into account in the loop. The values obtained in
this way are consistent with the full one-loop result of
Ref.~\cite{Anastasiou:2009rv}. The agreement of the model with experimental
data is then assessed through a $\chi^2$ test, described in detail in
App.~\ref{appendix:chisquared}. The latest electroweak precision
data are used, including the 2012 update of the $W$ mass. Fixing the Higgs mass to $m_h = 125$~GeV, the model is completely determined by the five parameters $\xi$, $\phi_L$, $\phi_R$, $R$ and $m_\rho$. Over the latter four a scan is performed for $\xi$ fixed to two representative values, namely 0.25 and 0.1.
The results are displayed in Fig.~\ref{fig:EWPT} for the left-handed compositeness angle $\phi_L$ versus the mass of the lightest top partner.
Note that the value of $R$ is bounded by the
requirement that $y < 4\pi$ and that we impose furthermore the
constraint $|V_{tb}| >0.77$~\cite{Group:2009qk}.
In Fig.~\ref{fig:spectrum}, the whole spectrum of composite fermions is shown for a sample of parameters passing the EWPT. The green points correspond to $B$, the red ones to $X^{5/3}$, and the blue points for each set of parameters denote the top partners $T$, $X^{2/3}$ and $\tilde{T}$, which cannot be properly distinguished one from another once the rotation in the physical basis is performed. At leading order in $v/f$, composite fermions within an electroweak doublet have, however, the same mass, so that the green points describe approximately the mass of the $(T, B)$ doublet and the red ones the mass of the $( X^{5/3}, X^{2/3} )$ doublet. The blue points far from the red and green regions in Fig.~\ref{fig:spectrum} can therefore be interpreted as singlets $\tilde{T}$.

There are two regions of the parameter space compatible with EWPT where in addition at least one of the top partners is light, as generically needed in order to have a light enough Higgs.
 The first region corresponds to low values of the top compositeness angle $\phi_L$, where the lightest top partner is typically the singlet $\tilde T$.
In this case the fermion bidoublet is always heavier than 1.5~TeV (see Fig.~\ref{fig:spectrum}) and decouples. Note that the right-handed top must then be very composite in order to yield the correct top Yukawa coupling $y_t$. The second region corresponds to large values of $\sin\phi_L$, for which the top-bottom doublet becomes fully composite.
In this second region, the `custodian' doublet $X$ is very light, having a mass well below a TeV. Since the $X$ doublet contains an exotic charge $5/3$ fermion (which turns out to be the lightest new fermion for large $\sin\phi_{L}$), this region is very sensitive to direct collider constraints, as will be discussed in the next section. An intermediate region with moderate values of $\sin\phi_L$ is also allowed by precision data, although all new fermions are rather heavy, above 1~TeV, so it could be difficult to obtain a light enough Higgs in this region.  A thorough discussion of the implications of top compositeness is contained in Ref.~\cite{Pomarol:2008bh} (see also Ref.~\cite{Lillie:2007hd}). A comment is however in order: in Ref.~\cite{Pomarol:2008bh}, when $t_{R}$ mixes with a $(\mathbf{1},\mathbf{1})$ of $SU(2)_{L}\times SU(2)_{R}$ (which is the case in MCHM5) a highly composite right-handed top is not accompanied by anomalously light fermionic resonances, because there is no custodial partner of the state mixing with $t_{R}$, the latter being $\tilde{T}$ in MCHM5. Indeed one can write from Eq.~\eqref{LO masses} $M_{\tilde{T}}=\Delta_{R}/\sin\phi_{R}\to \Delta_{R}$ when $\sin\phi_{R}\to 1$, implying that $\tilde{T}$ is not necessarily light in presence of a strongly composite $t_{R}$. However, EWPT select a light $\tilde{T}$ for large $\sin\phi_{R}$, as discussed for example in Ref.\cite{Gillioz:2008hs}.    

Note finally that the constraints on the parameter space from electroweak precision data can be significantly relaxed by extending the fermion sector of the model~\cite{Anastasiou:2009rv}.

Additional constraints on the model come from flavor
physics. Composite Higgs models generically give rise to
four-fermion operators which contribute to
flavor-changing processes and to electric dipole
moments. Low values of the compositeness scale $f$ as considered
in this paper are allowed if the strong sector is
flavor-symmetric, so that MFV can be implemented \cite{Redi:2011zi}.
In this case both flavor-changing processes and electric dipole
moments are inhibited, but the MFV assumption requires a large
degree of compositeness also for light quarks, which are therefore sizably coupled to the strong sector resonances, leading to a different phenomenology. Experimental constraints can be described in an effective formalism, in which four-fermion
operators arise after integrating out the vector resonances. The most relevant operator is
\begin{equation}
	\frac{g_\rho^2}{4 m_\rho^2} \left( \sin\phi_L \right)^4
		\left( \bar{q}_L \gamma^\mu t^a q_L \right)
		\left( \bar{q}_L \gamma_\mu t^a q_L \right),
\end{equation}
where $t^{a}$ are the generators of $SU(3)_{c}$, which imposes a constraint on the size of the mixing angle $\phi_L$.
From the most recent experimental dijet angular distributions \cite{Chatrchyan:2012bf}, the bound is
\begin{equation}
	\frac{g_\rho^2}{4 m_\rho^2} \left( \sin\phi_L \right)^4 \lesssim \frac{2\pi}{\left( 7.5~\textrm{TeV} \right)^2}
	\hspace{0.5cm} \Rightarrow \hspace{0.5cm}
	\left( \sin\phi_L \right)^2 \lesssim \frac{f}{1.5~\textrm{TeV}} \; ,
\end{equation}
or equivalently $\sin\phi_L \lesssim 0.6$ for $\xi = 0.25$ and
$\sin\phi_L \lesssim 0.7$ for $\xi = 0.1$. Similar bounds apply
to the compositeness of right-handed quarks. Minimal Flavor Violation with left-handed compositeness is in addition strongly constrained by EWPT \cite{Redi:2011zi}.

However, it has been recently pointed out \cite{Redi:2012uj} that it is possible to treat the top differently from the light quarks, thus deviating from MFV. Flavor bounds are still satisfied, but since the first two generations are mostly elementary the constraints from EWPT and from searches for compositeness are relaxed. In this setup left-handed and right-handed top compositeness are both viable, and the phenomenology is expected to be analogous to the case where the strong sector is flavor-anarchic, given that the light generations are mostly elementary.

\subsection{Constraints from searches for heavy
  fermions \label{sec:searches}}
\begin{figure}[b]
 \begin{center}
   \includegraphics[width=.45\textwidth]{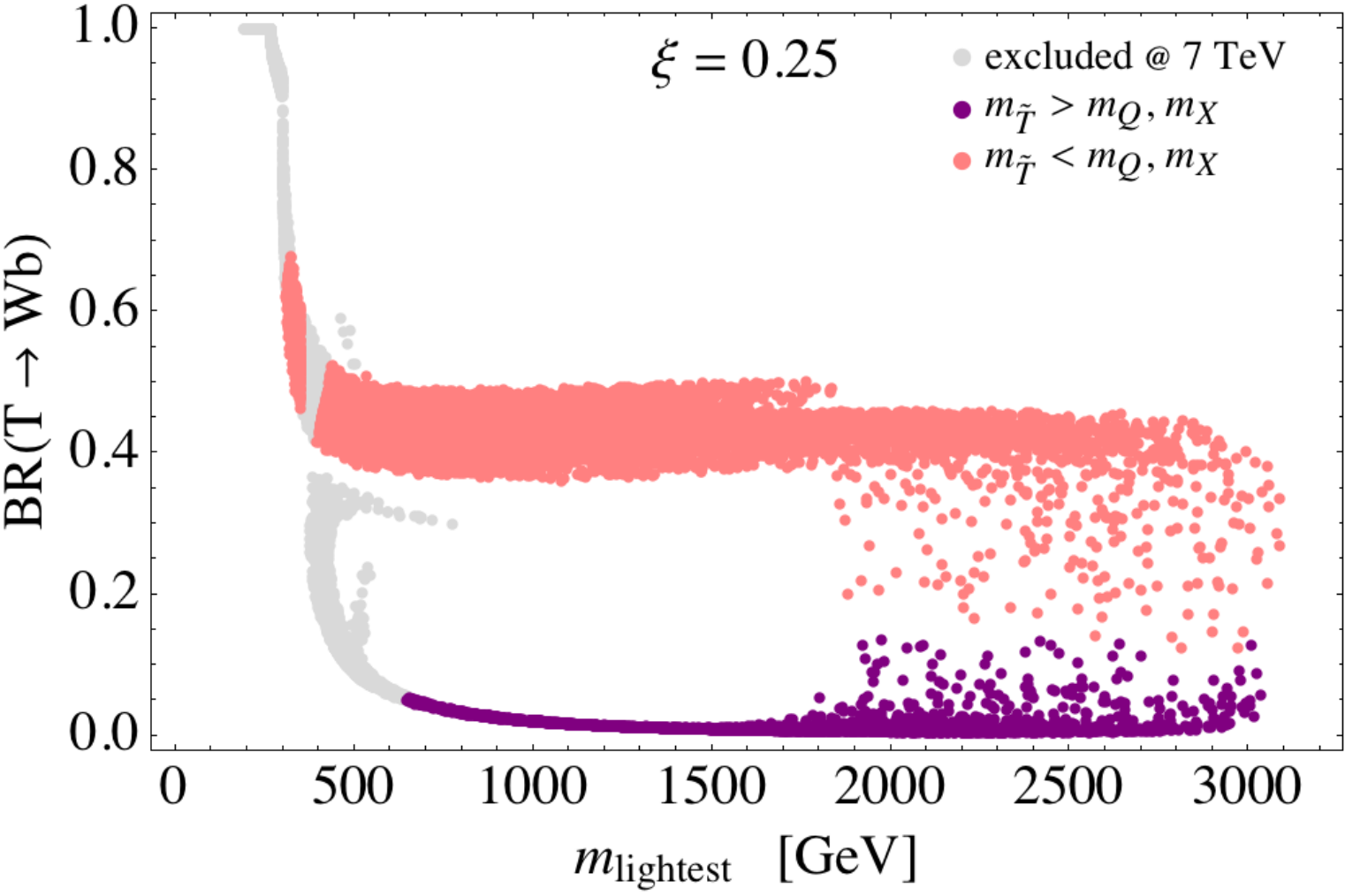}
\hspace*{0.5cm}
   \includegraphics[width=.45\textwidth]{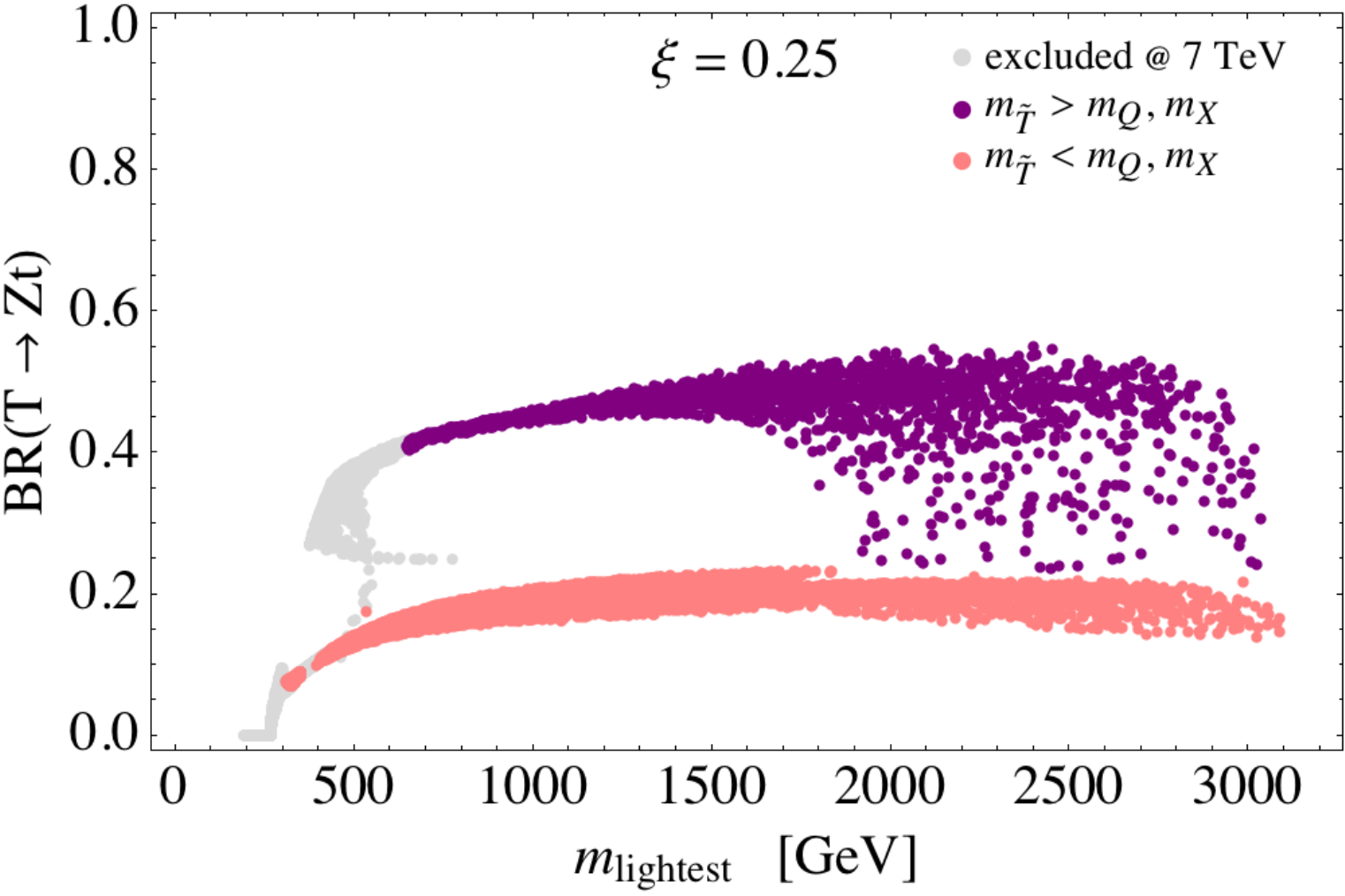} \\[0.5cm]
   \includegraphics[width=.45\textwidth]{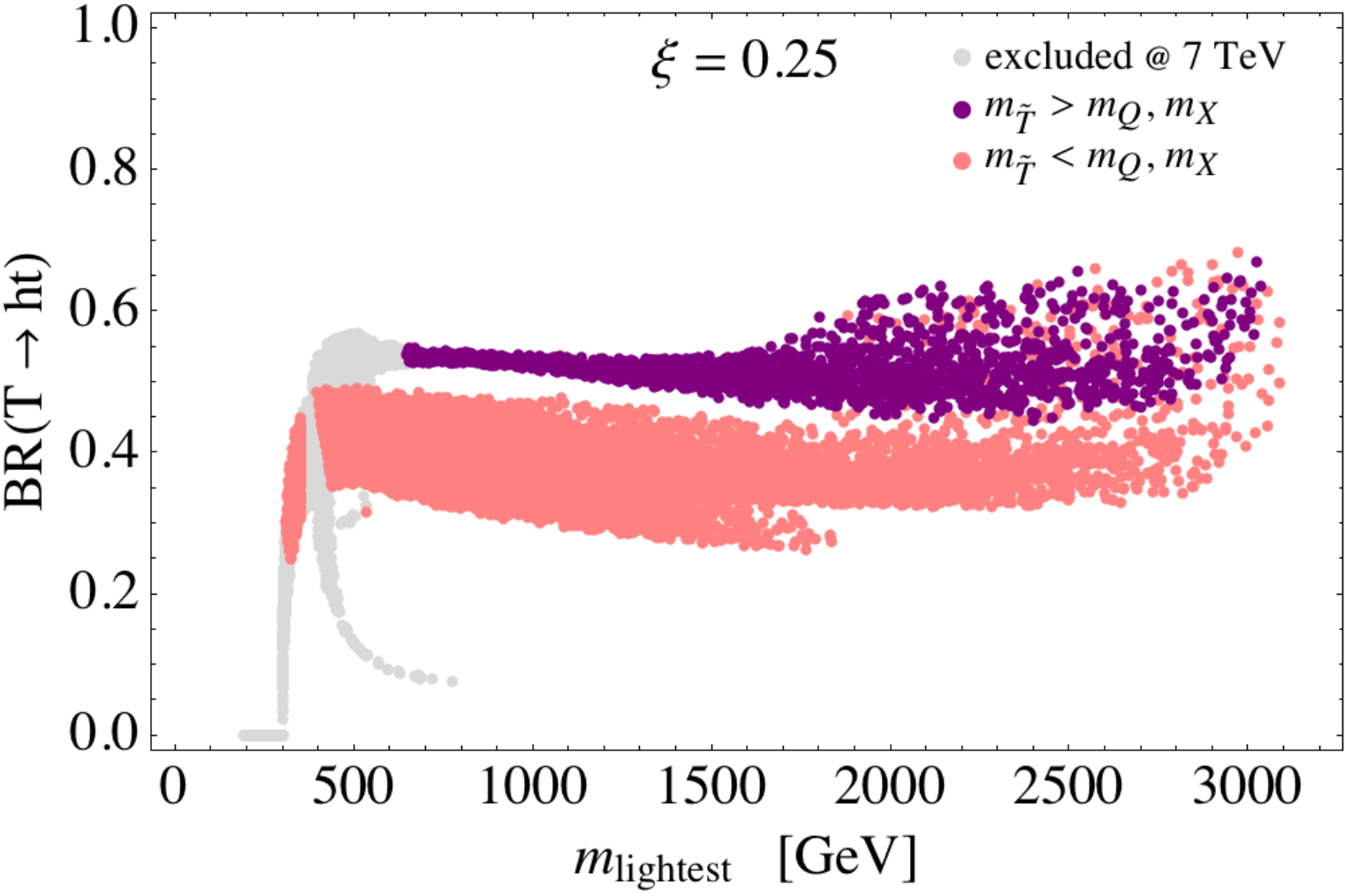}
 \end{center}
\vspace*{-0.4cm}
\caption{Branching ratios of the lightest top partner into $W^+b$
  (upper left), $Zt$ (upper right) and $ht$ (lower) as a function of
  its mass for $\xi=0.25$.  Points for which the singlet (doublet)
  is the lightest top partner are shown in pink/medium gray
  (purple/dark gray).}
\label{fig:tbranch}
\end{figure}
Expanding the composite Yukawa coupling we obtain the
leading interactions between one heavy fermion and two SM particles,
which mediate the decay of the heavy states 
\begin{align} \nonumber
\mathcal{L}_{\mathrm{Y}}=&\, ys_{L}c_{R}\left(\overline{b}_{L}\pi^{-}+\frac{\overline{t}_{L}}{\sqrt{2}}(h-i\pi^{0})\right)\tilde{T}_{R}+ y\frac{s_{R}}{\sqrt{2}}\overline{X}_{2/3L}\left(h+i\,\pi^{0}\right)t_{R} + \mathrm{h.c.} \\ \label{Yukawas}
&+\frac{y}{\sqrt{2}}s_{R}c_{L}\overline{T}_{L}\left(h-i\pi^{0}\right)t_{R} + ys_{R}c_{L}\overline{B}_{L}t_{R}\pi^{-} - ys_{R}\overline{X}_{5/3L}t_{R}\pi^{+} + \mathrm{h.c.}  \; , 
\end{align}
where we have already performed the rotations $\propto \phi_{L,R}\,$. From the Goldstone equivalence theorem then follow the leading order branching ratios (in the limit $M_{\psi}\gg m_{Z},m_{h}$)
\begin{align*}
\mathrm{BR}(\tilde{T}\to Wb)\,=&\, \frac{1}{2}\,,\qquad \mathrm{BR}(\tilde{T}\to Zt)\,=\,\mathrm{BR}(\tilde{T}\to ht)\,=\,\frac{1}{4}\,; \\
\mathrm{BR}(X^{2/3}\to Zt)\,=\,& \mathrm{BR}(X^{2/3}\to ht)\,=\,\frac{1}{2}\,,\qquad \mathrm{BR}(X^{5/3}\to Wt)\,=\,1\,; \\
\mathrm{BR}(T\to Zt)\,=\,& \mathrm{BR}(T \to ht)\,=\,\frac{1}{2}\,,\qquad \mathrm{BR}(B\to Wt)\,=\,1\,.
\end{align*}
However, in our analysis of the electroweak and collider constraints
we will keep all orders in $\xi$, by performing a full numerical
diagonalization of the mass matrix in the top sector and computing the
couplings of the mass eigenstates to gauge bosons and to the Higgs boson in the
unitary gauge. Complete formulae for the partial decay widths of heavy
fermions into SM fields are given in App.~\ref{App:partialwidths}. As we will see, for the relatively large
values of $\xi$ that we consider, sizeable corrections to the leading
approximations listed above arise. 
\begin{figure}[htb]
\begin{center}
\includegraphics[width=.5\textwidth]{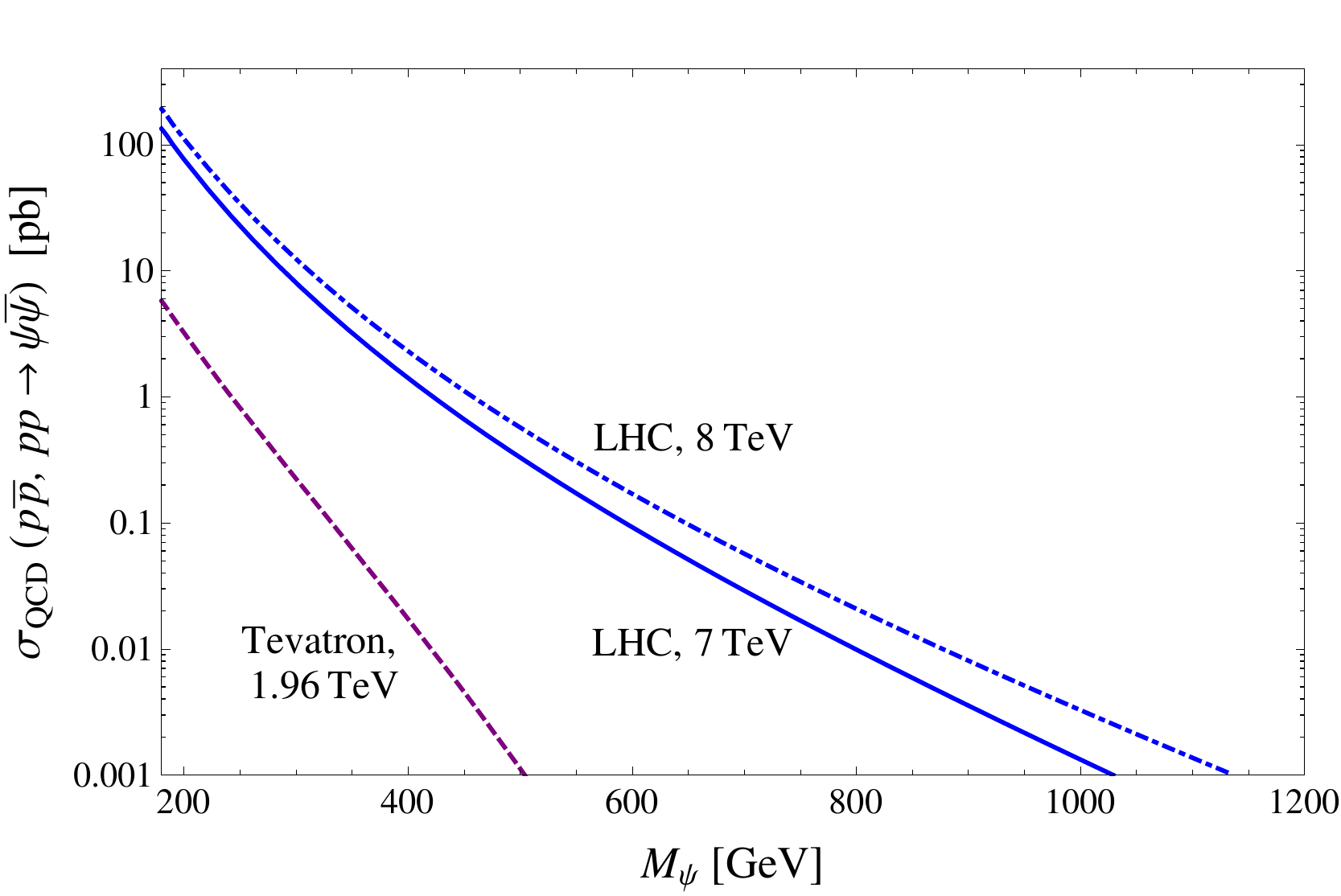}
\end{center}
\caption{Cross sections for QCD pair production of heavy fermions at approximate NNLO, at the Tevatron (dashed), at LHC7 (solid) and at LHC8 (dot-dashed). The cross sections were computed using HATHOR \cite{hathor}, and MSTW2008 PDFs.}
\label{fig:heavyprod}
\end{figure}
In Fig.~\ref{fig:tbranch} we show the branching ratios of the lightest
top partner as a function of its mass for $\xi=0.25$, for a set of points in parameter space which are compatible with EWPT. The pink points are the branching ratios in case the lightest top partner is the singlet. The purple
points correspond to the lightest partner being the doublet. Compared to the approximate formulae the branching
ratios into $ht$ are a bit enhanced while the ones into $Zt$ are somewhat reduced. \s

In certain regions of the parameter space, some fermionic resonances
can be very light, thus rendering constraints from direct searches for
heavy fermions at the LHC and Tevatron relevant. The experimental
collaborations have performed several searches for pair-produced heavy
fermions, with subsequent decay into the final states
$WbWb,\,ZtZt,\,WtWt\,$. Since pair-production of the heavy fermions is
a QCD process, the cross section $\sigma(pp,\,p\bar{p}\to
\psi\overline{\psi})\,$, with $\psi$ being a generic heavy fermion, only
depends on $M_{\psi}\,$. The constraint from {\it e.g.} a search for $\psi\overline{\psi}\to WbWb$ at the LHC will read
\begin{equation}
\sigma_{QCD}(pp\to \psi\overline{\psi})\times \mathrm{BR}(\psi\to Wb)^{2}\leq \sigma_{exp}\,,
\end{equation}
where $\sigma_{exp}$ is the upper bound on the cross section, as given
by the experiment for each value of the resonance mass. The QCD
pair-production cross sections were obtained at approximate
next-to-next-to-leading order (NNLO) \cite{hathor}, and are shown in Fig.~\ref{fig:heavyprod}. We remark that also single production of heavy fermions can give complementary, relevant constraints (see e.g. the fourth among Refs.~\cite{futuresearches} for a detailed analysis), however no such search has been published by ATLAS and CMS yet. Therefore, we do not discuss single production.

Note that the branching ratios are non-trivial only for the top partners
$T,X^{2/3}$ and $\tilde{T}$, whereas $B$ and $X^{5/3}$ decay with
unity branching ratio into $tW^{\mp}\,$. We summarize in Table
\ref{tab:exp searches} all the searches for pair-produced heavy
fermions that we included in our analysis. The analyses of $tWtW$
final states, although intended by the experiments to be searches for
heavy charge $-1/3$ quarks such as the $B$, apply straightforwardly
also to the $X^{5/3}$, which decays into the same final
state\footnote{Note that the decay products of $B\overline{B}$ and
  $X^{5/3}\overline{X}^{5/3}$ would have different spatial
  configurations. For example, same-sign leptons necessarily stem
  either from $X^{5/3}$ or from its antiparticle, while in the case of
  $B\overline{B}$ production each of the same-sign leptons arises from
  a different heavy particle. However, since in the current searches
  only basic cuts on single objects are applied, this kind of
  kinematic differences is expected to give negligible effects on the
  exclusion limits.}.  \s

The region of the parameter space corresponding to $\sin\phi_L \sim 1$ is the
most constrained by direct searches. The lightness of the $X^{5/3}$ fermion
in this case (see Fig.~\ref{fig:spectrum}) is prohibited by both Tevatron
and LHC searches in $W t W t$ final states. For a lower degree of
compositeness of the left-handed top quark, the lightest top partner is the
singlet $\tilde T$, which decays in all three final states $W b$, $Z t$ and
$h t$. The Tevatron only has enough sensitivity to exclude top partners
below 300~GeV, while the most stringent LHC constraints (\emph{i.e.} those based on the full
2011 luminosity) start at 350~GeV. This leaves a region of
the parameter space in the range $m_{\tilde T} \in [300, 350]$~GeV which is not
directly excluded by present constraints, see Fig.~\ref{fig:EWPT}.

In addition to the present exclusion limits, we show in Fig.~\ref{fig:EWPT}
an estimate of the reach of the LHC in 2012. 
The increase in energy enhances significantly the production
cross section of heavy fermion pairs (see Fig.\ref{fig:heavyprod}). On
the other hand, the present exclusion limits quoted by ATLAS and CMS
will be modified due to the changes in the background and to the
additional integrated luminosity. Backgrounds in searches for top
partners are dominated by top quark pair production, which is
increased by 42\% when going from 7 to 8~TeV c.m.~energy at the LHC. The
search strategy relies on a cut on the $t \bar{t}$ invariant mass,
whose distribution is not significantly affected by the increase in
energy, as explicitly checked using MadGraph~5 \cite{Alwall:2011uj}. The
upper limit on the top partner production cross section is therefore
softened in the Gaussian approximation by a factor $\sqrt{1.42} \cong
1.19$. The total luminosity of 15 fb$^{-1}$ expected to be attained in
2012 is nevertheless tightening the limit on the cross section,
lowering it by a square root factor of the luminosity in every channel. 
More refined searches after the LHC upgrade to 14~TeV will be needed in order
to explore the full parameter space~\cite{futuresearches}.

\begin{table}
		\centering
		\setlength{\tabcolsep}{5pt}
		\begin{tabular}{c|c|c|c|c} 
	 exp. & search & $L\,\,[\mathrm{fb}^{-1}]$  & range in $M_{\psi}\, [\mathrm{GeV}]$ & ref.  \\   
	\hline
CMS \cite{cms}	& $WbWb$ (1 lepton) & $4.7$  & $[400,625]$ & CMS-PAS-EXO-11-099 \\
	& $WbWb$ (2 leptons) & $5.0$ & $[350,600]$ & arXiv:1203.5410 \\
	& $WtWt$ & $1.14$ & $[350,550]$ & CMS-PAS-EXO-11-036 \\
	& $WtWt$ & $4.9$ & $[450,650]$ & arXiv:1204.1088 \\
	& $ZtZt$ & $1.14$ & $[250,550]$ & arXiv:1109.4985 \\
\hline
ATLAS \cite{atlas}	& $WbWb$ & 1.04 & $[250,500]$ & arXiv:1202.3076 \\
	& $WqWq$ & 1.04 & $[300,500]$ & arXiv:1202.3389 \\
	& $WtWt$ (1 lepton) & 1.04 & $[300,600]$ & arXiv:1202.6540 \\
	& $WtWt$ (2 leptons) & 1.04 & $[300,600]$ & arXiv:1202.5520 \\	
\hline
CDF \cite{cdf}	& $WbWb$ & 5.6 & $[180,500]$ & arXiv:1107.3875  \\
	& $WtWt$ & 4.8 & $[260,425]$ & arXiv:1101.5728  \\		
		\end{tabular}
		\caption{List of experimental searches for pair-produced heavy fermions that we included in our analysis of collider constraints.}
	\label{tab:exp searches}
\end{table}

\section{Single Higgs production in MCHM5} \label{singleprodMCHM5}
\begin{figure}[t]
\begin{center}
\includegraphics[width=.5\textwidth]{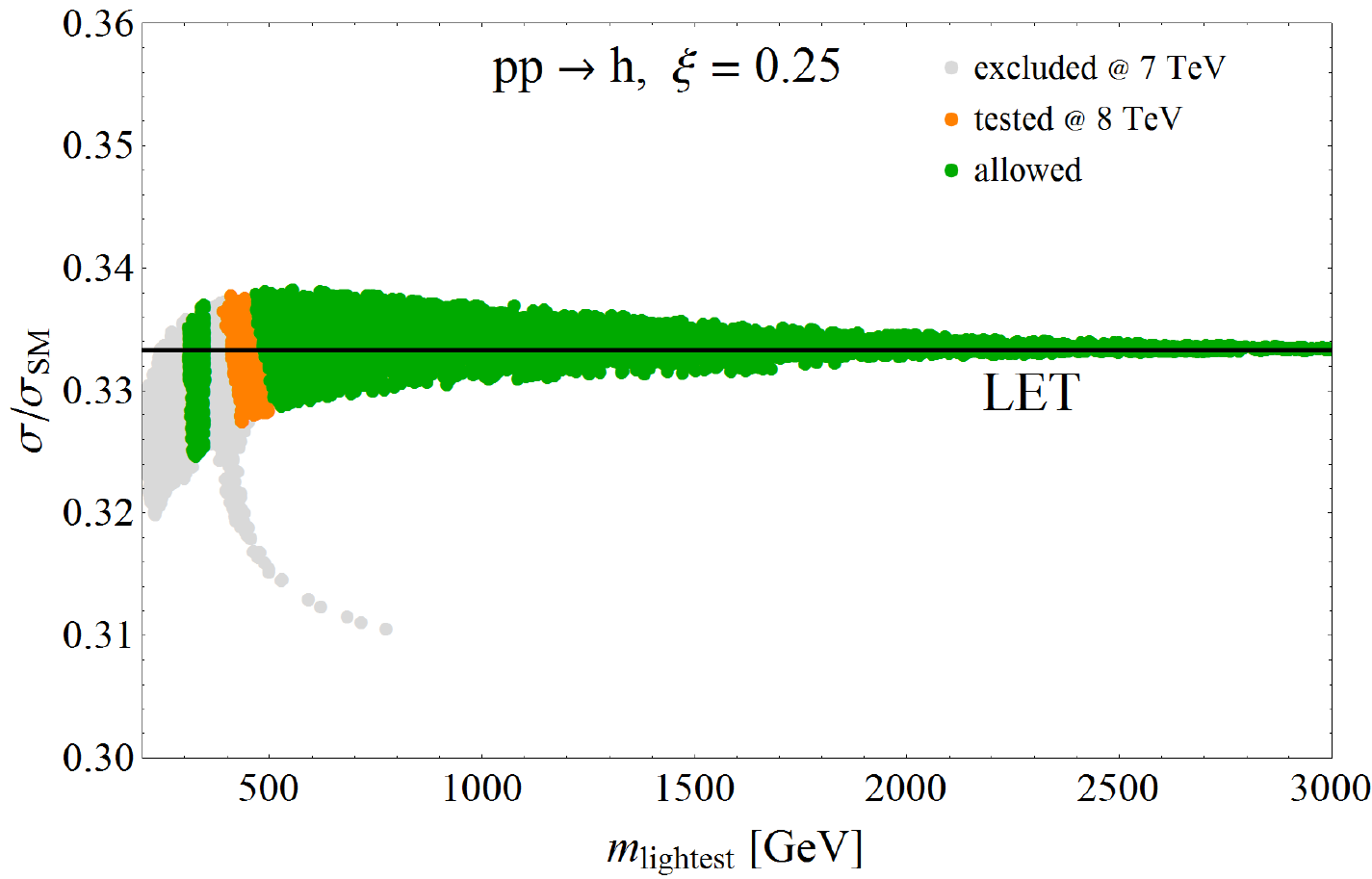}
\end{center}
\caption{The MCHM5 cross section for single Higgs production
  through gluon fusion (including the exact dependence on top and
  heavy fermion masses), normalized to the SM cross section (computed
  retaining the $m_{t}$ dependence), as a function of the mass of the
  lightest fermion resonance $m_{\rm{lightest}}$ for
  $m_{h}=125\,\mathrm{GeV}$. The compositeness parameter has been
  chosen $\xi=0.25$. Green/dark gray points are allowed, gray points are
  excluded by current collider constraints, whereas orange/fair gray
  points will be tested by LHC8 in 2012. For comparison, the cross
  section ratio computed with the LET, Eq.~\eqref{let singleprod}, is shown
  as a black line.}
\label{fig:singlehprod}
\end{figure}

The cross section for single Higgs production in MCHM5 can be readily
derived by noting that we can directly apply Eq.~\eqref{hngg} since
the Higgs kinetic term is canonically normalized. We therefore only
need the determinant of the mass matrix of top-like fermions in 
Eq.~\eqref{fermionmassmatrix}, which reads
\beq \label{det mass matrix}
\det {\cal M}^\dagger(H) {\cal M}(H) = \frac{M_0^4 y^2 f^2 \sin^2 \phi_L
  \sin^2 \phi_R}{8 \cos^2 \phi_L \cos^2 \phi_R} (M_0 + yf)^2 \sin^2
\left(\frac{2H}{f}\right) \;,
\eeq
and has the form of Eq.~\eqref{indep condition}. Thus we obtain $A_{1}=(2/v)(1-2\xi)/\sqrt{1-\xi}$ (where we have used $ \sin^{2}(\left\langle H\right\rangle/f)=\xi\,$), and 
\beq \label{let singleprod}
\frac{\sigma(pp\to h)}{\sigma(pp\to h)_{SM}} = \left(\frac{1-2\xi}{\sqrt{1-\xi}}\right)^{2}\,,
\eeq 
which is valid to all orders in $\xi$. Equation~\eqref{let singleprod}
is independent of the details of the fermion spectrum. While this
holds exactly only in the low-energy theorem limit, as discussed in
Section~\ref{sec:singlehprod} we expect that retaining the full mass
dependence will give corrections to the cross section at most of a few
percent. This is confirmed by a full computation of the 
cross section in which the dependence on the fermion masses is retained, as shown in Fig.~\ref{fig:singlehprod}. The figure shows the cross section of single Higgs production through gluon fusion including new fermionic resonances, normalized to the SM cross section computed with finite $m_{t}$, as a function of the mass of the lightest resonance.\footnote{We have
compared our results in the SM limit to the ones obtained with {\tt
  HIGLU} \cite{higlu}.} Note that the QCD
$K$-factors\footnote{The $K$-factor is defined as the ratio of the
  higher-order cross section to the leading order cross
  section.} cancel out under the assumption that the higher order
corrections are the same in both cases\footnote{This assumption is
  valid only at next-to-leading order (NLO) QCD. At NNLO QCD different
  mass scales play a role. Furthermore, the correct matching of the
  strong coupling constant in the effective and the full theory has to
  be performed. In Ref.~\cite{comphiggsnnlo}, however, it was shown
  that for parameters similar to ours the differences in the
  $K$-factors are of the order of a few percent only, so that the SM
  NNLO $K$-factor can safely be applied also to the MCHM5 case.} (see Ref.~\cite{comphiggsnnlo}). A parameter scan has been performed, selecting only points that satisfy EWPT. Among these, points that satisfy all current collider bounds are shown in green, whereas gray points are already excluded. In orange we show points currently allowed, but that will be excluded by searches for heavy fermions at the end of the LHC8 run if no excess is observed. The agreement with the prediction of the low-energy theorem in Eq.~\eqref{let singleprod}, shown as a black line, confirms that the cross section is to an excellent approximation independent of the details of the spectrum, and is fixed only by $\xi$. The sensitivity to the composite couplings is at most $2\% \times \sigma_{SM}$ for light top partners, in agreement with our previous estimate, and vanishes for heavy partners. We conclude that for single Higgs production the LET provides a very accurate cross section for any spectrum of the extra fermions.   

Finally we remark that the result in Eq.~\eqref{let singleprod} coincides with the one obtained considering only the Higgs nonlinearities, \emph{i.e.} rescaling the SM cross section by $c^{2}$, where $c$ is the correction to the top Yukawa in the limit where fermionic resonances are heavy and thus their effects negligible (see Table~\ref{coupvalues}). This is a consequence of the cancellation discussed above.

%

\subsection{Effect of non-minimal operators}
We can add to the minimal partial compositeness Lagrangian in Eq.~\eqref{PartComplagrangian} the following operators
\beq
\Delta\mathcal{L}=i\,y^{\prime}_{L}(\overline{\psi}_{L}\Sigma^{T})\slashed{D}(\Sigma \psi_{L}) + i\,y^{\prime}_{R}(\overline{\psi}_{R}\Sigma^{T})\slashed{D}(\Sigma \psi_{R}) \; ,
\eeq
where the covariant derivative reads $D_{\mu}=\partial_{\mu}-ig'X\,B_{\mu}\,\,(X=2/3)$. The most convenient way to discuss these operators is to perform the following field redefinition,
\begin{equation}
\psi_{L,R} \to {\zeta(x)}^{T}\psi_{L,R}\,,
\end{equation}
where $\zeta(x)$ was defined in Eq.~\eqref{xi definition}. Upon this transformation, the Lagrangian reads (omitting kinetic terms of elementary fields, and gauge interactions)
\begin{align} \nonumber
\mathcal{L}_{f}+\Delta\mathcal{L}\to &\,\, i\overline{\psi}_{L}\slashed{\partial} \psi_{L}+i\overline{\psi}_{L}\gamma^{\mu}\zeta(x) (\partial_{\mu}\zeta^{T}(x))\psi_{L}+ (L \to R) \\ \nonumber
& - yf (\overline{\psi}_{L}\Sigma_{0}^{T})(\Sigma_{0}\psi_{R}) - M_{0}\overline{\psi}_{L}\psi_{R}
+\mathrm{h.c.} \\ \label{redefined Lagr}
&+\, i\,y^{\prime}_{L}(\overline{\psi}_{L}\Sigma_{0}^{T})\slashed{D}(\Sigma_{0} \psi_{L}) + (L\to R) \\ \nonumber
&-\,\Delta_{L}\overline{\mathcal{Q}}_{L}\zeta^{T}(x)\psi_{R} - \Delta_{R}\overline{\psi}_{L}\zeta(x)\,\mathcal{T}_{R} + \mathrm{h.c.}
\end{align}
Thus we need to rescale the singlet $\tilde{T}$ to make it canonically
normalized, $\tilde{T}_{L,R}\to
\tilde{T}_{L,R}/\sqrt{1+y^{\prime}_{L,R}}\,\,.$ Let us now focus on
how the amplitude for $gg\to h$ is modified by the new operators. It
is easy to verify that the Higgs derivative interactions contained in
Eq.~\eqref{redefined Lagr} do not contribute to the amplitude for
single Higgs production, because they are antisymmetric in the fermion
fields \cite{Azatov:2011qy}. Therefore we can simply apply the
low-energy theorem. From the fermion mass matrix, which reads 
\begin{equation}
\mathcal{M}= \begin{pmatrix} 0 & \Delta_{L}\frac{1+\cos(H/f)}{2} & \Delta_{L}\frac{\cos(H/f)-1}{2} & \Delta_{L}\frac{\sin(H/f)}{\sqrt{2}}\,\frac{1}{\sqrt{1+y^{\prime}_{R}}} \\
-\frac{\sin(H/f)}{\sqrt{2}}\Delta_{R} & M_{0} & 0 & 0 \\
-\frac{\sin(H/f)}{\sqrt{2}}\Delta_{R}  & 0 & M_{0} & 0 \\
\frac{\cos(H/f)}{\sqrt{1+y^{\prime}_{L}}}\Delta_{R} & 0 & 0 & \frac{M_{0}+yf}{\sqrt{1+y_{L}^{\prime}}\sqrt{1+y_{R}^{\prime}}} \end{pmatrix}\,,
\end{equation}
we obtain
\beq
\det \mathcal{M}^{2} (H)= \frac{\Delta_{L}^{2}\Delta_{R}^{2}f^{2}M_{0}^{2}y^{2}}{8(1+y^{\prime}_{L})(1+y^{\prime}_{R})}\sin^{2}\left(\frac{2H}{f}\right)\,,
\eeq
which implies that the amplitude for $gg\to h$ is not sensitive to the
value of $y'_{L,R}$, see Eq.~(\ref{eq:matrixrel}). On the other hand, the Higgs derivative interactions in
Eq.~\eqref{redefined Lagr} contribute in general to the pair
production process, because they enter box diagrams. Therefore the
cross section for $gg\to hh$ will be sensitive to $y'_{L,R}\,$. In the
following section, however, we consider the minimal Lagrangian,
setting $y'_{L,R}=0\,$. 

Finally we comment about the contribution of the exotic state $X^{5/3}$ to the amplitudes for $gg\to h,hh$. In the field basis of Eq.~\eqref{redefined Lagr} the Higgs appears only in elementary/composite mixing terms and in derivative interactions (thus showing manifestly its pseudo-GB nature). Since $X^{5/3}$ does not mix with any elementary field, there is no contribution to the amplitudes from the mixing terms. On the other hand, it is easy to check explicitly that Higgs derivative interactions do not involve $X^{5/3}$. We conclude that the exotic state does not contribute at all to the amplitudes for single and double Higgs production via gluon fusion.

\section{Double Higgs production in MCHM5} \label{double Higgs prod MCHM5}
%
In this section we discuss the cross section for $pp\to hh$ first in
the LET approximation, and subsequently retaining the full dependence
on the masses of the fermions running in the loops.

\subsection{LET cross section}
From the determinant of the fermion mass matrix in Eq.~\eqref{det mass matrix} we can compute $A_{2}=(-2/v^{2})/(1-\xi)$, which determines the $hhgg$ coupling via fermion loops. This, together with the form of $A_{1}$ previously derived and with the expression of the $h^{3}$ coupling given in Table~\ref{coupvalues}, allows us to write down the amplitude for $gg\to hh$ at all orders in $\xi$:
\beq \label{C in mchm5}
C^{\scriptsize \mbox{LET}}_{MCHM5} (\hat{s}) = \frac{3m_h^2}{\hat{s}-m_h^2}
\left(\frac{1-2\xi}{\sqrt{1-\xi}}\right)^{2} -\frac{1}{1-\xi} \;.
\eeq
Thus analogously to single Higgs production, the LET cross section for
Higgs pair production is insensitive to the details of the heavy
fermion spectrum, and is fixed only by $\xi$. The corresponding $pp\to
hh$ cross section at LHC14, normalized to the SM cross section (also
computed in the infinite $m_{t}$ limit) was shown as a function of
$\xi$ in the left panel of Fig.~\ref{fig:let_models}. 

\subsection{Enhancement of the cross section}
\begin{figure}[t]\vspace*{-1.5cm}
\begin{center}
\includegraphics[width=0.6\linewidth]{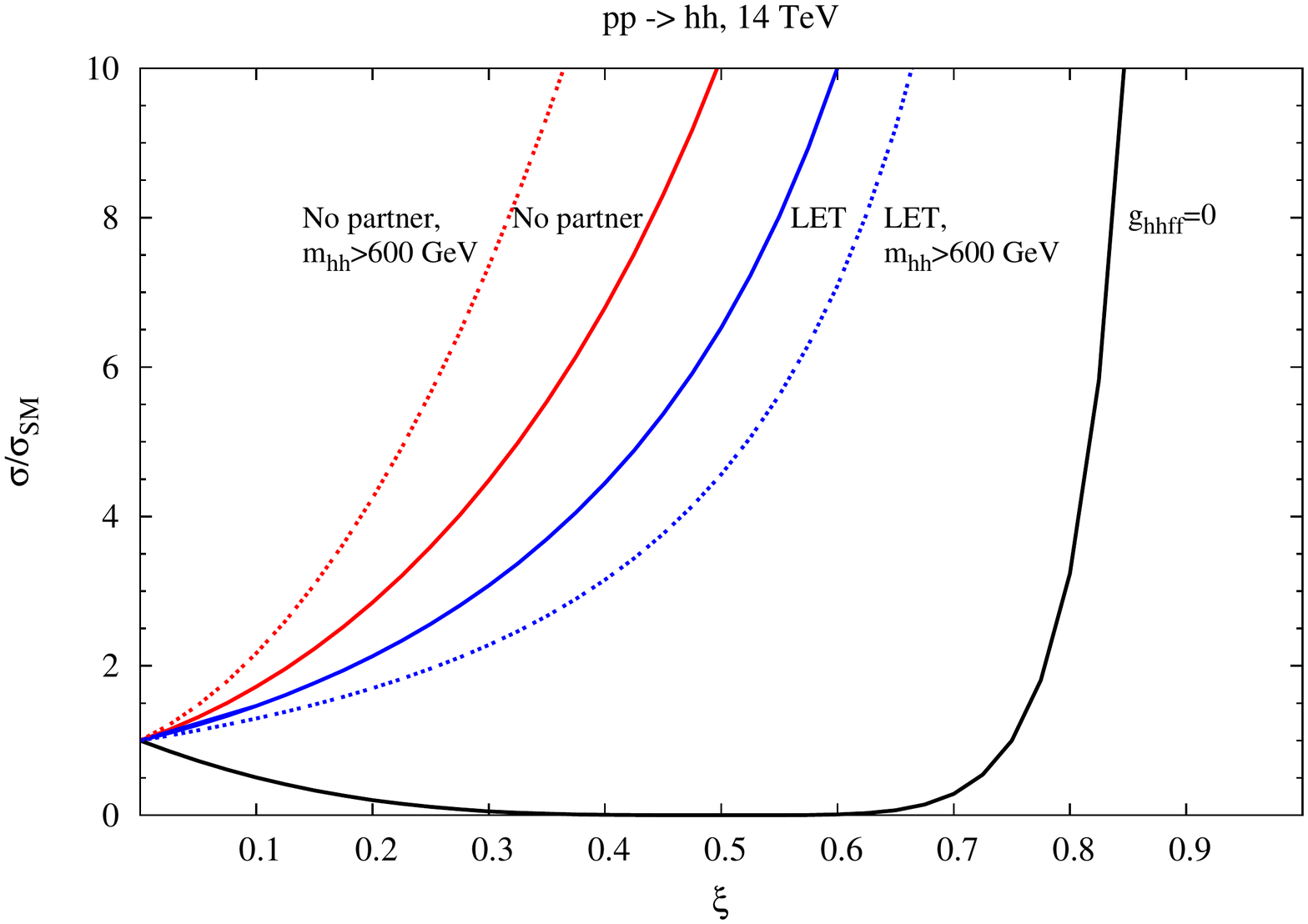}
\vspace*{-0.3cm}\caption{
The cross section for double Higgs production in MCHM5
  normalized to the SM as a function of $\xi$ for three different
  approximations. Red: in the limit of heavy top partners keeping the
  full top quark mass dependence. Blue: LET. Black: setting the two-Higgs
  two-fermion coupling to zero. The red/blue dotted lines show the same as the red/blue full lines
 after application of an invariant mass cut of $m_{hh} \ge 600$~GeV.}
\label{fig:xibehavior}
\end{center}\vspace*{-0.5cm}
\end{figure}

We have seen that for small values of $\xi$ single Higgs production in
the MCHM5 is suppressed compared to the SM while double Higgs production
is enhanced. The behavior of single Higgs production becomes clear from
the LET result given in Eq.~(\ref{let singleprod}). 
In double Higgs production the Higgs pair is either produced through Higgs bosons coupling to the gluons through triangle loops or through boxes. In the former case, in the SM we only have a diagram with a Higgs
subsequently decaying into two Higgs bosons, while in composite Higgs
models there is an additional triangle diagram due to the two-Higgs
two-fermion coupling. In the amplitude for Higgs pair production the parts coming from the triangle containing the triple Higgs coupling and from the box diagrams interfere destructively. 
In the MCHM5 amplitude where these two contributions are modified by $((1-2\xi)/\sqrt{1-\xi})^2$, the additional diagram with the two-Higgs two-fermion coupling proportional to $\xi$ can hence have order one effects so that it governs the total cross section. This can
be inferred from Fig.~\ref{fig:xibehavior} which shows the double
Higgs production MCHM5 cross section normalized to the SM as a function
of $\xi$ for three different approximations. The red line has been
obtained in the limit of heavy top partners keeping the full top quark
mass dependence, the blue line is the LET result, and the black line,
finally, is obtained by explicitly setting the two-Higgs two-fermion
coupling to zero. In this case the cross section ratio is given by  $((1-2\xi)/\sqrt{1-\xi})^4$ both for the LET and
for the approximation where the top quark mass dependence has been kept.
The dotted lines in the figure have been obtained by applying an invariant mass cut of $m_{hh} \ge 600$~GeV.
After application of the cut the discrepancy in the cross section results for the two approximations becomes even worse, see also the discussion in Section~\ref{sec:numerical}.\s

\subsection{Full 1-loop cross section}
In the triangle diagrams which contribute to double Higgs production the gluons couple to the total spin $S_z=0$ along the $z$-axis, whereas the box diagrams
involve $S_z=0$ and $S_{z}=2$ couplings. The amplitude for the process can
hence be expressed in terms of independent form factors $F_\triangle$,
$F_\Box$, $F_{\Box,5}$
associated with spin 0 and $G_\Box$, $G_{\Box,5}$ associated with spin 2. The total partonic cross section is given by
%
%
\begin{equation}
\begin{split}
\hat\sigma_{gg \to hh} =\,& \frac{\alpha_s^2}{1024 (2\pi)^3}\frac{1}{\hat{s}^2}
\int_{\hat{t}_-}^{\hat{t}_+} d\hat{t} \left[ \left|\sum_{i=1}^4\sum_{j=1}^4\left(g_{ h\bar{q}_i q_j}^2 G_{\Box}(m_i, m_j)+g_{h\bar{q}_i q_j, 5}^2 G_{\Box, 5}(m_i, m_j)\right)\right|^2\right.
\\
&\left.\,+ \,\,
\left|\sum_{i=1}^{4}\left(C_{i,\triangle}F_{\triangle}(m_i)+ \sum_{j=1}^{4}\left( g_{ h\bar{q}_i q_j}^2 F_{\Box}(m_i, m_j)+g_{
    h\bar{q}_i q_j, 5}^2 F_{\Box, 5}(m_i,
  m_j)\right)\right)\right|^2\right] \;, \label{eq:gghhcxn}
\end{split}
\end{equation}
with the integration limits
\beq
\hat{t}_{\pm} = - \frac{\hat{s}}{2} \left( 1 - 2
  \frac{m_h^2}{\hat{s}} \mp \sqrt{1-\frac{4 m_h^2}{\hat{s}}} \right) \;,
\label{t+-}
\eeq
where ${\hat {s}}$ denotes the partonic c.m. energy. 
The triangle and box form factors are given in App.~\ref{app:analyt}. The various
couplings appearing in Eq.~(\ref{eq:gghhcxn}) are also defined there. We have explicitely verified that in the SM limit our result agrees with Ref.~\cite{Plehn:1996wb}. The hadronic cross section is obtained by convolution with the parton
distribution function of the gluon in the proton, see Eq.~\eqref{hadr cross sect}.

\subsection{Numerical analysis \label{sec:numerical}}
\begin{figure}
\includegraphics[width=0.5\linewidth]{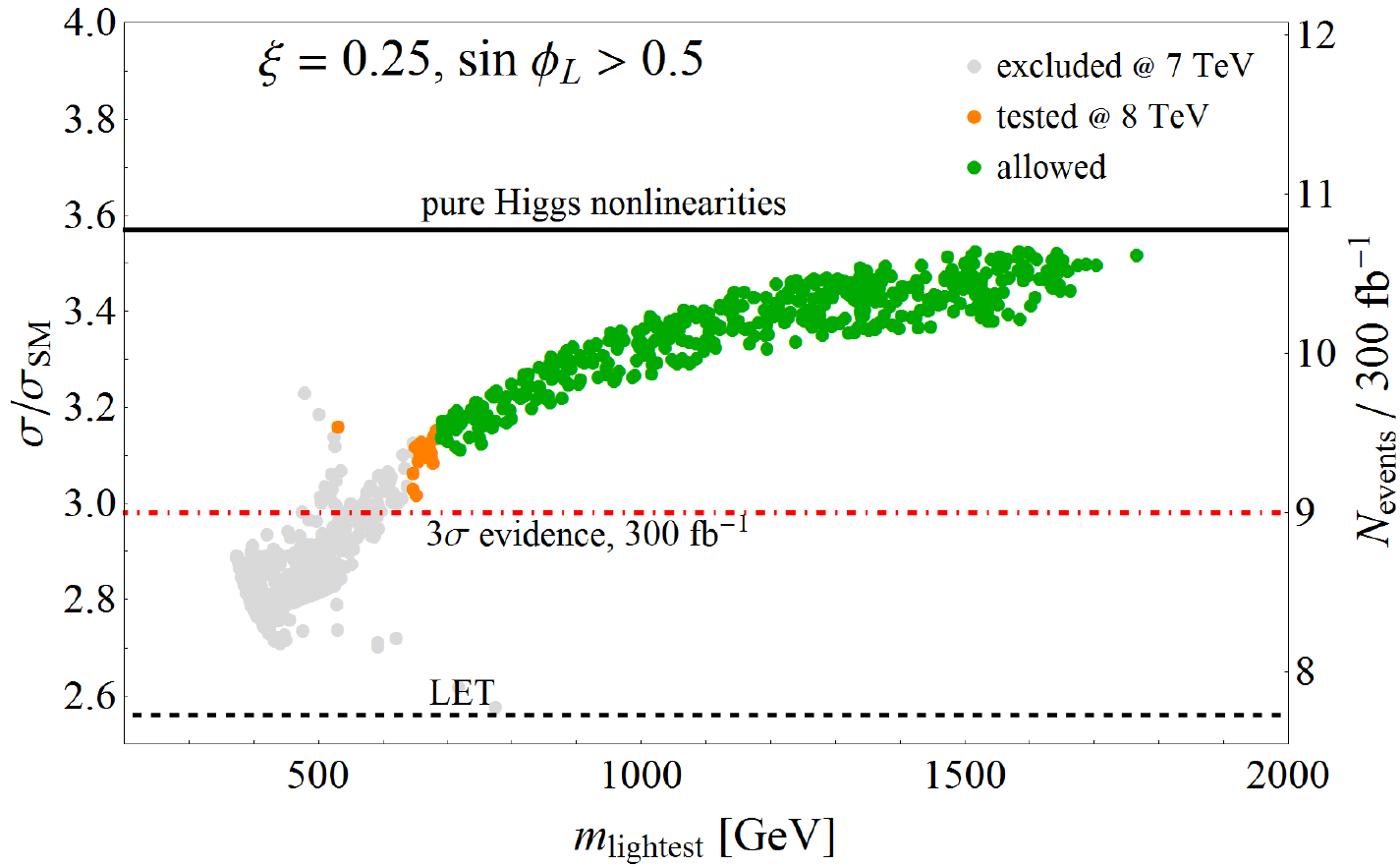}
\includegraphics[width=0.5\linewidth]{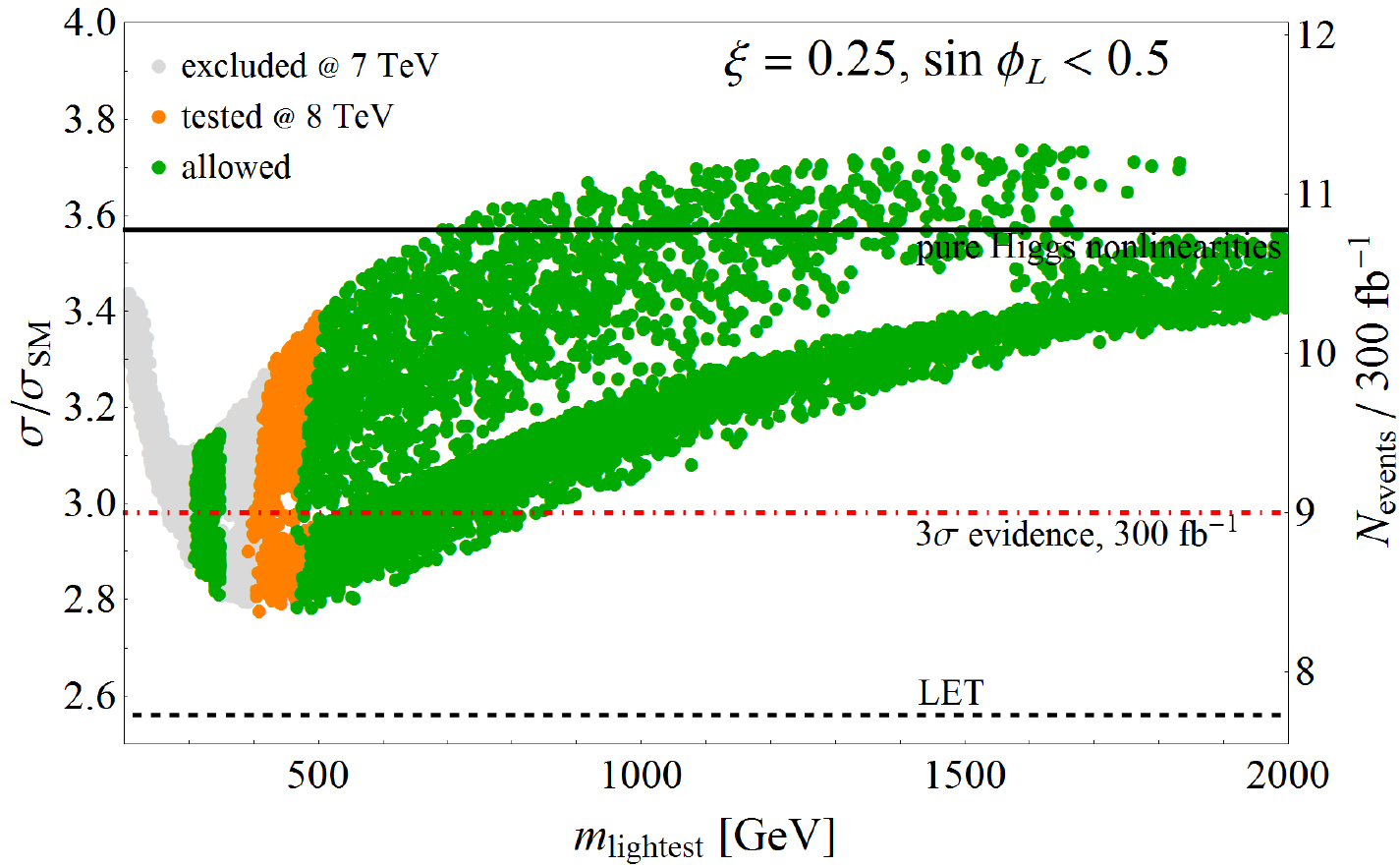}
\caption{The cross section for double Higgs production through gluon
  fusion normalized to the SM as function of the mass of the lightest
  resonance of the heavy top sector, for $m_h=125$~GeV. The compositeness parameter has
  been chosen $\xi=0.25$. Green/dark gray (gray) dots denote points
  which pass (do not pass) all current constraints, whereas
  orange/fair gray
  dots correspond to points that will be tested by LHC8. The left
  panel shows points for which $X^{2/3}$ is the lightest top partner
  (as a consequence of $t_{L}$ being largely composite), whereas for
  points in the right panel the lightest top partner is typically the
  singlet $\tilde{T}$. The black solid (dashed) line corresponds to
  the result in the limit of heavy top partners keeping the full top
  mass dependence  (to the LET result as in
  Fig.~\ref{fig:let_models}). The expected number of events in the
  $hh\to b\bar{b}\gamma\gamma$ final state after all cuts at LHC14
  with $L=300\,\mathrm{fb}^{-1}$ is also shown, along with the
  $3\sigma$ evidence threshold (dot-dashed line), see text for details.}
\label{fig:xi025}
\end{figure}

For the numerical analyis we have performed, after fixing $\xi$ to one of the benchmark values $\xi = 0.25$ or $\xi=0.1$, a scan in the parameter set $(\phi_L,\phi_R,R)$ and retained only the points which
fulfill the constraints from EWPT. By this we mean that there exists a value of $m_{\rho}\in [1.5\,\mathrm{TeV},4\pi f]$ such that the configuration $(\xi,\phi_L,\phi_R,R,m_{\rho})$ passes EWPT at $99\%$ CL. For this set of points we show in Fig.~\ref{fig:xi025} for $m_h=125$~GeV and $\xi=0.25$ the double Higgs production cross section normalized to the SM as a function of the lightest top partner mass. The dependence on the masses of the loop
particles has been fully taken into account. The black solid line shows the result in the limit of heavy partners, keeping only the top contribution (with full mass dependence) in the loop, while the black dashed line corresponds to the LET result in Fig.~\ref{fig:let_models}. The green (gray) dots are points which pass (do not pass) the current constraints from Tevatron and LHC data, whereas orange points will be tested by LHC8. 

Some comments are in order. First of all, we find a sizeable
dependence of the cross section on the spectrum of the heavy fermions with
$2.7\lesssim \sigma/\sigma_{SM}\lesssim 3.7$. We recall that both the
LET cross section and the cross section in the limit of heavy partners
only depend on $\xi$. The LET approximation, however, severely
underestimates the ratio $\sigma/\sigma_{SM}$, and this effect is even
worse if we refer directly to the cross section, since we are
consistently normalizing the LET cross section for MCHM5 to
$\sigma_{SM}(m_{t}\to \infty)$, which is $\sim 20\%$ smaller than the
full result. On the other hand, the result in the limit of heavy
partners, while keeping the full top mass dependence \cite{doubleh2},
overestimates the cross section in the region
$m_{\mathrm{lightest}}\lesssim 1\,\mathrm{TeV}$, which is compatible with a Higgs as light as $125\,\mathrm{GeV}$. For large values of the partner masses, of course, the cross section tends to the value obtained including
only top loops (with top couplings following the `trigonometric'
rescalings given in Table~\ref{coupvalues}). 

It should be noted that we have not taken into account higher-order QCD
corrections. They have been calculated at NLO for SM and MSSM Higgs
pair production in Ref.~\cite{Dawson:1998py} in the heavy top mass limit. 
However, they cannot be taken over here as we have the
additional diagram with the two-Higgs two-fermion coupling and more
seriously box diagrams with different loop particle masses. For heavy
loop particle masses we do not expect the corrections to be too
different from the SM case, so that they approximately cancel out in the
ratio of the two cross sections.

\begin{figure}[ht]
\includegraphics[width=0.5\linewidth]{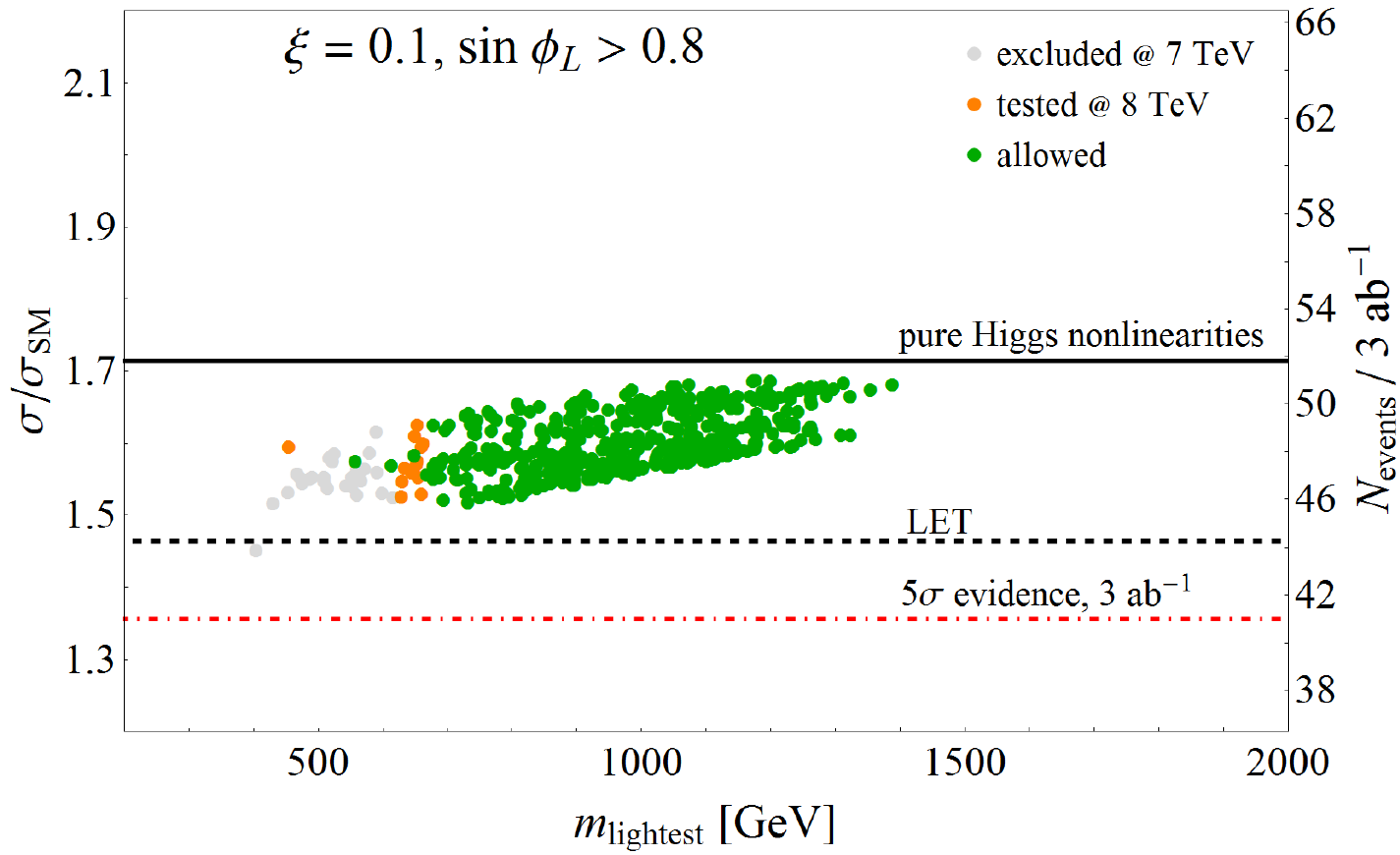}
\includegraphics[width=0.5\linewidth]{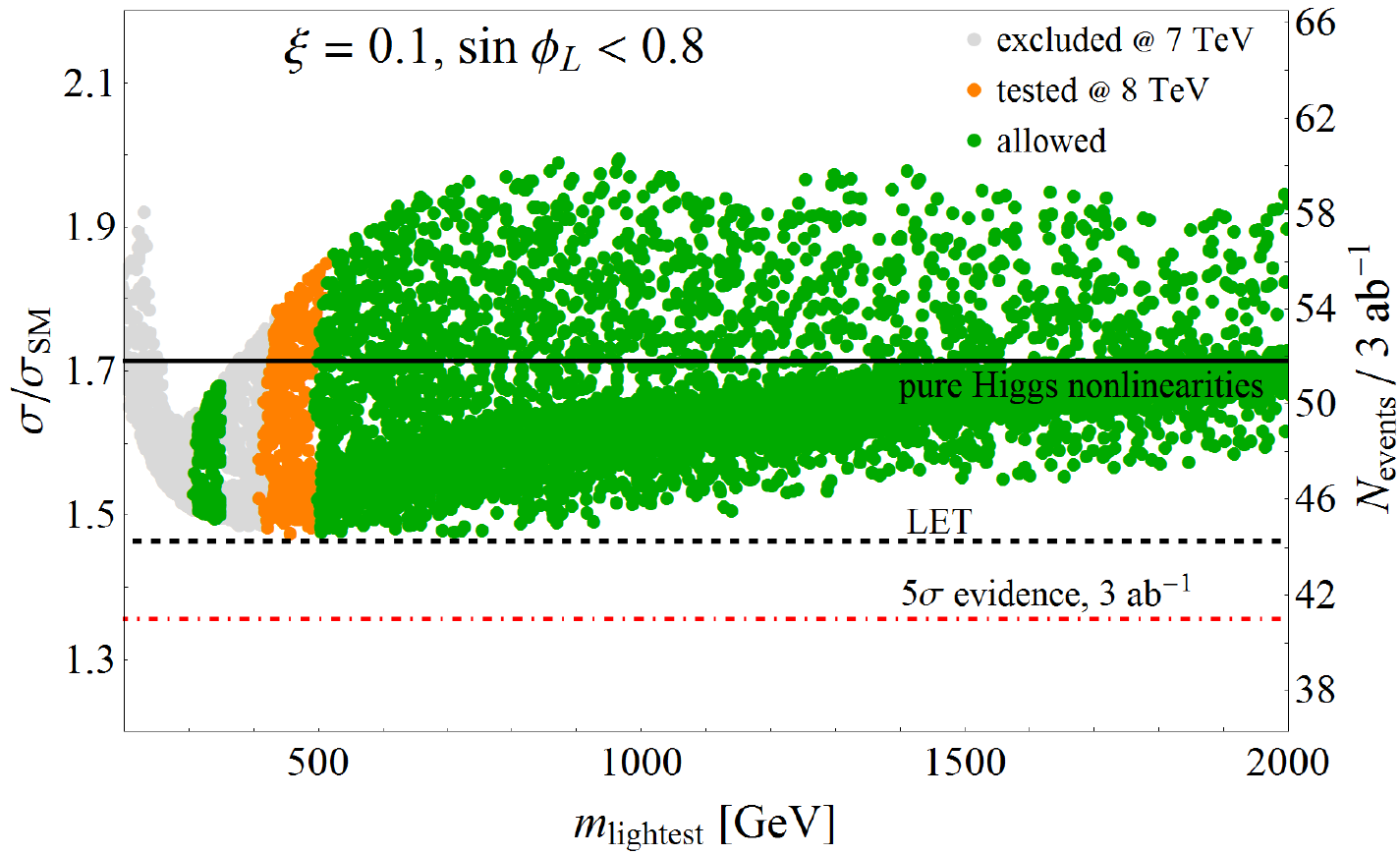}
\caption{Cross section for double Higgs production through gluon
  fusion normalized to the SM as function of the mass of the lightest
  resonance of the heavy top sector, for $\xi=0.1$ and $m_h=125$~GeV. 
  Points are split
  in the two panels depending on the degree of compositeness of
  $t_{L}$. The meaning of the dots and lines is the same as in
  Fig.~\ref{fig:xi025}, except that we assumed an integrated luminosity $L=3\,\mathrm{ab}^{-1}$ at LHC14. The dot-dashed line corresponds to the $5\sigma$ discovery threshold, see text for details.}
\label{fig:xi01}
\end{figure}
In Fig.~\ref{fig:xi01} we show the corresponding results for a lower value of
$\xi=0.1$, which corresponds to $f\simeq 800\,\mathrm{GeV}$. Due to the larger value of $f$, the cross section is less enhanced compared to the SM. Similarly to the case $\xi = 0.25$ the LET underestimates the cross section, although in a less dramatic way than in the previous case.  

To estimate the reach of the 14 TeV run of the LHC on double Higgs production, we focus on the final state $hh\to b\bar{b}\gamma\gamma$,
which was shown to be the most promising for a light Higgs boson \cite{Baur:2002rb,Baur:2002qd,Baur:2003gpa,Baur:2003gp}. 
Reference~\cite{Baur:2003gp} found that assuming a luminosity
$L=600\,\mathrm{fb}^{-1}$, 6 signal 
events could be obtained after all cuts, with a background of 11
events. We estimate the expected number of signal events for MCHM5 by
computing $\sigma(pp\to hh)\times \mathrm{BR}(hh\to
b\bar{b}\gamma\gamma)$ for each point in the parameter space (taking
into account the QCD production $K$-factor\footnote{As stated above the SM QCD corrections to double Higgs production cannot be translated trivially to the composite Higgs case. Assuming the top partners to be heavy we expect, however, the error not to be too large by applying the SM K-factor to MCHM5 double Higgs production. Concerning the diagram involving the two-Higgs two-fermion coupling we explicitly verified that it hardly changes the QCD corrections compared to the SM ones.} of $1.9$ and the
non-standard Higgs branching ratios) and multiplying it times the
acceptance for all cuts as computed in Ref.~\cite{Baur:2003gp} for the
SM. This rough approximation cannot of course replace a full analysis
of the effects of cuts in the MCHM5 case, which however goes
beyond the scope of this work. We therefore apply the simplified
procedure for an illustratory purpose. We also quote the number of
events needed for a $3\,(5)\sigma$ evidence with
$L=300\,(3000)\,\mathrm{fb}^{-1}$, based on the background estimate of
Ref.~\cite{Baur:2003gp} with the requirement of one $b$-tagged jet. Notice that this is likely to be conservative, because the analysis of reducible backgrounds (whose sum is larger than the irreducible $b\bar{b}\gamma\gamma$) performed in Ref.~\cite{Baur:2003gp} made use of efficiencies and misidentification probabilities, in particular for $b$-jets, that have since then been improved by ATLAS and CMS. We find that a $3\sigma$ excess can be obtained already
with $300\,\mathrm{fb}^{-1}$ if $\xi = 0.25$, except perhaps in some
regions of the parameter space with a very light top
partner. A $5\sigma$ discovery would be possible at the LHC luminosity
upgrade for a more moderate value $\xi = 0.1$.  

We note that in the recent Ref.~\cite{ContinoETAL} two $b$-tagged jets were required, and the efficiency and rejection probabilities for $b$-tagging were updated to current values. However, since we are only interested in a rough estimate of the LHC reach, we conservatively adopt the numbers of Ref.~\cite{Baur:2003gp}. Furthermore, a realistic analysis of the instrumental backgrounds relevant to $b\bar{b}\gamma\gamma$ would require a detailed knowledge of the detector properties, which is out of the reach of a theoretical analysis. See also Ref.~\cite{Dolan:2012rv} for a study of the $b\bar{b}\tau\tau$ final state, and Ref.~\cite{Zurichgghh} for an analysis of the $b\bar{b}W W \to b\bar{b}\ell\nu jj$ channel.

\begin{figure}
\begin{center}
\includegraphics[width=0.49\linewidth,angle=0]{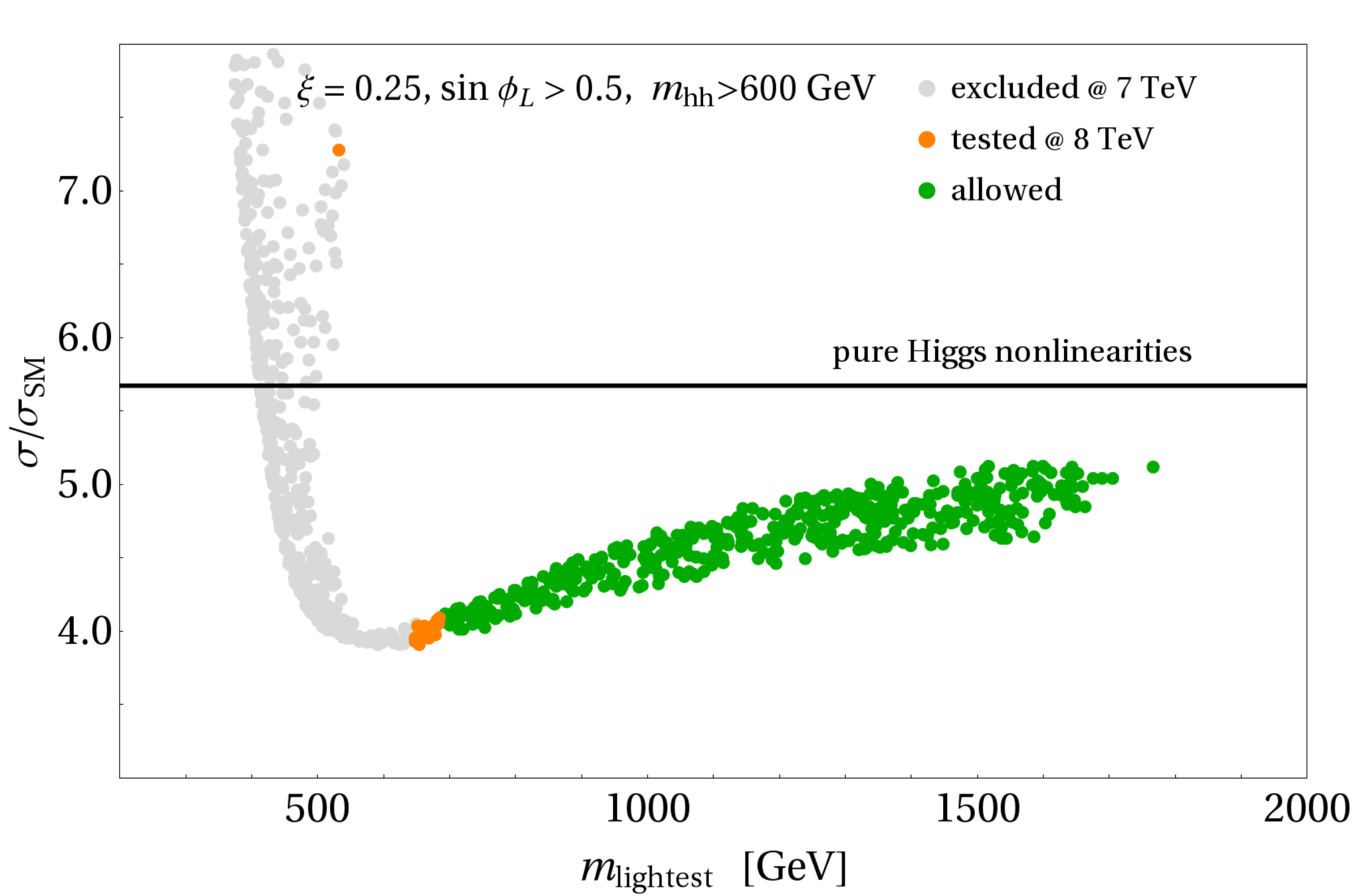}
\includegraphics[width=0.49\linewidth,angle=0]{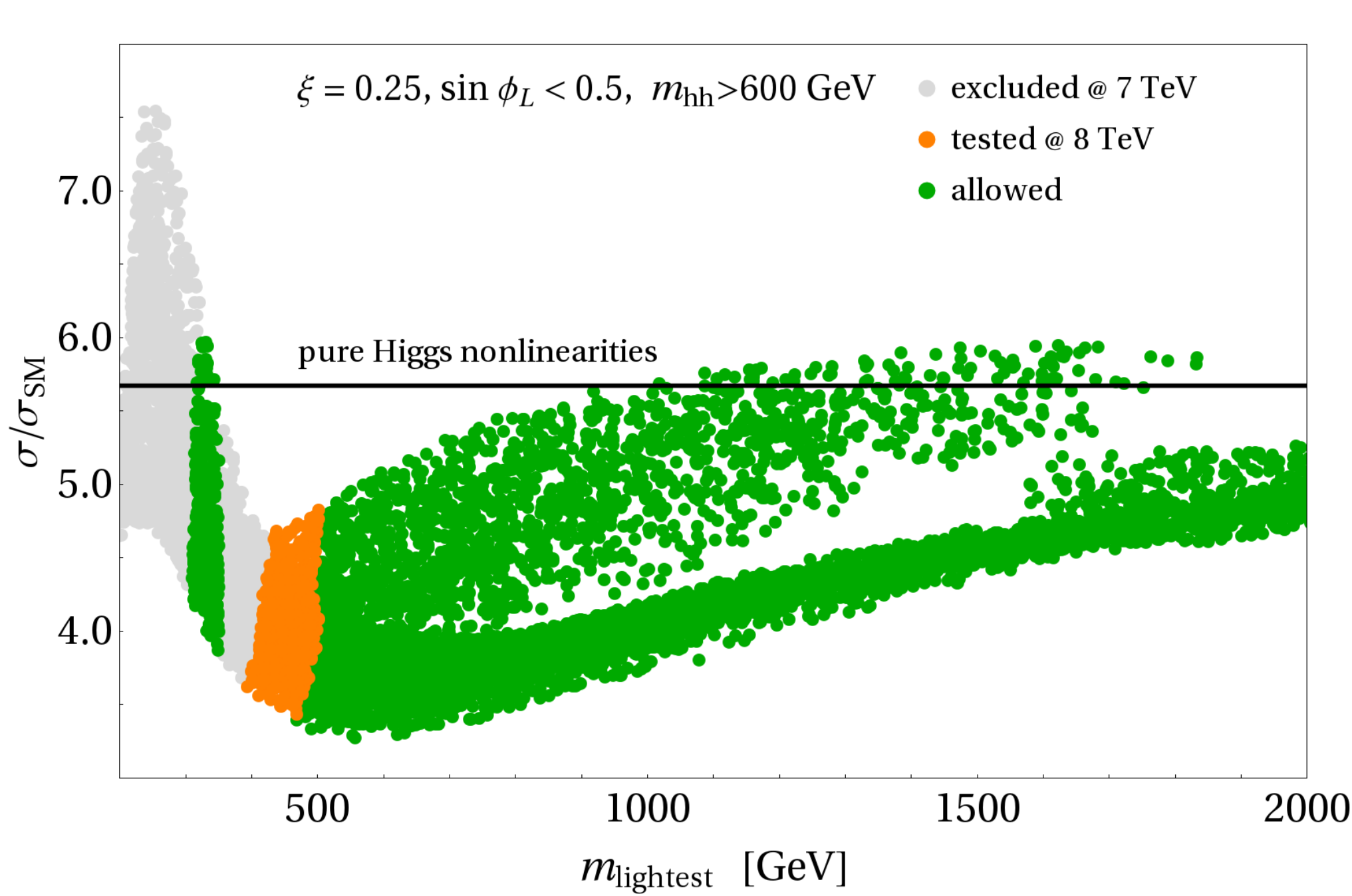}
\vphantom{blabla}
\vspace{-0.5cm}
\caption{The cross section for double Higgs production through gluon
  fusion after an invariant mass cut $m_{hh} \geq 600\,\mathrm{GeV}$, normalized to the SM for $m_h=125$~GeV, as function of the mass of the lightest resonance of the heavy top sector. The compositeness parameter has
  been chosen $\xi=0.25$. Green/dark gray (gray) dots denote points
  which pass (do not pass) all current constraints, whereas
  orange/fair gray
  dots correspond to points that will be tested by LHC8. The left
  panel shows points for which $X^{2/3}$ is the lightest top partner, whereas for the
  points in the right panel the lightest top partner is typically the singlet
  $\tilde{T}$. 
  The black solid line corresponds to
  the result obtained considering only Higgs nonlinearities, \emph{i.e.} in the limit of heavy top partners and keeping the full top
  mass dependence.}
\label{fig:invmasscut}
\end{center}
\end{figure}

Additionally, we studied if applying a cut on
the invariant mass $m_{hh}$ could be useful to measure deviations from the SM cross section.
Therefore, in Fig.~\ref{fig:invmasscut} we show the same as Fig.~\ref{fig:xi025} but after an invariant mass cut
 of $m_{hh} \ge 600$~GeV has been applied.
As can be
inferred from the plot the composite cross section is more enhanced
compared to the SM than without application of a cut, see also Fig.~\ref{fig:xi025}. On the other hand the absolute value of the cross section after cuts
becomes very small. The plots reveal, however, another 
interesting feature. While for masses of the lightest top partner
above 2~TeV the total cross section is reasonably well
approximated by the cross section where only Higgs nonlinearities are considered, see Fig.~\ref{fig:xi025}, this is not the case any more after application of cuts. This can be inferred from
Fig.~\ref{fig:invmasscut} by comparing the full result, given by the
points, to the black line, which is the ratio of the double Higgs production cross section considering only Higgs nonlinearities to the SM cross section (the full top dependence has been
included in both cases). So we conclude that not only the heavy top
partner limit in the total cross section of double Higgs production
is a rather bad approximation unless the top partners are really
heavy, but this approximation becomes even worse when a cut on $m_{hh}$ is
applied. The latter, however, may be relevant in the experimental
analyses to enhance the signal to background ratio and to extract
information on the couplings involved in the process.

\section{Conclusions \label{sec:conclusion}}

Models of electroweak symmetry breaking aiming at giving a rationale for the stability of the weak scale under radiative corrections predict an extended top sector at an energy scale typically below a TeV. Carrying color and electric charges, these top partners are naively expected to give significant corrections to the loop-induced couplings of the Higgs boson to massless photons and gluons. We examined this question in the context of composite Higgs models where the Higgs boson emerges as a Goldstone boson from a strongly-interacting theory. We first extended the well-known SM Higgs low-energy theorem that gives a simple way to estimate the contribution of heavy particles to the Higgs couplings to photons and gluons and we then checked the accuracy of this LET approximation to an explicit full one-loop computation taking into account the contributions of all fermionic resonances. We confirmed that in composite models there is actually a quite efficient cancellation for the contribution of the top partners to the {\it single} Higgs production cross section, which deviates by no more than a few percents from the result obtained taking into account the Higgs nonlinearities only. For single production, the LET provides a very accurate prediction of the cross section. The situation is, however, totally different for {\it double} Higgs production for which the LET approximation is not reliable any more, and deviates from the true result by up to 50\%. The top partners also significantly reduce the enhancement of the $gg \to hh$ cross section over the SM that was previously computed taking into account the strong dynamics effects only. This dependence on the top partner spectrum and couplings gives an indirect access to this sector that will complement the information gathered from direct searches as well as from electroweak and flavor precision data. 

The recent discovery of the Higgs boson puts the identification of its true nature on the immediate agenda of high-energy physics and a careful study of the top sector can bring invaluable information. If the Higgs is a composite object, then $t\bar{t}h$ and $gg\to hh$ will be important channels that can give access  to the top partners and where large deviations compared to the SM predictions are expected.

\subsubsection*{Acknowledgments} 
We thank A.~Azatov, R.~Contino, C.~Englert, M.~Farina, G.~Panico, D.~Pappadopulo, R.~Rattazzi, M.~Rauch, J.~Santiago, M.~Spira and A.~Wulzer for fruitful discussions, and J.~Serra for comments about the manuscript. M.~G. is supported by the Schweizer Nationalfonds and by the European Commission under the contract PITN-GA-2010-264564 {\it LHCPhenoNet}. R.~G. and M.~M. are supported by the DFG SFB/TR9 \emph{Computational Particle Physics}. R.~G. acknowledges financial
support from the Landesgraduiertenkolleg. The work of C.~G. and E.~S. has
been partly supported by the European Commission under the ERC Advanced Grant 226371 {\it MassTeV} and the contract
PITN-GA-2009-237920 {\it UNILHC}. E.~S. has been supported in part by the European Commission under the ERC Advanced Grant 267985 {\it DaMeSyFla}.

\section*{Appendices}
\begin{appendix}

\section{Derivation of the $hgg$, $hhgg$ and $h\gamma\gamma$ couplings in the SILH formalism} \label{appendix:silh gluon higgses}
In this section we derive the expressions of the couplings $hgg$, $hhgg$ and $h\gamma\gamma$ in the SILH formalism.

\subsection{The $hgg$ and $hhgg$ couplings}
Our starting point is Eq.~\eqref{hngg}. We also stress that since we are working in a general basis where $c_{r}\neq 0$, the relation between $\left\langle H\right\rangle$ and $v$ is non-trivial, as can be read off the $W$ boson mass term,
\beq \label{W mass}
m_{W}^{2}=\frac{g^{2}v^{2}}{4}\,,\quad \mbox{with} \quad  v^{2}=\left\langle H\right\rangle^{2}\left(1+\frac{c_{r}}{4}\frac{\left\langle H\right\rangle^{2}}{f^{2}}\right)\,.
\eeq
We assume for definiteness the presence of one or more vector-like top partners, which upon integration contribute to $c_{g}$, and identify the light mass eigenstate of the heavy fermion mass matrix $\mathcal{M}$ with the top quark, whose mass reads
\beq
m_{t}(H)=\frac{y_{t}H}{\sqrt{2}}\left(1-c_{y}^{(t)}\frac{H^{2}}{2f^{2}}\right)\,,
\eeq
where $c_{y}^{(t)}$ parameterizes the correction to the SM top Yukawa coupling. The
coefficients $A_{1,2}$ in Eq.~\eqref{hngg} can be related to $c_{y}^{(t)}$ and $c_{g}$ by
separating the contribution (to the $hgg$ and $hhgg$ coupling,
respectively) of the top quark, which involves $c_{y}^{(t)}$, from that of top partners, which is parameterized by $c_{g}\,$. The results are
\begin{align} \label{A1explicit}
\frac{1}{2}\left\langle H \right\rangle A_{1}\,=\,&\,\frac{1}{2}\left(\frac{\partial}{\partial\log H}\log \det \mathcal{M}^{2}(H)\right)_{H=v}=\, 1-c_{y}^{(t)}\frac{v^{2}}{f^{2}}+3c_{g}\frac{y_{t}^{2}}{m_{\rho}^{2}}v^{2}\,, \\ \label{A2explicit}
\frac{1}{2}\left\langle H \right\rangle^{2} A_{2}\,=\,&\,\frac{1}{2}\left(\left(\frac{\partial^{2}}{\partial (\log H)^{2}}-\frac{\partial}{\partial \log H}\right)\log\det \mathcal{M}^{2}(H)\right)_{H=v} =\, -1-c_{y}^{(t)}\frac{v^{2}}{f^{2}}+3c_{g}\frac{y_{t}^{2}}{m_{\rho}^{2}}v^{2}\,,
\end{align}
where we work at $\mathcal{O}(1/f^{2})$.\footnote{In the second and third term of each of Eqs.~(\ref{A1explicit}) and (\ref{A2explicit}) we have used the fact that the distinction between $\left\langle H\right\rangle$ and $v$ expressed by Eq.~\eqref{W mass} is higher order in $\xi$ there.} Note that the `implicit' expressions containing the determinant are in practice more useful than the explicit ones written in terms of $c_{y}^{(t)}$ and $c_{g}$, because using the former avoids diagonalizing the heavy fermion mass matrix, a rather complicated task in presence of multiple top partners. 

In Eq.~\eqref{hngg} we have assumed that $h$ has a canonical kinetic term. However, in the SILH Lagrangian the operators proportional to $c_{H}$ and $c_{r}$ correct the Higgs kinetic term as follows
\beq
\Delta\mathcal{L}_{h\,kin}= \frac{1}{2f^{2}}\left(c_{H}+\frac{c_{r}}{4}\right)(\left\langle H \right\rangle +h)^{2}\partial_{\mu} h\partial^{\mu} h\,,
\eeq
which also contains Higgs derivative interactions. At first order in $\xi$, these can be eliminated by the nonlinear redefinition \cite{silh}
\beq \label{nonlinear rescaling}
h\to h-\frac{\xi}{2}\left(c_{H}+\frac{c_{r}}{4}\right)\left(h+\frac{h^{2}}{v}+\frac{h^{3}}{3v^{2}}\right)\,,
\eeq
which leaves $h$ canonically normalized. Notice that in a nonlinear $\sigma$-model, the Higgs is canonically normalized at all orders, which corresponds to the relation $c_{H}=-c_{r}/4$. Performing the transformation in Eq.~\eqref{nonlinear rescaling}, we arrive at the effective coupling of the Higgs to one and two gluons, Eqs.~(\ref{hgg}) and (\ref{hhgg}), respectively. The invariance of these expressions under the reparameterization in Eq.~\eqref{reparam} can be easily verified by using Eqs.~(\ref{A1explicit}) and (\ref{A2explicit}), respectively.
\subsection{The $h\gamma\gamma$ coupling}
Starting from Eq.~\eqref{hgaga starting point}, recalling the expression of the $W$ boson mass in Eq.~\eqref{W mass} and taking into account the rescaling needed to make the Higgs kinetic term canonical, see Eq.~\eqref{nonlinear rescaling}, it is straightforward to obtain Eq.~\eqref{hgammagamma}. Similarly to Eq.~\eqref{A1explicit}, $A_{1}$ can be related to $c_{y}^{(t)}$ and $c_{\gamma}$ (assuming that all contributions to $c_{\gamma}$ come from the fermion sector)
\beq \label{A1 - cy and cgamma}
\frac{1}{2}\left\langle H \right\rangle A_{1}= \frac{1}{2}\left(\frac{\partial}{\partial \log H} \log \det \mathcal{M}^{2}(H)\right)_{H=v} = 1-c_{y}^{(t)}\frac{v^{2}}{f^{2}}+c_{\gamma}\frac{g^{2}}{m_{\rho}^{2}}v^{2}\frac{1}{2Q_{t}^{2}}\,.
\eeq
Plugging Eq.~\eqref{A1 - cy and cgamma} into Eq.~\eqref{hgammagamma}, the invariance under the reparameterization in Eq.~\eqref{reparam} becomes explicit. 
\section{The SILH coefficients for the Littlest Higgs and MCHM4} \label{LH and MCHM4}
In this section we give the minimal details which are needed to
compute the coefficients of the SILH Lagrangian relevant to Higgs
production via gluon fusion in the Littlest Higgs and in the minimal
composite Higgs model with fermions in the spinorial representation.
\subsection{Littlest Higgs}\label{AppB:LH}
The Littlest Higgs model \cite{ArkaniHamed:2002qy} is based on the
coset $SU(5)/SO(5)$. We consider here a variation where only one
$U(1)$ group is gauged, as this eliminates one source of custodial
breaking and thus relaxes the tension with EWPT suffered by the
original model. In Ref.~\cite{Csaki:2003si} it was shown that a scale as low as $f\sim 1.2\,\mathrm{TeV}$ is allowed in this case. This, however, leaves an extra singlet Goldstone boson in the spectrum, whose effects will be ignored in the following.\footnote{Additional sources of symmetry breaking are needed in order to give the singlet a potential.} The $\Sigma$ field reads
\begin{equation} \label{sigmafield}
\Sigma(x)=e^{2i\Pi/f}\Sigma_{0}\,,\qquad\Pi=\begin{pmatrix} \eta/(2\sqrt{5}) & H/\sqrt{2} & \varphi \\
H^{\dagger}/\sqrt{2} & -2\eta/\sqrt{5} & H^{T}/\sqrt{2} \\
\varphi^{\dagger} & H^{\ast}/\sqrt{2} & \eta/(2\sqrt{5}) 
\end{pmatrix}\,,\qquad \Sigma_{0}=\begin{pmatrix} & & \mathds{1}_{2} \\
& 1 & \\
\mathds{1}_{2} & &  
\end{pmatrix}\,,
\end{equation} 
where
$H$ is the Higgs doublet, $\varphi$ is a complex triplet and $\eta$ is
a singlet. An $SU(2)_{1}\times SU(2)_{2}\times U(1)_{Y}$ subgroup of
the global symmetry is gauged and is spontaneously broken at the scale $f$ to the diagonal $SU(2)_{L}\times U(1)_{Y}$. In Eq.~\eqref{sigmafield} we omitted the GBs that get eaten by the heavy $SU(2)$ triplet of vectors. The two-derivative Lagrangian reads 
\begin{equation}
\mathcal{L}=\frac{f^{2}}{8}\mathrm{Tr}\left[(D_{\mu}\Sigma)^{\dagger}D^{\mu}\Sigma\right]\,,\qquad D_{\mu}\Sigma = \partial_{\mu}\Sigma - i \sum_{j=1,2}g_{j}W_{j}^{a}(Q^{a}_{j}\Sigma+\Sigma Q_{j}^{a\,T})-ig'B_{\mu}(Y\Sigma+\Sigma Y)\,,
\end{equation}  
with the gauged generators given by
\begin{equation}
Q_{1}^{a}=\begin{pmatrix}
\sigma^{a}/2 & & \\
& & & \\
& & & 
\end{pmatrix}\,,\qquad Q^{a}_{2} = \begin{pmatrix}
& & \\
& & \\
& & -\sigma^{a\,\ast}/2 
\end{pmatrix}\,,\qquad Y=\mathrm{diag} \begin{pmatrix}
1/2, 1/2, 0, -1/2, -1/2
\end{pmatrix}\,.
\end{equation}
The SM fermions are taken to transform only under $SU(2)_{1}\times U(1)_{Y}$. 

The coefficients $c_{H}$ and $c_{r}$ receive contributions of three
different kinds. The first arises from the nonlinear $\sigma$ model
structure. Using
\beq
m_{W}^{2}=\frac{1}{2}g^{2}f^{2}\sin^{2}\left(\frac{\left\langle H\right\rangle}{\sqrt{2}f}\right)\,,\qquad g=\frac{g_{1}g_{2}}{\sqrt{g_{1}^{2}+g_{2}^{2}}}\,,
\eeq
we find $c_{H}^{\sigma}=1/6$ and $c_{r}^{\sigma}=-4\,c_{H}^{\sigma}=-2/3\,$.

The second contribution comes from integrating out the heavy vector triplet. The procedure has been described in detail in Ref.~\cite{Low:2009di}, and we simply apply it to the case under study, obtaining $c_{H}^{v}=1/4$ and $c_{r}^{v}= -1\,$, in agreement with Ref.~\cite{Low:2010mr}.

The third and last contribution arises from integrating out heavy scalars. Since we also need to compute $c_{6}$, we write down the scalar potential up to order $H^{6}\,$. Neglecting $g'$, the relevant terms are
\beq \nonumber
V=&\!\!\!c_{+}\left\{f^{2}\left|\varphi_{ij}+\frac{i}{4f}(H_{i}H_{j}+H_{j}H_{i})\right|^{2}-\frac{1}{6f^{2}}|H|^{6}+\frac{i}{2f}|H|^{2}\left(\varphi_{ij}H^{\ast}_{i}H^{\ast}_{j}-\varphi^{\ast}_{ij}H_{i}H_{j}\right)-\frac{4}{3}|\varphi_{ij}|^{2}|H|^{2} \right\} \\ \nonumber
+&\!\!\!c_{-}\left\{f^{2}\left|\varphi_{ij}-\frac{i}{4f}(H_{i}H_{j}+H_{j}H_{i})\right|^{2}-\frac{1}{6f^{2}}|H|^{6}-\frac{i}{2f}|H|^{2}\left(\varphi_{ij}H^{\ast}_{i}H^{\ast}_{j}-\varphi^{\ast}_{ij}H_{i}H_{j}\right)-\frac{4}{3}|\varphi_{ij}|^{2}|H|^{2}\right\}\,,
\eeq
where the coefficient $c_{+}$ receives contributions from $g_{1}$, whereas $c_{-}$ from $g_{2}$ and from the top Yukawa sector. In general, starting from a Lagrangian of the form 
\begin{equation} 
\mathcal{L}_{\Phi}=\,-\Phi^{a\,\ast}\square \Phi^{a}-(M^{2}-\beta_{2}|H|^{2})\Phi^{a\,\ast}\Phi^{a}+\left(\beta f \Phi^{a\,\ast}H^{T}\epsilon \frac{\sigma^{a}}{2}H + \mathrm{h.c.}\right)+ \left(\frac{\beta_{4}}{f}\Phi^{a\,\ast}H^{T}\epsilon\frac{\sigma^{a}}{2}H|H|^{2}+\mathrm{h.c.}\right)
\end{equation}
$(\epsilon = i\sigma^{2})$ and integrating out $\Phi$ one obtains $c_{H}^{s}=\beta^{2}f^{4}/(2M^{4})$ and  $c_{r}^{s}=2\beta^{2}f^{4}/M^{4}\,.$ In addition, there is a contribution to $c_{6}$,
\beq
c_{6}^{\Phi}=-\frac{1}{\lambda}\left(\frac{\beta\beta_{4}f^{2}}{M^{2}}+\frac{\beta^{2}\beta_{2}f^{4}}{2M^{4}}\right)\,,
\eeq
where $\lambda$ is the Higgs quartic coupling (after the triplet has been integrated out). In the Littlest Higgs case we make the identifications
\beq 
M^{2}=(c_{+}+c_{-})f^{2}\,,\qquad \beta = \frac{1}{\sqrt{2}}(c_{-}-c_{+})\,,\qquad 
\beta_{2}=\frac{4}{3}(c_{+}+c_{-})\,, \qquad
\beta_{4}=\frac{1}{\sqrt{2}}(c_{+}-c_{-})\,.
\eeq
Therefore we find 
\beq
c_{H}^{s}=\frac{1}{4}\left(\frac{c_{-}-c_{+}}{c_{-}+c_{+}}\right)^{2}\,,\qquad c_{r}^{s}=\left(\frac{c_{-}-c_{+}}{c_{-}+c_{+}}\right)^{2}\,.
\eeq
On the other hand, 
\beq
c_{6}=-\frac{1}{\lambda}\left(\frac{c_{+}+c_{-}}{6}\right)+c_{6}^{\Phi}= -\,\frac{2}{3}\,,
\eeq
where we have used the expression of the quartic coupling
$\lambda=c_{+}c_{-}/(c_{+}+c_{-})\,$. Notice that in general the
neutral component of $\varphi$ gets a nonzero VEV, which is strongly
constrained by EWPT. Small values of $f\sim 1\,\mathrm{TeV}$ in fact
require the approximate condition $c_{+}\simeq c_{-}$ to be satisfied, which makes the triplet VEV very small\footnote{When $c_{+}=c_{-}\,$ the potential does not contain any tadpole for $\varphi\,$.} \cite{Csaki:2003si}. We assume this condition to be realized, and therefore neglect effects due to the triplet VEV in our discussion.

Concerning the top sector, in addition to the doublet $q_{L}=\,(\,t_{L}\,,\,b_{L}\,)^{T}$ and to the singlet $t_{R}$ a pair of $SU(2)$-singlet fermions $\tilde{T}_{L},\tilde{T}_{R}\,$ with electric charge $Q=Y=2/3$ is introduced. The Yukawa Lagrangian then reads
\begin{equation}
-\mathcal{L}_{Y}=\frac{\lambda_{1}}{2}f\,\overline{t}_{R} \epsilon_{ijk}\epsilon_{ab}\chi_{i}\Sigma_{ja}\Sigma_{kb}+\lambda_{2}f \overline{\tilde{T}}_{R}\tilde{T}_{L} + \mathrm{h.c.}
\end{equation}
($i,j,k = 1,2,3$ and $a,b = 4,5$). Here $\chi$ is an $SU(3)$ triplet, $\chi = \,(\,b_{L}\,,\,t_{L}\,,\,\tilde{T}_{L}\,)^{T}.$ The fermion mass matrix in the Higgs background
\begin{equation}
H=\begin{pmatrix} 0 \\
H/\sqrt{2} \end{pmatrix}
\end{equation}
reads
\begin{equation}
-\mathcal{L}_{m}=\begin{pmatrix} \overline{t}_{R} & \overline{\tilde{T}}_{R} \end{pmatrix} 
 \mathcal{M} \begin{pmatrix} t_{L} \\
\tilde{T}_{L} \end{pmatrix} + \mathrm{h.c.}\,,\qquad \mathcal{M}=\begin{pmatrix} -\frac{i}{\sqrt{2}}\lambda_{1}f\sin\left(\frac{\sqrt{2}H}{f}\right) & \lambda_{1}f\cos^{2}\left(\frac{H}{\sqrt{2}f}\right) \\
0 & \lambda_{2}f \end{pmatrix}\,,
\end{equation}
implying
\begin{equation}
\det\mathcal{M}^{\dagger}\mathcal{M}=\frac{1}{2}\lambda_{1}^{2}\lambda_{2}^{2}f^{4}\sin^{2}\frac{\sqrt{2}H}{f}\,.
\end{equation}
This allows us to write the amplitude for $gg\to hh$ in the Littlest
Higgs, in the low-energy theorem limit as
\beq
C_{\textrm{LET}}^{\mathrm{LH}}(\hat{s})=\frac{3m_{h}^{2}}{\hat{s}-m_{h}^{2}}\left[1-\frac{3}{4}\xi\left(\frac{7}{3}+\left(\frac{c_{-}-c_{+}}{c_{-}+c_{+}}\right)^{2}\right)\right]-1-\frac{\xi}{4}\left(1+\left(\frac{c_{-}-c_{+}}{c_{-}+c_{+}}\right)^{2}\right)\,. \label{eq:cletlh}
\eeq
We also note that in this case it is easy to diagonalize explicitly the fermion mass matrix at $\mathcal{O}(1/f^{2})$, obtaining
\begin{equation}
-\mathcal{L}_{m}=m_{t}(H)\,\overline{t}_{R}t_{L}+m_{T}(H)\,\overline{T}_{R}T_{L} + \mathrm{h.c.}
\end{equation}
where
\begin{align} \label{t-mass}
m_{t}(H)\,=&\,\frac{y_{t}H}{\sqrt{2}}\left[1+\frac{H^{2}}{f^{2}}\left(-\frac{1}{3}+\frac{y_{t}^{2}}{4\lambda_{T}^{2}} \right)\right]\,, \\ \label{T-mass}
m_{T}(H)\,=&\,f \lambda_{T}\left(1-\frac{H^{2}}{f^{2}}\frac{y_{t}^{2}}{4\lambda_{T}^{2}}\right)\,,
\end{align}
with $y_{t}=\sqrt{2}\,\lambda_{1}\lambda_{2}/\sqrt{\lambda_{1}^{2}+\lambda_{2}^{2}}\,\,$ and $\,\lambda_{T}=\sqrt{\lambda_{1}^{2}+\lambda_{2}^{2}}\,$. In the notation of Eq.~\eqref{field-dep masses}, we have 
\begin{equation} \label{cyt and aT in LH}
c_{y}^{(t)}=\frac{2}{3}-\frac{y_{t}^{2}}{2\lambda_{T}^{2}}\,,\qquad a_{T}=-\frac{y_{t}^{2}}{4\lambda_{T}^{2}}\,.
\end{equation}
Thus $c_y^{(t)}-2 a_T = 2/3 = \mathrm{const.}$, as must be the case since the factorization in Eq.~\eqref{indep condition} holds.
\subsection{MCHM4}\label{AppB:mchm4}
Similarly to what we did for MCHM5, we can apply directly
Eq.~\eqref{hngg} to derive the $hgg$ and $hhgg$ effective couplings at
all orders in $\xi$. For more details about the model we refer the reader to
Ref.~\cite{singlehiggsindep}, the notation of which we adopt here. The composite fermions are embedded into $SO(5)$ spinors $\mathbf{4}_{1/6}$ as
\beq
\psi_{L}=\begin{pmatrix}q_{L\,1} \\-i\,T_{L} \\-i\,B_{L}\end{pmatrix}\,,\qquad \psi_{R}=\begin{pmatrix} Q_{R} \\ i\,t_{R\,1} \\
i\,b_{R\,1}\end{pmatrix}\,,
\eeq
where $q_{L\,1}=(t_{L\,1}, b_{L\,1})^{T}$ and $Q_{R}=(T_{R}, B_{R})^{T}$ are $SU(2)_{L}$ doublets, and $T_{L}, B_{L}, t_{R\,1}, b_{R\,1}$ are singlets. In addition, an elementary doublet $q_{L\,2}=(t_{L\,2}, b_{L\,2})^{T}$ and singlet $t_{R\,2}$ are present. The fermion Lagrangian reads
\beq
-\mathcal{L}_{f}= y\,\overline{\psi}_{L}\Gamma^{M}\Phi_{M}\psi_{R} + f\lambda_{q}\overline{q}_{L,2}Q_{R}+f\lambda_{t}\overline{T}_{L}t_{R\,2}+\mathrm{h.c.}\,,
\eeq
where $\Phi = f (0, 0, 0, \sin(H/f), \cos(H/f))\,,$ and $\Gamma^{M}$ are the Gamma matrices of $SO(5)\,$. The mass matrix reads
\beq
-\mathcal{L}_{m}= \begin{pmatrix} \overline{t}_{L\,1} & \overline{t}_{L\,2} & \overline{T}_{L} \end{pmatrix} \mathcal{M}\begin{pmatrix} t_{R\,1} \\ t_{R\,2} \\ T_{R} \end{pmatrix} + \mathrm{h.c.}\,,\qquad \mathcal{M}=f\begin{pmatrix} y\sin(H/f) & 0 & y\cos(H/f) \\
0 & 0 & \lambda_{q} \\
y\cos(H/f) & \lambda_{t} & -y\sin(H/f) \end{pmatrix}\,,
\eeq
from which we find 
\beq
\det\mathcal{M}^{\dagger}\mathcal{M}=\lambda_{q}^{2}\lambda_{t}^{2}y^{2}f^{6}\sin^{2}(H/f) \quad \mbox{and} \quad A_{1}=\frac{2}{v}\sqrt{1-\xi}\,,\qquad A_{2}=-\frac{2}{v^{2}}\;. 
\eeq
Finally recalling the expression of the Higgs trilinear coupling in MCHM4, $\mathcal{L}_{h^{3}}=-(m_{h}^{2}/2v)h^{3}\sqrt{1-\xi}\,$, we find the amplitude for Higgs pair production via gluon fusion in the low-energy theorem approximation
\beq
C_{\textrm{LET}}^{MCHM4}(\hat{s})=\frac{3m_{h}^{2}}{\hat{s}-m_{h}^{2}}(1-\xi)-1\,. \label{eq:cletmchm4}
\eeq
\section{The {\boldmath $\chi^2$} test for electroweak precision
  observables} \label{appendix:chisquared}

We discuss here in detail the $\chi^2$ test used to constrain the parameters of the MCHM5 described in section~\ref{sec:EWPT}. The best experimental determination of $\epsilon_1$, $\epsilon_3$ and $\epsilon_b$ still comes from the precision measurements at the $Z$ pole mass at LEP~\cite{ALEPH:2005ab}:
\begin{equation}
	\begin{array}{lcl}
		\epsilon_1^{(exp)} & = & \left( 5.4 \pm 1.0 \right) \cdot 10^{-3}, \\
		\epsilon_2^{(exp)} & = & \left( -8.9 \pm 1.2 \right) \cdot 10^{-3}, \\
		\epsilon_3^{(exp)} & = & \left( 5.34 \pm 0.94 \right) \cdot 10^{-3}, \\
		\epsilon_b^{(exp)} & = & \left( -5.0 \pm 1.6 \right) \cdot 10^{-3}, \\
	\end{array}
	\hspace{0.8cm}
	\rho = \left( \begin{array}{cccc}
		1 & 0.60 & 0.86 & 0.00 \\
		0.60 & 1 & 0.40 & -0.01 \\
		0.86 & 0.40 & 1 & 0.02 \\
		0.00 & -0.01 & 0.02 & 1
	\end{array} \right).
\end{equation}
Here $\rho$ is the correlation matrix between the $\epsilon_i$ obtained from the App.~E of Ref.~\cite{ALEPH:2005ab}, marginalizing over the three parameters $m_Z$, $\alpha_S(m_Z)$ and $\Delta\alpha^{(5)}_{had}(m_Z)$.\footnote{Alternatively, we could set the three extra parameters to their experimental best values. This would give slightly more stringent constraints.} The status of electroweak precision observables did not change since then, except for the mass of the $W$ boson. The latter was recently updated based on Tevatron results~\cite{Aaltonen:2012bp, Abazov:2012bv}, and the new world average is now~\cite{newWmass}
\begin{equation}
	m_W = 80.385 \pm 0.015~\textrm{GeV}.
\end{equation}
The parameter $\epsilon_2$ is the only one depending on the mass of the $W$ boson, through the term
\begin{equation}
	\epsilon_2 = \frac{s_0^2}{1- 2 s_0^2} \, \Delta r_w + \left[
          \textrm{terms independent of}~m_W \right] \;,
\end{equation}
where $s_0^2 = 0.23098$, and the measurement of $\Delta r_w$ is related to $m_W$ through
\begin{equation}
	\frac{\pi \,\alpha(0)}{\sqrt 2 \, m_Z^2 \, G_F} \left( 1 - \Delta\alpha - \Delta r_w \right)^{-1}
		= \frac{m_W^2}{m_Z^2} \left( 1 - \frac{m_W^2}{m_Z^2} \right).
\end{equation}
Here $\alpha(0)$ is the fine-structure constant and $G_F$ the Fermi
constant, both known to high accuracy. Furthermore, $\Delta\alpha$ accounts for the running of the electroweak coupling between the low energy limit and the $Z$-pole mass. The uncertainty associated to it is important, but the shift in $\Delta r_w$ induced by the new value of $m_W$ is independent of $\Delta\alpha$. The change of $m_W$ and consequently $\epsilon_2$ between the LEP data of 2006 and the present value is then
\begin{equation}
	\begin{array}{|c|c|c|}
		\hline
		& 2006 & 2012 \\
		\hline
		m_W & 80.425 \pm 0.034~\textrm{GeV} & 80.385 \pm 0.015~\textrm{GeV} \\
		\hline
		\epsilon_2 & \left( -8.9 \pm 1.2 \right) \cdot 10^{-3} & \left( -7.9 \pm 0.9 \right) \cdot 10^{-3} \\
		\hline
	\end{array}
\end{equation}
The experimental values for the $\epsilon_i$ used in this paper are therefore
\begin{equation}
	\begin{array}{lcl}
		\epsilon_1^{(exp)} & = & \left( 5.4 \pm 1.0 \right) \cdot 10^{-3}, \\
		\epsilon_2^{(exp)} & = & \left( -7.9 \pm 0.9 \right) \cdot 10^{-3}, \\
		\epsilon_3^{(exp)} & = & \left( 5.34 \pm 0.94 \right) \cdot 10^{-3}, \\
		\epsilon_b^{(exp)} & = & \left( -5.0 \pm 1.6 \right) \cdot 10^{-3}, \\
	\end{array}
	\hspace{0.8cm}
	\rho = \left( \begin{array}{cccc}
		1 & 0.80 & 0.86 & 0.00 \\
		0.80 & 1 & 0.53 & -0.01 \\
		0.86 & 0.53 & 1 & 0.02 \\
		0.00 & -0.01 & 0.02 & 1
	\end{array} \right),
\end{equation}
where we took into account the fact that $\epsilon_{1,3,b}$ and their covariances with $\epsilon_2$ are not affected by the new measurement of the $W$ mass.

On the theoretical side, the $\epsilon_i$ are predicted to take the values~\cite{Agashe:2005dk}
\begin{equation}
	\begin{array}{lcl}
		\epsilon_1^{(th)} & = & \left[ + 5.66 - 0.86 \log\left( m_h / m_Z \right) \right] \cdot 10^{-3}
			+ \Delta\epsilon_1^{\textrm{IR}} + \Delta\epsilon_1^{fermions} \\
		\epsilon_2^{(th)} & = & \left[ -7.11 + 0.16 \log\left( m_h / m_Z \right) \right] \cdot 10^{-3} \\
		\epsilon_3^{(th)} & = & \left[ + 5.25 +0.54 \log\left( m_h / m_Z \right) \right] \cdot 10^{-3} 
			+ \Delta\epsilon_3^{\textrm{IR}} + \Delta\epsilon_3^{\textrm{UV}} \\
		\epsilon_b^{(th)} & = & -6.48  \cdot 10^{-3} + \Delta\epsilon_b^{fermions}
	\end{array}
\end{equation}
where the first numbers are the Standard Model corrections and the
remaining contributions are given in Section~\ref{sec:EWPT} for the
MCHM5. For the computation, we used a top mass $m_{t}=173.3$~GeV and a Higgs mass of $m_h =
125$~GeV. The $\chi^2$ test is then defined as 
\begin{equation}
	\chi^2 \left( \xi, \phi_L, \phi_R, R, m_\rho \right) = \sum_{i,j} \left( \epsilon_i^{(th)} - \epsilon_i^{(exp)} \right) C^{-1}_{ij}
		\left( \epsilon_j^{(th)} - \epsilon_j^{(exp)} \right),
	\label{eq:chisquared}
\end{equation}
where $C^{-1}$ is the inverse of the covariance matrix
\begin{equation}
	C_{ij} = \Delta\epsilon_i^{(exp)} \rho_{ij} \, \Delta\epsilon_j^{(exp)}.
\end{equation}
As indicated in Eq.~(\ref{eq:chisquared}), the $\chi^2$ depends on
five parameters. However, $\xi$ carries a different meaning than the other parameters, since it provides a measure of the fine-tuning of the model. The absolute minimum of the $\chi^2$ is in particular obtained for a very small value of $\xi$, which is highly unnatural. We compute therefore the minimum of the $\chi^2$ for a fixed value of $\xi$, and require
\begin{equation}
	\left. \chi^2\left( \phi_L, \phi_R, R, m_\rho \right)
        \right|_{\xi} - \left. \chi^2_{min} \right|_{\xi} \leq 13.28 \;.
\end{equation}
The value on the right-hand side corresponds to a confidence level of 99\% with four degrees of freedom ($\phi_L$, $\phi_R$, $R$ and $m_\rho$). For the values quoted in section~\ref{sec:EWPT}, we have
\begin{equation}
	\left. \chi^2_{min} \right|_{\xi=0.25} \cong 0.98,
	\hspace{1cm}
	\left. \chi^2_{min} \right|_{\xi=0.1} \cong 0.85 \;.
\end{equation}
Note that the minimum of the $\chi^2$ in the MCHM5 is significantly lower than in the Standard Model, $\chi^2_{SM} = 5.03$, which is expected due to the larger number of fitting parameters.

\section{Partial decay widths of heavy fermions in MCHM5}\label{App:partialwidths}

In this section we collect the formulae for the partial decay widths of fermionic resonances in MCHM5. We start by defining the relevant couplings. We denote by $U_{L,R}$ the transformations that diagonalize the mass matrix in the top sector,
\begin{equation}
\mathcal{M}\to
U_{L}^{T}\,\mathcal{M}\,U_{R}=\mathrm{diag}(m_{1},m_{2},m_{3},m_{4}) \;,
\end{equation} 
where $\mathcal{M}$ is the mass matrix in the basis where the rotations in Eq.~\eqref{PCrotations} have already been performed.\footnote{The mass eigenstates $f_{iL,R}\,\,(i=1,\ldots,4)$ are ordered by decreasing mass, so $f_{4}$ is identified with the top, and $f_{3}$ with the lightest top partner.} On the other hand, $\bar{G}_{hf\bar{f}}$ is the Yukawa coupling matrix after the rotations in Eq.~\eqref{PCrotations}, and $G_{L,R}^{Z}$ are
the matrices containing the couplings of the fermions to the $Z$ boson, 
\begin{align}
G_{L}^{Z}\,=\,& \mathrm{diag}\left( \,\frac{1}{2}-\frac{2}{3}s^{2}_{w}\,,\,\,\, \frac{1}{2}-\frac{2}{3}s^{2}_{w}\,,\,\,\,  -\frac{1}{2}-\frac{2}{3}s^{2}_{w}\,,\,\,\, -\frac{2}{3}s^{2}_{w}\,\right)\,, \\
  G_{R}^{Z}\,=\,&\mathrm{diag} \left(\,-\frac{2}{3}s^{2}_{w}\,,\,\,\, \frac{1}{2}-\frac{2}{3}s^{2}_{w}\,,\,\,\, -\frac{1}{2}-\frac{2}{3}s^{2}_{w}\,,\,\,\, -\frac{2}{3}s^{2}_{w}\,\right)\,,
\end{align}
for left-handed and right-handed fields, respectively (the ordering of
the fields is understood to be that of Eq.~\eqref{fermionmassmatrix}). Notice that the rotations in Eq.~\eqref{PCrotations} leave $G_{L,R}^{Z}$ invariant, because they only mix states with the same EW quantum numbers. 

\subsection{Charge \boldmath{$2/3$} states}
We have for the lightest top partner $\psi$ the partial decay width
into a $Z$ boson and a top quark 

\begin{align} \nonumber
\Gamma(\psi\to Zt)=\,\frac{M_{\psi}}{32\pi}\sqrt{\zeta_{Zt}} \Big\{(\lambda_{Z L}^{2}+\lambda_{Z R}^{2})\left(\frac{M_{\psi}^{2}}{m_{Z}^{2}}\right) & \left[  \frac{m_{Z}^{2}(M_{\psi}^{2}+m_{t}^{2})+(M_{\psi}^{2}-m_{t}^{2})^{2}-2m_{Z}^{4}}{M_{\psi}^{4}}\right] \\
-&\,12\,\frac{m_{t}}{M_{\psi}}\lambda_{Z L}\lambda_{Z R}\Big\} \;,
\end{align}
where 
\begin{equation}
\zeta_{Zt}=1-2\frac{m_{t}^{2}+m_{Z}^{2}}{M_{\psi}^{2}}+\frac{(m_{t}^{2}-m_{Z}^{2})^{2}}{M_{\psi}^{4}}
\end{equation}
and
\begin{equation}
\lambda_{Z L}=g_{Z}(U_{L}^{T}G_{L}^{Z}U_{L})_{34}\,,\qquad \lambda_{Z R}=g_{Z}(U_{R}^{T}G_{L}^{Z}U_{R})_{34}\,,
\end{equation}
where $g_{Z}\equiv g/\cos\theta_w$.
From the Yukawa Lagrangian Eq.~\eqref{Yukawas} we can extract the
leading-order couplings of $\tilde{T},T,X^{2/3}\,$. They are given by
\begin{equation} \label{ZcouplingsET1}
\lambda_{Z L}^{\tilde{T}}=\frac{y}{\sqrt{2}} s_{L}c_{R}\frac{m_{Z}}{M_{\tilde{T}}}\,,\quad \lambda_{Z R}^{\tilde{T}}=0\,; \qquad \lambda_{Z L}^{X^{2/3}}=0\,, \quad \lambda_{Z R}^{X^{2/3}}= \frac{y}{\sqrt{2}}s_{R}\frac{m_{Z}}{M_{X^{2/3}}}\,;
\end{equation}
\begin{equation} \label{ZcouplingsET2}
\lambda_{Z L}^{T}=0\,,\quad \lambda_{Z R}^{T}=\frac{y}{\sqrt{2}}s_{R}c_{L}\frac{m_{Z}}{M_{T}}\,.
\end{equation}
On the other hand for the decay $\psi \to Wb$ we find (neglecting $m_{b}$)
\begin{equation}
\Gamma(\psi\to Wb)=\frac{M_{\psi}}{32\pi}\lambda_{W L}^{2}\left(\frac{M_{\psi}^{2}}{m_{W}^{2}}\right)\left(1-3\frac{m_{W}^{4}}{M_{\psi}^{4}}+2\frac{m_{W}^{6}}{M_{\psi}^{6}}\right)\,,
\end{equation}
where $\lambda_{W L} = (g/\sqrt{2})(U_{L})_{13}\,$.
The leading order couplings read
\begin{equation} \label{WcouplingsET}
\lambda_{W L}^{\tilde{T}}=\, y s_{L}c_{R}\frac{m_{W}}{M_{\tilde{T}}}\,; \qquad \lambda_{W L}^{X^{2/3}}= 0\,;\qquad \lambda_{W L}^{T}= 0\,. 
\end{equation}
For the decay $\psi\to ht$ we find
\begin{equation}
\Gamma(\psi\to
ht)=\frac{M_{\psi}}{32\pi}\sqrt{\zeta_{ht}}\Big[(\lambda_{hL}^{2}+\lambda_{hR}^{2})\left(1+\frac{m_{t}^{2}}{M_{\psi}^{2}}-\frac{m_{h}^{2}}{M_{\psi}^{2}}\right)+4\frac{m_{t}}{M_{\psi}}\lambda_{hL}
\lambda_{hR}\Big] \;,
\end{equation}
where
\begin{equation}
\zeta_{ht}=1-2\frac{m_{t}^{2}+m_{h}^{2}}{M_{\psi}^{2}}+\frac{(m_{t}^{2}-m_{h}^{2})^{2}}{M_{\psi}^{4}}
\end{equation}
and 
\begin{equation}
\lambda_{hL} = -y (U_{L}^{T}\bar{G}_{hf\bar{f}}\,U_{R})_{43}\; ,\qquad \lambda_{hR}= -y (U_{L}^{T}\bar{G}_{hf\bar{f}}\,U_{R})_{34}\,.
\end{equation}
The leading order couplings read
\begin{equation} \label{hcouplingsET1}
\lambda_{h L}^{\tilde{T}}= \frac{y}{\sqrt{2}} s_{L}c_{R}\,,\quad \lambda_{h R}^{\tilde{T}}=0\,; \qquad \lambda_{h L}^{X^{2/3}}=0\,, \quad \lambda_{h R}^{X^{2/3}}= \frac{y}{\sqrt{2}}s_{R}\,;
\end{equation}
\begin{equation}  \label{hcouplingsET2}
\lambda_{h L}^{T}=0\,,\quad \lambda_{h R}^{T}= \frac{y}{\sqrt{2}}s_{R}c_{L}\,.
\end{equation}

\subsection{\boldmath{$B$} and \boldmath{$X^{5/3}$}}

For $\chi = X^{5/3},B$ we find for the decay width into a $W$ boson
and a top quark
\begin{align} \nonumber
\Gamma(\chi\to W^{\pm}t)=\,\frac{M_{\chi}}{32\pi}\sqrt{\zeta_{Wt}} \Big\{(\lambda_{\chi L}^{2}+\lambda_{\chi R}^{2})\left(\frac{M_{\chi}^{2}}{m_{W}^{2}}\right) & \left[  \frac{m_{W}^{2}(M_{\chi}^{2}+m_{t}^{2})+(M_{\chi}^{2}-m_{t}^{2})^{2}-2m_{W}^{4}}{M_{\chi}^{4}}\right] \\
-&\,12\,\frac{m_{t}}{M_{\chi}}\lambda_{\chi L}\lambda_{\chi R}\Big\} \;,
\end{align}
where 
\begin{equation}
\zeta_{Wt}=1-2\frac{m_{t}^{2}+m_{W}^{2}}{M_{\chi}^{2}}+\frac{(m_{t}^{2}-m_{W}^{2})^{2}}{M_{\chi}^{4}}
\end{equation}
and 
\begin{equation}
\lambda_{X^{5/3} L} = \frac{g}{\sqrt{2}}(U_{L})_{34}\,,\quad \lambda_{X^{5/3} R} = \frac{g}{\sqrt{2}}(U_{R})_{34} 
\end{equation}
\begin{equation}
\lambda_{B L} = \frac{g}{\sqrt{2}}(U_{L})_{24}\,,\quad \lambda_{B R} = \frac{g}{\sqrt{2}}(U_{R})_{24}\,. 
\end{equation}
The leading order couplings are
\begin{equation} \label{X53BcouplingsET}
\lambda_{X^{5/3}L}=0\,,\quad \lambda_{X^{5/3}R}= y s_{R}\frac{m_{W}}{M_{5/3}}\,;\qquad \lambda_{BL}=0\,,\quad \lambda_{BR}= ys_{R}c_{L}\frac{m_{W}}{M_{B}}\,.
\end{equation}
The leading order expressions of the heavy fermion masses are reported in Eq.~\eqref{LO masses}. 

The formulae for the partial decay widths contained in this appendix reduce to those given in Ref.\cite{BiniETAL} when the approximations in Eqs.~(\ref{ZcouplingsET1}), (\ref{ZcouplingsET2}), (\ref{WcouplingsET}), (\ref{hcouplingsET1}), (\ref{hcouplingsET2}), (\ref{X53BcouplingsET}) are made.

\section{Analytical results for the $gg\to hh$ cross section in MCHM5 \label{app:analyt}}
We present here the analytical result for the partonic gluon fusion
cross section into two Higgs bosons, $\hat{\sigma}_{gg \to hh}$, in MCHM5.

\subsection{Notation}
The four-momenta of the gluons are denoted by $p_1$ and
$p_2$, and the four-momenta of the Higgs bosons by $p_3$ and $p_4$. All momenta are taken incoming. The Mandelstam variables $\hat{s},\hat{t},\hat{u}$ are given by
\begin{center}\begin{tabular}{c c c}
$\hat{s}=\left(p_1+p_2\right)^2$\hspace{1cm}&\hspace{1cm}$\hat{t}=\left(p_1+p_3\right)^2$\hspace{1cm}&\hspace{1cm}$\hat{u}=\left(p_2+p_3\right)^2\,.$
\end{tabular}\end{center}
The scalar integrals are defined as
\begin{eqnarray}
&& C_{ij}(m_1^2, m_2^2, m_3^2) =\nonumber\\
&&\qquad \int \f{\mathrm{d}^4 q}{i\pi^2}\f{1}{\left(q^2-m_1^2\right)(\left(q+p_i\right)^2-m_2^2)(\left(q+p_i+p_j\right)^2-m_3^2)}\\
 && D_{ijk}(m_1^2, m_2^2, m_3^2, m_4^2)=\nonumber\\
&&\qquad \int \f{\mathrm{d}^4 q}{i\pi^2}\f{1}{\left(q^2-m_1^2\right)(\left(q+p_i\right)^2-m_2^2)(\left(q+p_i+p_j\right)^2-m_3^2)
 (\left(q+p_i+p_j+p_k\right)^2-m_4^2)}\nonumber\;.
\end{eqnarray}
 The analytic expressions can be found in Refs.~\cite{'tHooft:1978xw}. 
They have been evaluated numerically in our code with the help of LoopTools \cite{looptools}.

\subsection{Tensor basis and projectors}
The tensor basis, which has been given in Ref.~\cite{Glover:1987nx}, reads
\begin{align}
 A_1^{\mu\nu}&=g^{\mu\nu}-\f{p_1^{\nu}\,p_2^{\mu}}{\left(p_1\cdot p_2\right)} \label{A1}\\
A_2^{\mu\nu}&=g^{\mu\nu}+\f{p_3^2\, p_1^{\nu}\, p^{\mu}_2}{p_{\sssty{T}}^2\left(p_1\cdot p_2\right)}
-\f{2\left(p_3\cdot p_2\right)p_1^{\nu}\,p_3^{\mu}}{p_{\sssty{T}}^2\left(p_1\cdot p_2\right)}
-\f{2\left(p_3\cdot p_1\right)p_3^{\nu}\,p_2^{\mu}}{p_{\sssty{T}}^2\left(p_1\cdot p_2\right)}
+\f{2p_3^{\mu}\,p_3^{\nu}}{p_{\sssty{T}}^2}\label{A2}
\end{align}
\begin{equation*}
\hspace{1.5cm}\text{with}\hspace{1cm} p_{\sssty{T}}^2=2\f{\left(p_1\cdot p_3\right)\left(p_2\cdot p_3\right)}{\left(p_1\cdot p_2\right)}-p_3^2
\end{equation*}
\begin{equation}
\text{and} \hspace{1cm} A_1\cdot A_2=0 \hspace{1cm}\text{and}
\hspace{1cm}A_1\cdot A_1=A_2\cdot A_2=2 \; . \label{projrel}
\end{equation}

\subsection{Triangle form factor}
The triangle form factor can be cast into the form
\begin{equation}
 F_{\triangle}(m)=2\left[2 m +\left(4 m^3-\hat{s}\, m\right)C_{12}(m^2, m^2, m^2)\right]
\end{equation}
and can be found in Ref.~\cite{Plehn:1996wb}. In the limit of large quark mass $m\gg \sqrt{\hat{s}} \sim m_h$ the triangle form factor simplifies to $F_{\triangle}=2 \hat{s}/(3 m)$. This is equivalent to applying the low-energy theorem.
The corresponding amplitude is given by 
\begin{equation}
 \mathcal{A}_{\triangle}=\frac{\alpha_s}{4 \pi}A_1^{\mu\nu}\epsilon_{\mu}^{a}\epsilon_{\nu}^{b}\delta_{ab}
 \sum_{i=1}^{4} \underbrace{\left(\frac{1}{\hat{s}-m_{h}^2}g_{hhh}\,
   g_{h\bar{q}_iq_i}+2\,g_{hh\bar{q}_i q_i}\right)}_{C_{i, \triangle}}F_{\triangle}(m_i)\;. \label{amptria}
\end{equation}
The couplings $g_{h\bar{q}_iq_i}$ and $g_{hh\bar{q}_i q_i}$ are the diagonal elements
obtained from the Higgs coupling matrices $y\,G_{hf\bar{f}}$ in Eq.~(\ref{eq:yukcoup}) and $y/(2\, f)\, G_{hhf\bar{f}}$ 
in Eq.~(\ref{eq:2h2qcoup}), respectively, after rotation to the mass eigenstate basis. The triple Higgs coupling $g_{hhh}$ 
is given in the MCHM5 by 
\begin{equation}\label{tripHig}
 g_{hhh}=\frac{3\, m_{h}^2}{v} \frac{1-2 \xi}{\sqrt{1-\xi}}\;.
\end{equation}
In the SM limit, in Eq.~\eqref{amptria} there is no sum over heavy top
partners, and we are only left with the top quark contribution with the
Higgs coupling to the tops given by  $g_{h\bar{t} t}=m_{t}/v$,
and $g_{hh\bar{t}t}=0$. The triple Higgs coupling in the SM limit can
be obtained from Eq.~\eqref{tripHig} by setting $\xi=0$.
\subsection{Box form factors}
In the box diagrams we can have spin $S_z=0$ and 2 gluon gluon
couplings. The matrix elements can therefore be written in terms of
two gauge invariant form factors. Furthermore, we have divided the
form factors for the boxes into the parts which do not 
involve a $\gamma_5$ and the parts which are proportional to two
$\gamma_5$ matrices. Couplings with a $\gamma_5$ arise only for Higgs
couplings to two different fermions (but of same flavour). The
diagrams including only one 
$\gamma_5$ vanish because of the sign flip of the coupling when the direction of the fermion line changes.  
The form factors have been calculated with FeynCalc
\cite{Mertig:1990an} and checked against Ref.~\cite{Figy:2011yu}. For
the limit $m_1\to m_2$ they agree with Ref.~\cite{Plehn:1996wb}. 
The form factors are UV-finite as the coefficients in front of the
UV-divergent one- and two-point functions $A_0$
and $B_0$ are anti-symmetric in $m_i$ and $m_j$ and vanish upon
summation over $i,j$. \s

We introduce the following abbreviations
\begin{alignat}{2}
 C_{12} &\equiv C_{12}(m_1^2, m_1^2, m_1^2)    \qquad  \qquad     	C_{13}&&\equiv C_{13}(m_1^2, m_1^2. m_2^2) \nonumber \\
 C_{14} &\equiv C_{14}(m_1^2, m_1^2, m_2^2)    \qquad  \qquad      C_{23}&&\equiv C_{23}(m_1^2, m_1^2, m_2^2) \nonumber \\
 C_{24} &\equiv C_{24}(m_1^2, m_1^2, m_2^2)    \qquad  \qquad      C_{34}&&\equiv C_{34}(m_1^2, m_1^2, m_2^2) \nonumber \\
 D_{123}&\equiv D_{123}(m_1^2, m_1^2, m_1^2, m_2^2)  \qquad D_{132}&&\equiv D_{132}(m_1^2,m_1^2, m_2^2, m_2^2) \nonumber \\
 D_{213}&\equiv D_{213}(m_1^2, m_1^2, m_1^2, m_2^2) \;.
\end{alignat}
The box form factors $F_\Box$ and $G_\Box$ associated with
spin 0 and spin 2, respectively, are then given by (contributions
which cancel by summing up all contributions are omitted)
\begin{eqnarray}
\label{eq:Aresults}
& &  \!\!\!\!\! \!\!   \!\!\!\!\! \!\! F_\Box(m_i,m_j)   \nonumber \\[0.1cm]  
&=&  \frac{2}{\hat{s}}  \Big[ 2\hat{s}   + 4 m_i^2 \hat{s}\, C_{12} +  \hat{s} ( (m_i+m_j)( 2 m_i^2(m_i+m_j)- m_i \hat{s})- m_i^2 (\hat{t}+\hat{u})  )(D_{123}+D_{132}+D_{213}) \nonumber \\[0.1cm]  &+&
 (m_h^2 - (m_i+m_j)^2)\big[ (\hat{t} - m_h^2) (C_{13} + C_{24}) +
  (\hat{u} - m_h^2) (C_{23} + C_{14}) \\[0.1cm]
&-&
 (\hat{t} \hat{u} - m_h^4 +\hat{s}(m_j^2-m_i^2)) D_{132} \big]  
\Big]  \nonumber
 \end{eqnarray}
 \begin{eqnarray}  
& & \!\!\!\!\! \!\!   \!\!\!\!\! \!\!    G_\Box(m_i,m_j)  \nonumber  \\ 
&=&  \frac{1}{\hat{t} \hat{u}  - m_h^4 } \Big[   (\hat{t}^2 + \hat{u}^2 - (4m_j^2+4 m_i m_j)(\hat{t}+\hat{u}) +4 (m_j-m_i)(m_i+m_j)^3+ 2 m_h^4)   \hat{s} C_{12} \nonumber \\[0.1cm]  &+&  
( m_h^4 + \hat{t}^2 - 2\hat{t}(m_i+m_j)^2)((\hat{t}-m_h^2)(C_{13} + C_{24}) - \hat{s}\hat{t} D_{213})\nonumber \\[0.1cm]  &+&  
( m_h^4 + \hat{u}^2  - 2\hat{u}(m_i+m_j)^2)((\hat{u}-m_h^2)(C_{23} + C_{14}) - \hat{s}\hat{u} D_{123})\nonumber \\[0.1cm]  &-&  (\hat{t}^2 + \hat{u}^2- 2 m_h^4)(\hat{t}+\hat{u}- 2(m_i+m_j)^2)C_{34} \nonumber \\[0.1cm]  &-& 
(\hat{t}+\hat{u}- 2(m_i+m_j)^2) ( (\hat{t} \hat{u}  - m_h^4)(m_i^2+m_j^2) + \hat{s} (m_i^2-m_j^2)^2)(D_{123}+D_{132}+D_{213})
 \Big] \nonumber \\[0.1cm] 
 \nonumber  \end{eqnarray}
and \begin{equation}
     F_{\Box,5}(m_i, m_j)=-F_{\Box}(m_i,-m_j), \hspace*{1cm} G_{\Box,5}(m_i, m_j)=-G_{\Box}(m_i, -m_j).
    \end{equation}
Here $F_{\Box,5}$ and $G_{\Box,5}$ denote the spin 0 and 2 box form
factors which are proportional to the Higgs couplings to quarks, $g_{
    h\bar{q}_i q_j, 5}$, containing a $\gamma_5$ matrix. 
In the large quark mass limit for $m_i=m_j$ the form factors reduce to
$F_{\Box}=- 2 \hat{s}/(3 m_i^2)$ and $G_{\Box}=0$.\footnote{For
  $m_i=m_j$ the couplings in front of the form factors $F_{\Box, 5}$
  and $G_{\Box, 5}$ vanish, so that in this case these form factors
  are not needed.} 
The spin 0 and spin 2 box amplitudes read
\begin{equation}
\mathcal{A}_{0,\Box}=\frac{\alpha_s}{4\, \pi}\epsilon_{\mu}^{a}\epsilon_{\nu}^{b}\delta_{ab} 
A_1^{\mu\nu}\left(\sum_{i=1}^4\sum_{j=1}^4  
  g_{ h\bar{q}_i q_j}^2 F_{\Box}(m_i, m_j)+
  \sum_{i=1}^4\sum_{j=1}^4 g_{
    h\bar{q}_i q_j, 5}^2 F_{\Box, 5}(m_i, m_j)\right)
\end{equation}
and
\begin{equation} 
\mathcal{A}_{2,\Box}=\frac{\alpha_s}{4\, \pi}\epsilon_{\mu}^{a}\epsilon_{\nu}^{b}\delta_{ab} 
A_2^{\mu\nu}\left(\sum_{i=1}^4\sum_{j=1}^4  g_{h\bar{q}_i q_j}^2
  G_{\Box}(m_i, m_j) + \sum_{i=1}^4\sum_{j=1}^4
  g_{h\bar{q}_i q_j, 5}^2 G_{\Box, 5}(m_i, m_j)\right).
\end{equation}
The couplings $g_{h \bar{q}_i q_j}$ and $g_{h \bar{q}_i q_j, 5}$ are given by
\begin{align}
g_{h \bar{q}_i q_j}&=\frac{y}{2} \left(\tilde{G}_{hff, ij}+\tilde{G}_{hff, ji}\right)\\
g_{h \bar{q}_i q_j, 5}&=\frac{y}{2} \left(\tilde{G}_{hff, ij}-\tilde{G}_{hff, ji}\right)\; ,
\end{align}
where $\tilde{G}_{hff, ij}$ denotes the ($i$th, $j$th) matrix 
element of the coupling matrix of Eq.~(\ref{eq:yukcoup}) in the mass
eigenstate basis.
Note that $g_{h \bar{q}_i q_j, 5}$ is antisymmetric in $i$ and $j$ and hence 
changes sign, if incoming and outgoing fermions in the vertex are interchanged. 
\par
The complete amplitude of the process is given by
\begin{equation}
 \mathcal{A}(gg\to hh)=\mathcal{A}_{\triangle}+\mathcal{A}_{0,\Box}+\mathcal{A}_{2,\Box}\;.
\end{equation}
\subsection{Expansion of the form factors \label{app:expansion}}
In the case where top partners are neglected, we can perform explicitly the expansion of the form factors in $1/m_{t}^{2}$, \emph{i.e.} for small external momenta (see Ref.~\cite{Hoogeveen:1985tf}), and go beyond the leading order, the latter corresponding to the LET result. We find
\begin{align}
 F_{\triangle}&=\frac{\hat{s}}{m_t}\left(\frac{2}{3}+\frac{7}{180}\frac{\hat{s}}{m_t^2}\right)\,,\\
F_{\Box}&=\frac{\hat{s}}{m_t^2}\left(-\frac{2}{3}-\frac{7}{30}\frac{m_h^2}{m_t^2}\right)\,,\\
G_{\Box}&=\frac{\hat{s}}{m_t^4}\frac{11}{90}\left(\frac{m_h^4-\hat{t}\hat{u}}{\hat{s}}\right)\;.
\end{align}
The partonic cross section then reads
 \begin{equation}
\begin{split}
\hat\sigma_{gg \to hh} =\,& \frac{\alpha_s^2}{1024 (2\pi)^3}\frac{1}{\hat{s}^2}
\int_{\hat{t}_-}^{\hat{t}_+} d\hat{t} \left[ \left|C_{\triangle}F_{\triangle}+ C_{\Box} F_{\Box}\right|^2+\left|C_{\Box} G_{\Box}\right|^2\right] \;
\end{split}
\end{equation}
with $\hat{t}_{\pm}$ given by Eq.~\eqref{t+-} and
\begin{equation}
\left|C_{\triangle}F_{\triangle}+ C_{\Box}
  F_{\Box}\right|^2+\left|C_{\Box} G_{\Box}\right|^2 =
\frac{\hat{s}^{2}}{v^{4}}\frac{4}{9}(c_{\triangle}-c_{\square})^2\left[1+\frac{1}{m_{t}^{2}(c_{\triangle}-c_{\square})}\left(c_{\triangle}\frac{7}{60}\hat{s}-c_{\square}\frac{7}{10}m_{h}^{2}\right)\right] \label{eq:exp1}
\end{equation}
\begin{equation}
c_{\triangle}=\frac{3m_{h}^{2}}{\hat{s}-m_{h}^{2}}\left(\frac{1-2\xi}{\sqrt{1-\xi}}\right)^{2}-4\xi\,,\qquad
c_{\square} = \left(\frac{1-2\xi}{\sqrt{1-\xi}}\right)^{2}\,, \label{eq:exp2}
\end{equation}
%
where we used \begin{equation}
    C_{\triangle}=\left(\frac{1}{\hat{s}-m_{h}^2}g_{hhh}\,
   g_{h\bar{t}t}+2\,g_{hh\bar{t}t}\right)\equiv \frac{m_{t}}{v^{2}}c_{\triangle}\,,\qquad\qquad C_{\Box}=g_{h\bar{t}t}^2\equiv \frac{m_{t}^{2}}{v^{2}}c_{\square}\;, 
    \end{equation}
with the couplings given by $g_{h\bar{t}t}=(m_t/v)(1-2\xi)/\sqrt{1-\xi}$, $g_{hh\bar{t}t}=-2 m_t \xi/v^{2}$ and $g_{hhh}$ as defined in Eq.~\eqref{tripHig}. The leading term in $1/m_{t}^{2}$ corresponds to the LET result, see Eqs.~(\ref{eq:partoniccxn}) and (\ref{C in mchm5}).

\end{appendix}
%
%
%

\end{document}